\documentclass{article} 

\usepackage[utf8]{inputenc}
\usepackage[T1]{fontenc}
\usepackage{lmodern}

\usepackage{booktabs}
\usepackage{tabularx}
\usepackage{multirow}
\newlength{\Oldarrayrulewidth}
\newcommand{\Cline}[2]{%
  \noalign{\global\setlength{\Oldarrayrulewidth}{\arrayrulewidth}}%
  \noalign{\global\setlength{\arrayrulewidth}{#1}}\cline{#2}%
  \noalign{\global\setlength{\arrayrulewidth}{\Oldarrayrulewidth}}}
  
\usepackage{amsfonts}
\usepackage{amsmath}
\usepackage{amssymb}
\usepackage{amsthm}
\usepackage[dvipsnames]{xcolor}
    \definecolor{darkgreen}{rgb}{0,0.5,0}
    \definecolor{darkblue}{rgb}{0,0,0.6}
    \definecolor{purple}{rgb}{0.4,.2,0.7}
\usepackage[margin = 2.5cm]{geometry}
\usepackage{graphicx}

\usepackage[numbers,sort]{natbib}

\usepackage[hyperfootnotes = false, colorlinks = true, linkcolor = darkblue, citecolor = purple]{hyperref}
\usepackage{subcaption}
\usepackage{ytableau}

\usepackage{csquotes}

\newcommand{\fnl}{f_{\rm NL}}
\newcommand{\cs}{{\cal{S}}}
   
\usepackage{tikz}
\usetikzlibrary{decorations.pathmorphing}
\usetikzlibrary{decorations.markings}
\definecolor{mathred}{RGB}{180,44,37}
\definecolor{mathblue}{RGB}{39,94,190}
\tikzset{>=latex} 
\tikzset{ photon/.style={decorate, decoration={snake}, draw=black}}
\tikzset{ heavy scalar/.style={line width=1.8pt}}

\usepackage{comment}

\usepackage{physics}

\renewcommand{\dd}{\mathrm{d}}
\newcommand{\be}{\begin{equation}}
\newcommand{\ee}{\end{equation}}
\newcommand{\bea}{\begin{eqnarray}}
\newcommand{\eea}{\end{eqnarray}}

\def\tr{\mathrm{tr}}

\usetikzlibrary{calc}

\begin{document}

\begin{flushright}
MIT-CTP/6107
\end{flushright}

\thispagestyle{empty}
\begin{center}
    ~\vspace{5mm}

  \vskip 2cm 
  
   {\LARGE \bf 
   On the Numerical Integration of One-Loop Cosmological Collider Signals
   }

   \vspace{0.5in}
     
   {\bf Aidan Herderschee$^{1,2}$ and Qianshu Lu$^{1,3}$}

    \vspace{0.2in}

   {\it and an appendix with} {\bf Michael Borinsky$^{4}$}

    \vspace{0.5in}
 
  $^{1}$Institute for Advanced Study, Princeton, NJ 08540, USA \\[1mm]
  $^{2}$MIT Center for Theoretical Physics - a Leinweber Institute, \\ Massachusetts Institute of Technology, Cambridge, MA 02139, USA\\[1mm]
  $^{3}$Center for Cosmology and Particle Physics, New York University, New York, NY 10003, USA \\[1mm]
  $^{4}$Perimeter Institute for Theoretical Physics, Waterloo, ON N2L 2Y5, Canada 
                
    \vspace{0.5in}

    \vspace{0.5in}

\end{center}
  
\vspace{0.5in}

\begin{abstract} 
In a broad class of inflationary models, the leading non-Gaussian bispectrum arises from one-loop rather than tree-level processes. However, realistic one-loop contributions remain largely unexplored for phenomenologically relevant masses beyond the simplest bubble diagrams. This is insufficient for observational purposes because, for generic masses and couplings, triangle contributions need not be suppressed relative to their bubble counterparts.
We present a numerical method for evaluating scalar one-loop in-in diagrams contributing to the inflationary bispectrum at fully general external momenta and masses. First, the Witten--Feynman parameterization reduces de Sitter loop integrals to generalized Euler--Mellin integrals governed by Symanzik graph polynomials. Unfortunately, the exponents of the polynomials in the integrand become complex for sufficiently heavy masses, leading to sign problems when evaluating the integral using standard numerical techniques. To remedy this problem, we introduce the reduced Schwinger method, which evaluates a highly oscillatory subintegral analytically to obtain a Gauss hypergeometric kernel, leaving the remaining integrals to standard numerical quadrature. We validate the algorithm by reproducing known analytic results for the tree-level bispectrum in terms of ${}_3F_2$ functions, and then apply it to the one-loop bubble and triangle contributions at general kinematics. This yields the first direct numerical evaluation of the complete one-loop bispectrum, valid across the full kinematic range. Using these numerical results, we construct bispectrum templates and compare them with CMB data. 

\end{abstract}

\vspace{1in}

\pagebreak

\setcounter{tocdepth}{3}
{\hypersetup{linkcolor=black}\tableofcontents}

\section{Introduction}\label{sec:introduction}

Inflation remains the most empirically successful framework for generating
primordial perturbations~\cite{Guth:1980zm,Linde:1981mu,Albrecht:1982wi,Starobinsky:1982ee}, yet the particle physics governing the inflationary
epoch is almost entirely unknown. 
The inflationary potential scale $V^{1/4}$ could be as high as $10^{16}$ GeV~\cite{Lyth:1996im}. Inflation thus provides a unique opportunity to probe high-energy physics well beyond the reach of terrestrial colliders and
tabletop experiments. Cosmological fluctuations in the form of observables such as the cosmic microwave background (CMB), large-scale structure (LSS), and 21-cm radiation constitute a direct snapshot of the dynamics and interactions present during this era, and the statistical properties of these fluctuations, in particular their non-Gaussian correlations, encode information about the inflationary Lagrangian that the power spectrum alone cannot capture. This observation has motivated the \emph{cosmological collider} program~\cite{Maldacena:2002vr,Creminelli:2004yq,Chen:2006nt,Cheung:2007st,Chen:2009we,Chen:2009zp,Arkani-Hamed:2015bza},
which aims to use the bispectrum and higher-point functions as spectroscopic tools for probing the mass spectrum and spin content of particles present during inflation~\cite{Noumi:2012vr,Lee:2016vti,Chen:2016uwp,Kumar:2017ecc,Kumar:2019ebj,Meerburg:2016zdz}.

The reach of this program, however, depends critically on the theoretical templates used to analyze the data. A specific template coherently combines the expected signal across many momentum configurations while averaging down the noise, making the search far more sensitive than a completely model-independent search for departures from Gaussian $\Lambda$CDM.\footnote{The template bank cannot be made arbitrarily large because of the look-elsewhere effect~\cite{Gross:2010qma}. Even if the data contain only noise, searching more distinguishable shapes increases the chance that one will have a large accidental correlation with the data. However, a bank of $10^2$--$10^3$ physically motivated, independent templates incurs only a modest statistical penalty of requiring a $\sim 6\sigma$ local signal to achieve a global $5\sigma$ detection.} The three-point correlation function of the curvature perturbation $\zeta$ at the end of inflation is conventionally parameterized as
\begin{equation}
    \langle \zeta(\vec{k}_1) \zeta(\vec{k}_2) \zeta(\vec{k}_3)\rangle \equiv (2\pi)^3\delta^{(3)}(\vec{k}_1+\vec{k}_2+\vec{k}_3)\frac{18}{5} \fnl (2\pi^2 A_s)^2 \frac{\cs(k_1, k_2, k_3)}{k_1^2 k_2^2 k_3^2},\label{eq:bispecdef}
\end{equation}
where $A_s \approx 2.1\times 10^{-9}$ is the observed amplitude of the dimensionless power spectrum~\cite{Planck:2018vyg}, $k_i=|\vec{k}_i|$, and $\cs(k_1, k_2, k_3)$ is the shape of the bispectrum~\cite{Babich:2004gb, Fergusson:2008ra,Senatore:2009gt}. Searching for a particular particle model in cosmological data thus requires an accurate $\cs$ template across the full physical kinematic space. This space obeys the triangle inequality $2\max(k_1,k_2,k_3)\leq k_1+k_2+k_3$ and lies within the dynamical range of $k$ probed by the data. Such a template allows the amplitude $\fnl$ to be constrained or measured. Moreover, shape functions of different microphysical origins can have a small ``inner product''~\cite{Babich:2004gb,Fergusson:2008ra,Senatore:2009gt}, so a constraint on $\fnl$ for one shape template does not in general apply to another. It is therefore crucial to develop full kinematic shape function templates from different underlying microphysics~\cite{Sohn:2024xzd,Suman:2025tpv,Kumar:2026ogn,Kumar:2026dih}.

Remarkably, a significant obstruction to further theoretical progress has been computational. Witten diagrams\footnote{Throughout this work, we use ``Witten diagrams'' synonymously
with the de~Sitter in-in diagrams contributing to late-time correlators.} in de Sitter space require nested integrals over bulk vertex times involving products of Bessel functions, making them difficult to evaluate analytically or numerically. At tree level, however, substantial progress has been made. The cosmological bootstrap program~\cite{Arkani-Hamed:2018kmz,Baumann:2019oyu,Baumann:2020dch,Arkani-Hamed:2023bsv,Arkani-Hamed:2023kig,Sleight:2019mgd,Sleight:2019hfp,Qin:2022fbv,Jazayeri:2022kjy,Qin:2023ejc,Aoki:2023wdc,Xianyu:2023ytd,Liu:2024str, Aoki:2024uyi, Xianyu:2025lbk} has produced powerful analytic results. On the numerical side, \texttt{CosmoFlow}~\cite{Werth:2023pfl,Pinol:2023oux,Werth:2024aui} provides a general-purpose numerical framework, and Ref.~\cite{An:2017hlx} enables numerical evaluation of complex topologies unexplored by analytic methods. Together, these developments have enabled searches for a wide range of cosmological collider signals in real cosmological data as of August, 2026.

\begin{figure}[t]
\centering
\begin{tikzpicture}[
    thick,
    inflaton/.style={solid, thick},
    vertex/.style={circle, fill=black, inner sep=1.5pt},
    bginsert/.style={draw, circle, inner sep=3pt, thick},
]

\fill[gray!15] (-4,3) ellipse (3.15cm and 0.62cm);
\draw[dashed, thick] (-4,3) ellipse (3.15cm and 0.62cm);

\coordinate (V1) at (-5.0, 0.4);
\coordinate (V2) at (-3.0, 0.4);

\coordinate (B1) at ({-4 + 3.15*cos(150)}, {3 + 0.62*sin(150)});
\coordinate (B2) at ({-4 + 3.15*cos(230)}, {3 + 0.62*sin(230)});
\coordinate (B3) at ({-4 + 3.15*cos(340)}, {3 + 0.62*sin(340)});

\draw[thin, solid, gray!70] (B1) -- (B2) -- (B3) -- cycle;

\draw[inflaton] (B1) -- (V1);
\node[anchor=east] at ($(B1)!0.35!(V1) + (-0.12,0)$) {$\vec{k}_1$};
\draw[inflaton] (B2) -- (V1);
\node at ($(B2)!0.55!(V1) + (0.3,0)$) {$\vec{k}_2$};
\draw[inflaton] (B3) -- (V2);
\node[anchor=west] at ($(B3)!0.5!(V2) + (0.12,0)$) {$\vec{k}_3$};

\fill (B1) circle (2pt);
\fill (B2) circle (2pt);
\fill (B3) circle (2pt);

\draw[heavy scalar] (V1) to[bend left=65] (V2);
\draw[heavy scalar] (V1) to[bend right=65] (V2);
\node at (-4, 1.25) {$\sigma$};
\node at (-4, -0.5) {$\sigma$};

\node[vertex] at (V1) {};
\node[vertex] at (V2) {};

\draw[thick] (V2) -- ($(V2)+(20pt,0)$);
\node[anchor=west] at ($(V2)+(20pt,0)$) {$\dot{\phi}_{0}$};
\draw plot[mark=x,mark size=3pt] coordinates {($(V2)+(20pt,0)$)};

\node[gray, anchor=west] at (-7.0, 0.2) { bulk};

\node at (-7.7, 3) {$(a)$};
\node at (-4, 3.95) {$\mathcal{I}^+\;\;(\eta \to 0^-)$};


\node at (0.3, 3) {$(b)$};

\fill[gray!15] (4,3) ellipse (3.15cm and 0.62cm);
\draw[dashed, thick] (4,3) ellipse (3.15cm and 0.62cm);

\coordinate (W1) at (4.0, 1.5);
\coordinate (W2) at (5.3, 0);
\coordinate (W3) at (2.7, 0);

\coordinate (C1) at ({4 + 3.15*cos(150)}, {3 + 0.62*sin(150)});
\coordinate (C3) at ({4 + 3.15*cos(30)},  {3 + 0.62*sin(30)});
\coordinate (Cm) at (4, {3 - 0.62});

\draw[thin, solid, gray!70] (C1) -- (Cm) -- (C3) -- cycle;

\draw[inflaton] (C1) -- (W3);
\node[anchor=east] at ($(C1)!0.5!(W3) + (-0.1,0)$) {$\vec{k}_1$};
\draw[inflaton] (Cm) -- (W1);
\node[anchor=west] at ($(Cm)!0.5!(W1) + (0.15,0)$) {$\vec{k}_2$};
\draw[inflaton] (C3) -- (W2);
\node[anchor=west] at ($(C3)!0.5!(W2) + (0.1,0)$) {$\vec{k}_3$};

\fill (C1) circle (2pt);
\fill (Cm) circle (2pt);
\fill (C3) circle (2pt);

\draw[heavy scalar] (W1) -- (W2);
\draw[heavy scalar] (W2) -- (W3);
\draw[heavy scalar] (W3) -- (W1);
\node at (4.0, -0.2) {$\sigma$};

\node[vertex] at (W1) {};
\node[vertex] at (W2) {};
\node[vertex] at (W3) {};

\draw[thick] (W1) -- ($(W1)+(0,-20pt)$);
\node[anchor=north] at ($(W1)+(0,-20pt)+(0,-0.1)$) {$\dot{\phi}_{0}$};
\draw plot[mark=x,mark size=3pt] coordinates {($(W1)+(0,-20pt)$)};

\draw[thick] (W2) -- ($(W2)+(20pt,0)$);
\node[anchor=west] at ($(W2)+(20pt,0)$) {$\dot{\phi}_{0}$};
\draw plot[mark=x,mark size=3pt] coordinates {($(W2)+(20pt,0)$)};

\draw[thick] (W3) -- ($(W3)+(-20pt,0)$);
\node[anchor=east] at ($(W3)+(-20pt,0)$) {$\dot{\phi}_{0}$};
\draw plot[mark=x,mark size=3pt] coordinates {($(W3)+(-20pt,0)$)};

\end{tikzpicture}
\caption{One-loop Witten diagrams contributing to the inflationary bispectrum from the interaction $(\partial^\eta \phi)(\partial_\eta \phi)\,\sigma^2$. Three external inflaton legs with momenta $\vec{k}_1$, $\vec{k}_2$, $\vec{k}_3$ originate from the late-time boundary $\mathcal{I}^+$ ($\eta \to 0^-$) of de Sitter space, forming a momentum triangle on the boundary slice. (a) In the bubble diagram, the legs $\vec{k}_1$ and $\vec{k}_2$ attach to a quartic bulk vertex, while $\vec{k}_3$ attaches to a second vertex carrying a background inflaton insertion ($\dot{\phi}_{0}$). The two bulk vertices are connected by a one-loop bubble of the heavy scalar $\sigma$ (thick lines). (b) In the triangle diagram, each of the three bulk vertices carries one external inflaton leg and one background insertion, and the vertices are connected by three $\sigma$ propagators forming a closed loop.}
\label{fig:bub_tri}
\end{figure}
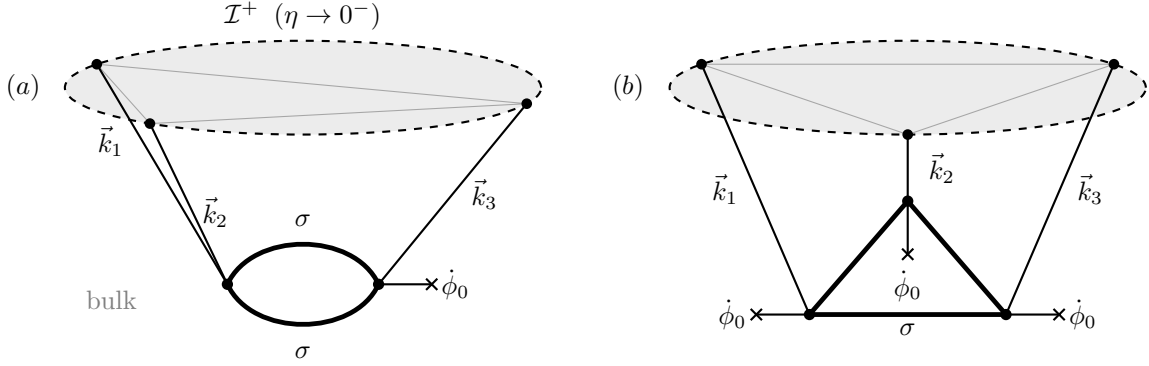

The situation at loop level is qualitatively different. Although the bootstrap has been extended to loop integrands~\cite{Baumann:2024mvm,Bhowmick:2025mxh} and loop integrals~\cite{Xianyu:2022jwk,Qin:2023bjk,Qin:2023nhv, XianyuQinTriangle}, closed-form results at general external kinematics remain limited to special cases, and no general-purpose numerical method currently exists for computing loop-level cosmological correlators. Most previous work has focused on the bubble contribution to the bispectrum. Analytic approximations in the squeezed limit were developed in Refs.~\cite{Chen:2016uwp,Chen:2016hrz,Chen:2016nrs}, while a numerical evaluation based on Pauli--Villars regularization was performed in Ref.~\cite{Wang:2021qez} for a gauge boson with a chemical potential. For principal-series masses, analytic series expansions in the squeezed ratio were subsequently derived in Refs.~\cite{Xianyu:2022jwk,Qin:2024gtr}, and related one-loop diagrams have been investigated more recently in Refs.~\cite{Zhang:2025nzd,Bodas:2025vpb,Aoki:2026vbc,Liu:2026jzn}. However, the triangle contribution remains largely unexplored compared with the bubble. For generic masses and spin, only squeezed ratio series expansion of the signal component is available~\cite{Qin:2023bjk, XianyuQinTriangle, You:2026xoq}.\footnote{For a conformally coupled scalar, whose propagators simplify the integrand dramatically, the triangle was evaluated analytically at generic kinematics in Ref.~\cite{Pimentel:2026kqc}, and a substantially more compact representation was subsequently obtained in Ref.~\cite{Henn:2026lfz}.} Thus, despite substantial progress on the bubble, the full one-loop bispectrum has remained unavailable except for models with specific masses or in specific kinematic regimes.

With tree-level calculations under substantial control and loop contributions usually subleading, one might ask whether one-loop results are phenomenologically necessary. A concrete motivation arises when a symmetry forbids couplings linear in the additional particle, as for a scalar charged under an internal symmetry or for any fermion~\cite{Bodas:2025vpb,Aoki:2026vbc}. In such theories, the leading cosmological-collider signal first appears at one loop. Computing the triangle contribution to the bispectrum is therefore not merely a formal exercise, but essential for constructing realistic templates for a broad class of models. 

To enable such calculations, we develop a general-purpose algorithm for the direct numerical evaluation of one-loop de Sitter correlators. The algorithm proceeds in two stages. In the first stage, we use the Witten--Feynman (WF) parameterization~\cite{Herderschee:2025znl} to reduce de Sitter Witten diagrams to generalized Euler--Mellin integrals~\cite{Gelfand:1990bua,MR3010271,MR3189470,Matsubara-Heo:2023ylc,Grimm:2024tbg} of the form
\begin{equation}
\int \prod_{i} dx_{i}, x_{i}^{\nu_{i}-1} \prod_{j} P_{j}(x_{i})^{-\delta_{j}} \ ,
\end{equation}
where the $P_j$ are polynomials and the exponents $\nu_i$ and $\delta_j$ depend on the spacetime dimension and particle masses. The WF parameterization generalizes the familiar Feynman parameterization of flat-space amplitudes~\cite{Cardona:2017tsw} and have a polynomial structure governed by the same Symanzik graph polynomials familiar from the amplitudes literature~\cite{Heckelbacher:2022hbq}.

In the second stage, we consider the actual numerical integration 
of the generalized Euler integrals produced by WF parameterization. Numerical evaluation of the integral is complicated by the appearance of an annoying sign problem. For sufficiently heavy masses, the exponents of the generalized Euler--Mellin integrand acquire imaginary parts and the integrand oscillates in the logarithms of the Schwinger variables. Crucially, for the integrals we consider, the difficulty is largely concentrated along a single direction in Schwinger space, which we call the common-scale direction. Our new integration method, which we term the reduced Schwinger method,\footnote{\label{fableusedec}The reduced Schwinger method was developed by the AI model Claude Fable 5 (Anthropic). An earlier, unreleased draft of this paper used the tropical Monte Carlo method of Appendix~\ref{sec:tropdecomp} as its primary integration method. When we asked the model to devise an independent integration method to cross-check those results, it developed the reduced Schwinger method, which proved efficient enough to become the primary method in this work.} isolates this direction through an exact change of variables and integrates it analytically. The result is a Gauss hypergeometric kernel that can be evaluated efficiently even though the original integrand was highly oscillatory. The remaining integration directions exhibit only bounded oscillation and are therefore suitable for deterministic quadrature. An independent cross-check is provided by the tropical Monte Carlo method, which we adapt to de Sitter integrals in Appendix~\ref{sec:tropdecomp}. We then combine our results for the bubble and triangle to make full one-loop templates which we compare with Planck data. 

To illustrate the power of this algorithm, we apply it to a toy model containing a massive real scalar $\sigma$ coupled to the inflaton through $c(\partial^{\eta}\phi)(\partial_{\eta}\phi)\sigma^{2}/\Lambda^2$.\footnote{Such couplings arise naturally from the K\"ahler potential in supersymmetric completions~\cite{Baumann:2011nk}, where operators of the form $\Phi^{\dagger}\Phi\,\Sigma^{\dagger}\Sigma/\Lambda^{2}$ generate derivative portals between the inflaton and heavy GUT-sector multiplets. Related effective field theory analyses of couplings of the inflaton or curvaton to heavy sectors can be found in Refs.~\cite{Kumar:2017ecc,Kumar:2019ebj,Assassi:2013gxa}.} We evaluate the full one-loop inflationary bispectrum at completely general external momenta and principal-series masses for this model for the first time. The leading contributions arise from the bubble and triangle diagrams shown in Fig.~\ref{fig:bub_tri}. Importantly, we show that the triangle need not be smaller than the bubble despite containing an additional interaction vertex and massive propagator~\cite{Chen:2016uwp,Lu:2019tjj,Wang:2019gbi}, so both diagrams are required for the full bispectrum prediction. We regulate the UV divergence of the bubble dimensionally and extract its divergent and finite parts separately through a subtraction at the level of the integrand. To cross-check the result, we reproduce the known analytic tree-level bispectrum expressed in terms of ${}_{3}F_{2}$ functions~\cite{Qin:2023ejc} and check the method against an analytic one-loop four-point result for the bubble~\cite{Xianyu:2022jwk}. We then evaluate the one-loop bubble and triangle contributions to the bispectrum and compare them with Planck data. 

Unfortunately, we find that the bispectrum templates produced by the toy model we consider have significant overlap with the equilateral template and therefore do not provide a useful signal of novel physics. In principle, we could simply project out the equilateral overlap to isolate the novel contribution, but this does not seem well-motivated from a physics standpoint for the model we consider.\footnote{A somewhat more physical prescription would be to choose a renormalization scheme that minimizes the overlap. However, this is equivalent to fine-tuning the Wilson coefficient of $(\partial\phi)^4$, which also seems unjustified.} Nonetheless, we are hopeful that the method proposed here could produce novel templates in other well-motivated models. 

This paper is organized as follows. In Sec.~\ref{sec:wittrop}, we review the WF parameterization, deriving the Schwinger representation of the de Sitter propagator, showing how Witten diagrams reduce to generalized Euler--Mellin integrals, and explaining how the $i\epsilon$ prescription determines the integration cycle and contour deformation. In Sec.~\ref{sec:redschwinger}, we develop the reduced Schwinger method, including the treatment of UV divergences in dimensional regularization by subtraction at the level of the reduced integrand. In Sec.~\ref{sec:treeanalyt}, we apply the reduced Schwinger method to the tree-level bispectrum as an illustrative cross-check against known analytic results. In Sec.~\ref{sec:bubble}, we compute the one-loop bubble contribution for general kinematics. In Sec.~\ref{sec:triangle}, we compute the one-loop triangle contribution, which arises from the same derivative portal but involves a topologically distinct diagram with different analytic structure. In Sec.~\ref{sec:CMBdata}, we use our numerical results to construct bispectrum templates valid over the full kinematic range and compare them with \textit{Planck} CMB data. We conclude in Sec.~\ref{sec:disc} with a discussion of extensions to higher loops, higher-point functions, and spinning exchanges. Appendix~\ref{sec:reviewinin} reviews the in-in formalism, Appendix~\ref{sec:tropdecomp}, written in collaboration with Michael Borinsky, presents the tropical Monte Carlo cross-check, and Appendix~\ref{sec:analyticcomparison} validates the reduced Schwinger method against the analytic bubble result in a particular kinematic limit.

\section{Witten--Feynman parameterization for de Sitter correlators}\label{sec:wittrop}

In this section, we review the derivation of the Witten--Feynman parameterization. The schematic procedure is the following. We first derive a Schwinger representation of the de Sitter two-point function of the massive scalar, which puts all spatial momentum dependence in a Gaussian. When all bulk propagators are written in the Schwinger form, the spatial loop momentum can then be trivially integrated. For the remaining integral over the Schwinger parameters, the $i\epsilon$ prescription of the propagators fixes the appropriate integration cycle, i.e., the contour on which these parameter integrals should be evaluated. Finally, we convert the Schwinger parameter integrals into Witten--Feynman form, where the integrand is a product of polynomials and thus conducive to the reduced Schwinger method for numerical integration which we will discuss in Sec.~\ref{sec:redschwinger}.

\subsection{Schwinger representation of the de Sitter two-point function}

In this section, we derive a Schwinger-parameter representation for the massive scalar two-point function in de Sitter space, and then obtain its momentum-space counterpart by Fourier transform. Our presentation closely follows the strategy of~\cite{Herderschee:2025znl}, where Schwinger-like representations provide the main computational input for WF parameterization.

We work in conformal coordinates, in which the de Sitter metric takes the form
\begin{equation}
ds^{2}=\frac{-d\eta^{2}+d\vec{x}^{2}}{\eta^{2}} \ ,
\end{equation}
where $\eta\in(-\infty,0)$ is the conformal time and $\vec{x}\in\mathbb{R}^{d}$ are the spatial coordinates. We set the Hubble scale to $H=1$ throughout. We consider the time-ordered bulk two-point function $G(x_1,x_2)$, which satisfies the Klein--Gordon equation with a covariant delta-function source,
\begin{equation}
(\nabla^{2}-m^{2})G(\eta_{1},\vec{x}_{1};\eta_{2},\vec{x}_{2})=i\frac{\delta(\eta_{1}-\eta_{2})\delta^{d}(\vec{x}_{1}-\vec{x}_{2})}{\sqrt{-g}}
\end{equation}
supplemented by the requirement that at short distances the correlator reproduces the flat-space singularity structure and early-time boundary conditions determined by the choice of vacuum. In Lorentzian signature, the choice of operator ordering is conveniently implemented through an $i\epsilon$ prescription. For example, in Minkowski space, the time-ordered two-point function of a massless scalar is given by 
\begin{equation}
\langle 0|\,T\{\phi(x_{1})\phi(x_{2})\}\,|0\rangle
=\lim_{\epsilon\to 0^{+}}\frac{\Gamma\!\left(\frac{d-1}{2}\right)}{4\pi^{(d+1)/2}}\,
\frac{1}{\big(s(x_{1},x_{2})^{2}+i\epsilon\big)^{\frac{d-1}{2}}} \, ,
\end{equation}
where $s$ is the geodesic separation between operator insertions. Concretely, the time-ordered, anti-time-ordered, and Wightman functions differ by how one shifts the time coordinate (or equivalently the invariant separation) into the complex plane. 

The above conditions fix the de Sitter invariant two-point function to take the standard hypergeometric form
\cite{Gibbons:1977mu,Bros:1994dn,Bros:1995js,Chernikov:1968zm,Bunch:1978yq} 
\begin{equation}\label{desitterprop}
G(\eta_{1},\vec{x}_{1};\eta_{2},\vec{x}_{2})=\frac{\Gamma(\Delta_{+})\Gamma(\Delta_{-})}{(4\pi)^{\frac{d+1}{2}}\Gamma((d+1)/2)}{}_{2}F_{1}(\Delta_{+},\Delta_{-},\frac{d+1}{2},\chi_{\pm})\, ,
\end{equation}
where 
\begin{equation}\label{originaiepsilon}
\begin{split}
\chi_{\pm}&=\frac{(\eta_{1}+\eta_{2})^{2}-(\vec{x}_{1}-\vec{x}_{2})^{2}\mp i \textrm{sgn}(\eta_{1}-\eta_{2})\epsilon}{4\eta_{1}\eta_{2}}  \ ,\\
\Delta_{\pm}&=\frac{1}{2}(d\pm \sqrt{d^{2}-4m^{2}}) \ .
\end{split}
\end{equation}
For timelike separation, the invariant cross-ratio, $\chi_{\pm}$, lies on a branch cut of ${}_{2}F_{1}$, and the $i\epsilon$ specifies on which side of the cut the correlator is defined, i.e., which Green's function (time-ordered vs.\ Wightman, etc.) one is computing. The $i\epsilon$ prescription given in Eq.~\eqref{originaiepsilon} corresponds to the choice of operator orderings
\begin{equation}\label{etimeprop}
\begin{split}
\langle \phi(\eta_{1},\vec{x}_{1})\phi(\eta_{2},\vec{x}_{2})\rangle&=G_{m}^{\textrm{dS}}(\chi_{+}) \ , \\
\langle \phi(\eta_{2},\vec{x}_{2})\phi(\eta_{1},\vec{x}_{1})\rangle&=G_{m}^{\textrm{dS}}(\chi_{-}) \ .
\end{split}
\end{equation}
The $i\epsilon$ prescription corresponding to the (anti-)time-ordered two-point correlator is
\begin{equation}\label{timeordered}
\chi_{\pm}^{T}=\frac{(\eta_{1}+\eta_{2})^{2}-(\vec{x}_{1}-\vec{x}_{2})^{2}\mp i \epsilon}{4\eta_{1}\eta_{2}} \ ,
\end{equation}
such that 
\begin{equation}\label{timeprop}
\begin{split}
\langle T \{\phi(\eta_{1},\vec{x}_{1})\phi(\eta_{2},\vec{x}_{2})\}\rangle&=G_{m}^{\textrm{dS}}(\chi_{+}^{T}) \ , \\
\langle \bar{T} \{\phi(\eta_{1},\vec{x}_{1})\phi(\eta_{2},\vec{x}_{2})\}\rangle&=G_{m}^{\textrm{dS}}(\chi_{-}^{T}) \ . \\
\end{split}
\end{equation}
All of the orderings in Eqs.~\eqref{etimeprop} and \eqref{timeprop} appear in in-in (Schwinger--Keldysh) computations, as reviewed in Appendix~\ref{sec:reviewinin}.

The key step, and the input needed for WF parameterization in Ref.~\cite{Herderschee:2025znl}, is to rewrite the propagator in a Schwinger-like form which exponentiates the dependence on $\vec{x}_{12}^{2}$. For $G(\chi_{+}^{T})$, the relevant expression of the propagator is
\begin{equation}
G(\chi_{+}^{T})=N_{\Delta_{+}}'(\eta _1 \eta _2)^{\Delta _+}\int_{0}^{\infty} d\beta\,d\kappa \frac{ (\beta -i \kappa )^{\Delta _+-\frac{d}{2}-\frac{1}{2}}}{\beta ^{\frac{d}{2}+\frac{1}{2}} \kappa ^{2 \Delta _+-d+1}} \exp \left(-\frac{4 \eta _1 \eta _2}{\kappa }+\frac{i \left(\vec{x}_{12}^{2}-\left(\eta _1-\eta _2\right){}^2+i \epsilon \right)}{\beta }\right) 
\end{equation}
where 
\begin{equation}
N'_{\Delta_{+}}=\frac{\pi ^{-\frac{d}{2}-\frac{1}{2}} (-i)^{d-\Delta _+} 2^{2 \Delta _+-d-1}}{\Gamma \left(\Delta _+-\frac{d}{2}+\frac{1}{2}\right)} \ .
\end{equation}
In this representation, the dependence on $\vec{x}_1-\vec{x}_2$ enters only through a Gaussian factor, so the Fourier transform to momentum space is immediate. Defining
$G^{++}(\vec{k},\eta_1,\eta_2)=\int d^{d}\vec{x}\,e^{-i\vec{k}\cdot\vec{x}}\,G(\chi_{+}^{T})$,
we obtain
\begin{equation}\label{eq:schwingermomentum}
G^{++}(\vec{k},\eta_{1},\eta_{2})=N_{\Delta_{+}}(\eta _1 \eta _2)^{\Delta _+} \int_{0}^{\infty} d\beta d\kappa \frac{  \exp \left(-\frac{i \beta  \vec{k}^{2}}{4}-\frac{i \left(\eta _1-\eta _2\right){}^2}{\beta }-\frac{4 \eta _1 \eta _2}{\kappa }- \epsilon \beta\right) }{\kappa ^{2 \Delta _+-d+1} \beta^{1/2}(\beta -i \kappa )^{\frac{d}{2}+\frac{1}{2}-\Delta _+}  }
\end{equation}
where 
\begin{equation}
N_{\Delta_{+}}=\frac{2^{2 \Delta _+-d-1} e^{\frac{1}{4} i \pi  \left(2 \Delta _+-d\right)}}{\sqrt{\pi } \Gamma \left(\Delta _+-\frac{d}{2}+\frac{1}{2}\right)}
\end{equation}
with $\vec{k}^{\,2}$ appearing only in the Gaussian kernel. The position-space regulator $e^{-\epsilon/\beta}$ has been replaced by the momentum-space prescription $e^{-\epsilon\beta}$. Both define the same $\epsilon\to0^{+}$ limit, and the latter enforces convergence at large $\beta$. The propagator notation in Eq.~\eqref{eq:schwingermomentum} follows the standard convention used in in-in correlator computations. The $i\epsilon$ prescription is essential here as well because it enforces convergence in the large-Schwinger-parameter region. Note that $G^{--}(\vec{k},\eta_{1},\eta_{2})$ is simply related by complex conjugation to $G^{++}(\vec{k},\eta_{1},\eta_{2})$.

As a consistency check, we have verified numerically that this momentum-space Schwinger representation reproduces the standard closed-form expressions for the de Sitter two-point function in $D=d+1=4$~\cite{Chernikov:1968zm,Bunch:1978yq,Qin:2023ejc},
\begin{equation}\label{explicitmomentumspacbesse}
\begin{split}
G_{D=4}^{++}(\vec{k},\eta_{1},\eta_{2})=&\frac{1}{4} \pi  \left(\eta _1 \eta _2\right){}^{3/2} e^{-\pi  \widetilde{\nu} } \theta \left(\eta _1-\eta _2\right) H_{i \widetilde{\nu} }^{(1)}\left(-|\vec{k}| \eta _1\right) H_{-i \widetilde{\nu} }^{(2)}\left(-|\vec{k}| \eta _2\right)\\
&+\frac{1}{4} \pi  \left(\eta _1 \eta _2\right){}^{3/2} e^{-\pi  \widetilde{\nu} } \theta \left(\eta _2-\eta _1\right) H_{i \widetilde{\nu} }^{(1)}\left(-|\vec{k}| \eta _2\right) H_{-i \widetilde{\nu} }^{(2)}\left(-|\vec{k}| \eta _1\right)
\end{split}
\end{equation}
where 
\begin{equation}
\widetilde{\nu}=\sqrt{m^{2}-9/4}>0 \ .
\end{equation}
The Schwinger representation is far from obvious from the explicit momentum-space form of the two-point function in Eq.~\eqref{explicitmomentumspacbesse}.\footnote{A related plane-wave representation of massive mode functions was introduced in Refs.~\cite{Belrhali:2026ygh,Belrhali:2026jqe}.}

In principle, we also need to consider the late-time limit of the bulk two-point function when computing late-time correlation functions. These limits correspond to external propagators in the Witten diagram. However, in the inflationary applications of interest, we are computing late-time correlation functions of inflatons. In $D=4$, inflatons are effectively massless in slow-roll, so the late-time two-point functions take the very simple form\footnote{We note that Eq.~\eqref{ex:extprop} is the Fourier transform of the finite, late-time part of the two-point function in Eq.~\eqref{desitterprop}. Equation~\eqref{desitterprop} also contains a divergent contribution in the limit $m\to 0$ that goes as a constant times $m^{-2}$. This contribution is independent of the separation of the two points (it is the familiar zero-mode divergence of the massless de Sitter propagator), so it only contributes at $\vec{k}=0$ and drops out of the correlators at nonzero momentum considered here.}
\begin{equation}\label{ex:extprop}
G^{\pm}(\vec{k},\eta)=\lim_{\epsilon \to 0^{+}}\frac{1}{2|\vec{k}|^{3}}(1\mp i |\vec{k}|\eta)e^{\pm i |\vec{k}|\eta+\epsilon \eta} \ .
\end{equation}
As a result, the external legs are already exponentiated (in the sense relevant for WF parameterization), so there is no need to introduce Schwinger parameters for them.\footnote{When dimensional regularization is required, as for the bubble in Sec.~\ref{sec:bubble}, the inflaton can be continued to $d$ spatial dimensions with a $d$-dependent mass chosen so that Eq.~\eqref{ex:extprop} retains its exponential form up to a power of $(-\eta)$. In Sec.~\ref{sec:bubble} we explain why the external legs and the vertex time measures can nevertheless be kept in their $d=3$ form for practical calculations without loss of generality.}

\subsection{Witten diagrams as generalized Euler--Mellin integrals}\label{sec:wittendiagramgeul}

With the Schwinger representation of the momentum-space bulk two-point function in hand, we can now reorganize Witten diagrams into generalized Euler--Mellin integrals. The basic point is that once every internal line is written in Schwinger form, all spatial loop-momentum integrals become Gaussian and can be carried out trivially. What remains is a finite-dimensional integral over Schwinger parameters, whose structure mirrors the familiar Feynman representation of flat-space Feynman integrals, but now with additional dependence on the conformal times associated with bulk vertices.

Concretely, for a given diagram, we write each internal propagator using the Schwinger representation in Eq.~\eqref{eq:schwingermomentum}. The component of the exponent that depends on the $\eta_{v}$ variables can be read off from Eqs.~\eqref{eq:schwingermomentum} and \eqref{ex:extprop}. The component of the exponent independent of the $\eta_{v}$ variables is nontrivial because it depends on loop momenta, which we denote by $\vec\ell_i$. However, since the integrand is Gaussian in every $\vec\ell_i$, these integrations are straightforward. For example, consider the $++$ bubble diagram with incoming momentum $\vec{k}$ and loop momentum $\vec{\ell}$. The diagram is simply a product of two bulk propagators. Using the Schwinger form from Eq.~\eqref{eq:schwingermomentum}, we have
\begin{equation}
\begin{split}
    &\int_{-\infty}^{\infty}\frac{d^{d}\vec{\ell}}{(2\pi)^d}\,G^{++}(\vec{\ell},\eta_{1},\eta_{2})G^{++}(\vec{k}-\vec{\ell},\eta_{1},\eta_{2}) \\
    &= (N_{\Delta_{+}})^2\int_{0}^{\infty} d\beta_1 d\kappa_1d\beta_2 d\kappa_2 \;(\eta _1 \eta _2)^{2\Delta _+} \\
    &\quad \times \frac{  \exp \left(-\frac{i \left(\eta _1-\eta _2\right){}^2}{\beta_1 }-\frac{4 \eta _1 \eta _2}{\kappa_1 }- \epsilon \beta_1-\frac{i \left(\eta _1-\eta _2\right){}^2}{\beta_2 }-\frac{4 \eta _1 \eta _2}{\kappa_2 }- \epsilon \beta_2\right)}{(\kappa_1 ^{2 \Delta _+-d+1} \beta_1^{1/2}(\beta_1 -i \kappa_1 )^{\frac{d}{2}+\frac{1}{2}-\Delta _+})(\kappa_2 ^{2 \Delta _+-d+1} \beta_2^{1/2}(\beta_2 -i \kappa_2 )^{\frac{d}{2}+\frac{1}{2}-\Delta _+}  )  }
    \\
    &\quad \times  \int_{-\infty}^{\infty}\frac{d^{d}\vec{\ell}}{(2\pi)^d} \exp \left(-\frac{i \beta_1  \vec{\ell}^{2}}{4}-\frac{i \beta_2  (\vec{k}-\vec{\ell})^{2}}{4}\right)\;.\label{eq:bubble}
\end{split}
\end{equation}
We indeed see that the loop momentum integral is independent of the bulk time variables $\eta_v$ and can be evaluated on the branch obtained from $\beta_i\to\beta_i-i0$,
\begin{equation}
\int_{-\infty}^{\infty}\frac{d^{d}\vec{\ell}}{(2\pi)^d} \exp \left(-\frac{i \beta_1  \vec{\ell}^{2}}{4}-\frac{i \beta_2  (\vec{k}-\vec{\ell})^{2}}{4}\right) = \frac{1}{(2\pi)^d}\exp\left(-\frac{i}{4}\frac{\beta_1\beta_2 \vec{k}^2}{\beta_1+\beta_2}\right)\left(\frac{4\pi}{i(\beta_1+\beta_2)}\right)^{d/2}.\label{eq:bubble_loop}
\end{equation}
The kinematic terms in the resulting exponential that are independent of the bulk times $\eta_v$ are governed by the following standard graph polynomials.
\begin{equation}
\textrm{kinematic component of the exponent independent of $\eta_{v}$ at $\epsilon=0$:} \quad \frac{-i}{4}\frac{\mathcal{F}(\beta_{i})}{\mathcal{U}(\beta_{i})}
\end{equation}
where $\mathcal{U}(\beta_i)$ and $\mathcal{F}(\beta_i)$ are the Symanzik (Feynman) polynomials associated with the graph.\footnote{See~\cite{Weinzierl:2022eaz} for a nice review.} We collect a few representative examples here for pedagogy. For the bubble that we just computed, we have $\vec{k}=\vec{k}_1=-\vec{k}_2$ and 
\begin{equation}
\label{eq:bubble_diagram}
\begin{tikzpicture}[baseline={([yshift=-.5ex]current bounding box.center)},every node/.style={font=\scriptsize}]
\pgfmathsetmacro{\r}{1.9}

\pgfmathsetmacro{\yTop}{0.8*\r}
\draw ({-2.0*\r},\yTop) -- ({ 2.0*\r},\yTop);

\tikzset{decoration={snake,amplitude=.4mm,segment length=1.5mm,post length=0mm,pre length=0mm}}

\filldraw ({1.3*\r},\yTop) circle (1pt) node[above=0pt]{$2$};
\filldraw ({ -1.3*\r},\yTop) circle (1pt) node[above=0pt]{$1$};

\filldraw (0:\r/2.2)   circle (1pt) node[right=0pt]{};
\filldraw (180:\r/2.2) circle (1pt) node[above=0pt]{};

\draw [heavy scalar] (0,0) circle (\r/2.2);

\draw [thick] (0:\r/2.2)   -- ({ 1.3*\r},\yTop);
\draw [thick] (180:\r/2.2) -- ({-1.3*\r},\yTop);

\filldraw (90:\r/1.8)  circle (0pt) node{$\beta_{1}$};
\filldraw (270:\r/1.8) circle (0pt) node{$\beta_{2}$};

\coordinate (SE) at ([xshift=-2mm,yshift=2mm]current bounding box.south east);
\draw[->,thick] (SE) -- ([yshift=8mm]SE) node[above] {$\eta$};

\end{tikzpicture}
\end{equation}
\begin{equation}\nonumber
\begin{split}
\mathcal{F}_{\textrm{bubble}}&=\beta_{1}\beta_{2}\vec{k}_{1}^{2}\ , \\
\mathcal{U}_{\textrm{bubble}}&=\beta_{1}+\beta_{2}\ . \\ 
\end{split}
\end{equation}
For the triangle, one finds
\begin{equation}
\label{eq:triangle_diagram}
\begin{tikzpicture}[baseline={([yshift=-.5ex]current bounding box.center)},every node/.style={font=\scriptsize}]
\pgfmathsetmacro{\r}{1.9}

\pgfmathsetmacro{\yTop}{0.8*\r}

\draw (-2*\r,\yTop) -- (2*\r,\yTop);

\tikzset{decoration={snake,amplitude=.4mm,segment length=1.5mm,post length=0mm,pre length=0mm}}

\filldraw (0*\r,\yTop) circle (1pt) node[above=0pt]{$2$};
\filldraw ( 1.4*\r,\yTop) circle (1pt) node[above=0pt]{$1$};
\filldraw ( -1.4*\r,\yTop) circle (1pt) node[above=0pt]{$3$};

\filldraw (-30:\r/1.7)   circle (1pt) node[right=0pt]{};
\filldraw (90:\r/1.7) circle (1pt) node[above=0pt]{};
\filldraw (210:\r/1.7) circle (1pt) node[below=0pt]{};

\draw [thick] (90:\r/1.7) -- (-0*\r,\yTop);
\draw [thick] (-30:\r/1.7)   -- ( 1.4*\r,\yTop);
\draw [thick] (210:\r/1.7) -- ( -1.4*\r,\yTop);

\draw [heavy scalar] (90:\r/1.7) -- (-30:\r/1.7);
\draw [heavy scalar] (210:\r/1.7) -- (90:\r/1.7);
\draw [heavy scalar] (210:\r/1.7) -- (-30:\r/1.7);

\filldraw (30:0.4*\r)  circle (0pt) node{$\beta_{1}$};
\filldraw (150:0.4*\r) circle (0pt) node{$\beta_{2}$};
\filldraw (270:0.4*\r) circle (0pt) node{$\beta_{3}$};

\coordinate (SE) at ([xshift=-2mm,yshift=2mm]current bounding box.south east);
\draw[->,thick] (SE) -- ([yshift=8mm]SE) node[above] {$\eta$};

\end{tikzpicture}
\end{equation}
\begin{equation}\nonumber
\begin{split}
\mathcal{F}_{\textrm{triangle}}&=\beta_{2}\beta_{3}\vec{k}_{3}^{2}+\beta_{1}\beta_{3}\vec{k}_{1}^{2}+\beta_{1}\beta_{2}\vec{k}_{2}^{2}\ , \\
\mathcal{U}_{\textrm{triangle}}&=\beta_{1}+\beta_{2}+\beta_{3} \ . \\ 
\end{split}
\end{equation}
Finally, for the box, one finds 
\begin{equation}
\pgfmathsetmacro{\r}{1.9}
\begin{tikzpicture}[baseline={([yshift=-.5ex]current bounding box.center)},every node/.style={font=\scriptsize}]

\pgfmathsetmacro{\yTop}{0.8*\r}
\draw (-2*\r,\yTop) -- (2*\r,\yTop);

\filldraw (45:\r/2)  circle (1pt) node[above=0pt]{};
\filldraw (135:\r/2) circle (1pt) node[above=0pt]{};
\filldraw (225:\r/2) circle (1pt) node[below=0pt]{};
\filldraw (-45:\r/2) circle (1pt) node[below=0pt]{};

\draw [heavy scalar] (-45:\r/2) -- (225:\r/2) -- (135:\r/2) -- (45:\r/2) -- (-45:\r/2);

\filldraw (-1.5*\r,\yTop) circle (1pt) node[above=0pt]{$1$};
\filldraw (-0.8*\r,\yTop) circle (1pt) node[above=0pt]{$2$};
\filldraw ( 0.8*\r,\yTop) circle (1pt) node[above=0pt]{$3$};
\filldraw ( 1.5*\r,\yTop) circle (1pt) node[above=0pt]{$4$};

\draw [thick] (225:\r/2) -- (-1.5*\r,\yTop);
\draw [thick] (135:\r/2) -- (-0.8*\r,\yTop);
\draw [thick] (45:\r/2)  -- ( 0.8*\r,\yTop);
\draw [thick] (-45:\r/2) -- ( 1.5*\r,\yTop);

\filldraw (0:0.45*\r)   circle (0pt) node{$\beta_{3}$};
\filldraw (90:0.45*\r)  circle (0pt) node{$\beta_{2}$};
\filldraw (180:0.45*\r) circle (0pt) node{$\beta_{1}$};
\filldraw (270:0.45*\r) circle (0pt) node{$\beta_{4}$};

\filldraw (45:0.85*\r)   circle (0pt) node[left=0pt]{};
\filldraw (135:0.85*\r)  circle (0pt) node[right=0pt]{};
\filldraw (-45:0.85*\r)  circle (0pt) node[left=0pt]{};
\filldraw (-135:0.85*\r) circle (0pt) node[right=0pt]{};

\coordinate (SE) at ([xshift=-2mm,yshift=2mm]current bounding box.south east);
\draw[->,thick] (SE) -- ([yshift=8mm]SE) node[above] {$\eta$};

\end{tikzpicture}
\end{equation}
\begin{equation}\nonumber
\begin{split}
\mathcal{F}_{\textrm{box}}&=\beta_{1}\beta_{3}(\vec{k}_{2}+\vec{k}_{3})^{2}+\beta_{2}\beta_{4}(\vec{k}_{3}+\vec{k}_{4})^{2}\\
&+\beta_{1}\beta_{2} \vec{k}_{2}^{2} +\beta_{2}\beta_{3} \vec{k}_{3}^{2}+\beta_{3}\beta_{4} \vec{k}_{4}^{2}+\beta_{4}\beta_{1} \vec{k}_{1}^{2}\ , \\
\mathcal{U}_{\textrm{box}}&=\beta_{1}+\beta_{2}+\beta_{3}+\beta_{4} \ . \\ 
\end{split}
\end{equation}
We note that the combinatorial structure of these polynomials has been extensively studied due to their relevance for flat-space Feynman integrals.

At this stage, for an arbitrary Witten diagram associated with a de Sitter correlator, the remaining integral over the Schwinger parameters will contain products of polynomials and an exponential. An appropriate integration contour must be identified by analyzing the exponent, as we discuss in Sec.~\ref{sec:WF-ieps}. We now organize the Schwinger-parameter integral into a fully polynomial form, assuming such a contour has been chosen.

Notice that the exponent and all polynomial factors in the integrands in Eqs.~\eqref{eq:bubble} and \eqref{eq:bubble_loop} are homogeneous under an overall rescaling of all integration parameters, 
\begin{equation}
    \beta_i\rightarrow \lambda \beta_i,\; \kappa_i\rightarrow\lambda  \kappa_i,\; \eta_i\rightarrow \lambda \eta_i.
\end{equation}
The exponent has degree one. This common scaling exists for arbitrary Witten diagrams for de Sitter correlators. Separating the common scale from the projective ratios is a gauge choice (a total $GL(1)$ in the language of WF parameterization, since the rescaling is by a positive real number). We will now insert in the integral the identity (which fixes this choice)
\begin{equation}\label{eq:deltafunction}
1=\int_0^{\infty} d\lambda \,\delta\!\left(\lambda-\sum_{e\in E^{\textrm{int}}}\mathbf{m}_{e}\beta_{e}-\sum_{e\in E^{\textrm{int}}}\mathbf{n}_{e}\kappa_{e}+\sum_{v\in V}\mathbf{q}_{v}\eta_{v}\right)\,,
\end{equation}
where the $(\mathbf{m}_{e},\mathbf{n}_{e},\mathbf{q}_{v})$ are real numbers chosen so that the linear combination in the delta function is positive on the region of integration. Taking them nonnegative with at least one nonzero is one such choice, but in Secs.~\ref{sec:treeanalyt}--\ref{sec:triangle} we make a region-dependent choice after the domain has been split into sectors. We then rescale all integration variables
\begin{equation}
\begin{split}
\beta_e &\rightarrow \lambda\,\beta_e \ ,\\
\kappa_e &\rightarrow \lambda\,\kappa_e \ ,\\
\eta_v &\rightarrow \lambda\,\eta_v \ .
\end{split}
\end{equation}
The rescaling symmetry of the integral ensures that the $\lambda$ dependence of the integral will be in the form
\begin{equation}
    \int_0^{\infty} d\lambda\, \lambda^a \exp\left(-\lambda^b (\cdots)\right),
\end{equation}
where $(\cdots)$ denote $\lambda$-independent terms. Contributions to $a$ come from rescaling the polynomial and monomial factors in the integral, the integration measures $d\beta_{e}$, $d\kappa_{e}$ and $d\eta_{v}$, the vertex monomials $(-\eta_{v})^{p_{v}}$ that carry the vertex measure and the derivative legs, and a $\lambda^{-1}$ from factoring $\lambda$ out of the delta function. Performing the $\lambda$ integral converts the exponential into polynomial form,
\begin{equation}
    \int_0^{\infty} d\lambda\, \lambda^a \exp\left(-\lambda^b (\cdots)\right) = \frac{1}{b}(\cdots)^{-\frac{1+a}{b}}\Gamma\left[\frac{1+a}{b}\right],
\end{equation}
leaving a residual delta function that removes one of the remaining integration variables. The resulting expression can be recognized as a generalized Euler--Mellin integral after choosing a gauge in which all but one of the coefficients are set to zero in the sector variables. 

In the computations relevant for inflationary correlators, bulk interactions typically involve derivatives, which introduce additional powers of the conformal times $\eta_v$ at the vertices and hence modify the exponents of the polynomials and monomials in the integral. For this reason, we postpone writing the fully explicit Euler form until later sections, where we treat derivative vertices systematically and keep track of the associated $\eta_v$-dependence.

\subsection{Deriving the integration cycle from the \texorpdfstring{$i\epsilon$}{i epsilon} prescription}\label{sec:WF-ieps}

The analysis in the previous subsections yields a finite-dimensional integral representation for de Sitter correlators. Concretely, after expressing each internal bulk two-point function in the Schwinger form of Eq.~\eqref{eq:schwingermomentum} and carrying out the Gaussian integrals over spatial loop momenta, any Witten diagram reduces to an integral over the remaining Schwinger parameters (together with the vertex times $\eta_v$), with an oscillatory phase inherited from Lorentzian kinematics. We will now discuss how the $i\epsilon$ prescription for the propagators determines the class of admissible integration cycles in the WF representation. To do this we will consider the form of the WF integral before inserting the delta function of Eq.~\eqref{eq:deltafunction} and removing the exponential factor.

Before turning to the WF integrals themselves, it is useful to isolate the basic mechanism by which an
$i\epsilon$ prescription selects an integration cycle. Consider the one-dimensional toy analogue
\begin{equation}\label{toymodel}
f \;=\; \lim_{\epsilon\to 0^{+}} \int_{0}^{\infty} \! dx \; e^{\, i x^{2}-\epsilon x}\,,
\end{equation}
which evaluates to
\begin{equation}\label{eq:toymodel-value}
f \;=\; \sqrt{\frac{\pi}{4}}\, e^{\, i\pi/4}\,.
\end{equation}
For fixed $\epsilon>0$, the factor $e^{-\epsilon x}$ renders the integral absolutely convergent, and
Eq.~\eqref{toymodel} is unambiguous. The point of the $i\epsilon$ prescription, however, is not merely to
define the integral for $\epsilon>0$, but to specify a \emph{limit} as $\epsilon\to0^{+}$, where the
integrand becomes purely oscillatory and direct numerical evaluation suffers from severe cancellations.

A standard practical substitute for sending $\epsilon$ to zero is to trade the infinitesimal damping
for an equivalent choice of contour that is convergent already at $\epsilon=0$. In the present example,
one may rotate the contour into the upper half-plane as follows.
\begin{equation}\label{eq:contour1}
\text{Contour 1:}\qquad x \;=\; e^{i\theta}x'\,,\qquad x'\in\mathbb{R}^{+}\,,\qquad 0<\theta<\frac{\pi}{2}\,.
\end{equation}
Setting $\epsilon=0$ after this deformation, the exponent becomes
$
i x^{2}= i e^{2i\theta} x'^{2}
$
, whose real part is $-\sin(2\theta)\,x'^{2}<0$ for $0<\theta<\pi/2$. Hence, the integral is exponentially
damped at large $x'$ and is manifestly convergent. Moreover, within this range of $\theta$ the value is
independent of $\theta$, because the contour can be varied continuously without encountering any loss of
convergence or any singularity of the integrand.

It is important, however, that a contour deformation, which changes the integration cycle, is not synonymous with an arbitrary complex change of variables, which describes the same cycle using different variables. The correct statement is that two cycles yield the same integral only if they are
continuously deformable into one another through a family of admissible cycles on which the integral
remains convergent. Even in this elementary example, apparently similar
rotations can land on a different homology class. For instance, consider
\begin{equation}\label{eq:contour2}
\text{Contour 2:}\qquad x \;=\; e^{i\theta}x'\,,\qquad x'\in\mathbb{R}^{+}\,,\qquad \pi<\theta<\frac{3\pi}{2}\,.
\end{equation}
Along \eqref{eq:contour2} one again has $\Re(i x^{2})<0$, so the integral converges, but the value is
\emph{not} equal to $f$. Instead, it differs by an overall minus sign as follows.
\begin{equation}
\int_{0}^{\infty}\!dx\; e^{i x^{2}} \Big|_{\text{Contour 2}}
\;=\;
-\sqrt{\frac{\pi}{4}}\, e^{\, i\pi/4}\,.
\end{equation}
The reason is not that contour methods fail, but that \eqref{eq:contour2} is \emph{not} continuously
connected to the original cycle $[0,\infty)$ (or to \eqref{eq:contour1}) through convergent contours.
Any attempted interpolation necessarily passes through angles $\theta$ for which the integrand grows
exponentially at infinity. Concretely, for
\begin{equation}\label{eq:bad-contour}
\text{Bad Contour:}\qquad x \;=\; e^{i\theta}x'\,,\qquad x'\in\mathbb{R}^{+}\,,
\qquad \frac{\pi}{2}<\theta<\pi\ \ \text{or}\ \ \frac{3\pi}{2}<\theta<2\pi\,,
\end{equation}
one has $\Re(i x^{2})= -\sin(2\theta)\,x'^{2} >0$, so the integrand grows like
$\exp\!\big(+|\sin(2\theta)|\,x'^{2}\big)$ and the integral diverges. Thus the space of admissible
cycles is separated by ``walls'' of nonconvergence, and crossing such a wall changes the homology class of
the cycle and may change the value of the integral.

The lesson, which will reappear in the WF representation, is that an $i\epsilon$ prescription specifies the integration cycle. Any
alternative cycle is legitimate only insofar as it can be reached from the original one by a continuous
deformation along which convergence is maintained throughout.

We now turn to the actual WF integrals. For an all-$+$ Schwinger--Keldysh assignment, evaluating the integral over the loop momentum variables gives the schematic structure
\begin{equation}\label{eq:WF-structure-ieps-AH}
\begin{split}
\lim_{\epsilon\to 0^{+}}&\int^{0}_{-\infty}\prod_{v} d\eta_v \;
\int_{0}^{\infty}\prod_{e\in E^{\textrm{int}}} d\beta_e\, d\kappa_e \;
\Big(\cdots\Big)\,
\exp\!\Big[i\,\mathcal{V}(\beta_e,\kappa_e,\eta_{v},\vec{k}_{i})\Big]\, \\
&\times \exp\!\Big[-\sum_{e=(v,v')\in E^{\textrm{int}}} \frac{4\eta_{v}\eta_{v'}}{\kappa_{e}}-\epsilon\sum_{e\in E^{\textrm{int}}}\beta_e+\epsilon \sum_{v\in V}\eta_{v}\Big] \, ,
\end{split}
\end{equation}
where $\mathcal{V}$ is the Lorentzian phase, which collects the external-leg energies, the $(\eta_{v}-\eta_{v'})^{2}/\beta_{e}$ terms of Eq.~\eqref{eq:schwingermomentum}, and the term generated by the Gaussian integrations and $(\cdots)$ denotes additional polynomial prefactors, which do not affect the exponential damping at large parameters. Just as in the toy example in Eq.~\eqref{toymodel}, finite $\epsilon$ damps the large-parameter directions, but the resulting integral rapidly becomes numerically unstable as $\epsilon$ is taken closer to zero. The $i\epsilon$ prescription does not regulate ultraviolet divergences, which we treat separately in Sec.~\ref{sec:bubble}. The remedy is to consider contour deformations such as Eq.~\eqref{eq:contour1}. 

To illustrate the procedure, we first consider contact diagrams with no internal propagators, so there are no Schwinger parameters. Consider the interaction term 
\begin{equation}
\mathcal{L}_{\textrm{int}}\ni (g^{\eta\eta}\partial_{\eta}\phi\partial_{\eta}\phi)^{2} \, ,
\end{equation}
which in $d=3$ leads to the Witten diagram contribution
\begin{equation}
\begin{split}
\int_{-\infty}^{0} d\eta \prod_{i=1}^{4}\partial_{\eta}G^{+}(\vec{k}_{i},\eta)&=\lim_{\epsilon\to 0^{+}}\frac{2^{-4}}{\prod_{i}|\vec{k}_{i}|}\int_{-\infty}^{0} d\eta\,(-\eta)^{4}e^{i\sum_i|\vec{k}_{i}|\eta+\epsilon \eta } \\
&=\frac{i^{3} 2^{-4}\Gamma(5)}{\prod_{i}|\vec{k}_{i}|}\left(\sum_{i} |\vec{k}_{i}|\right)^{-5}
\end{split}
\end{equation}
We again note that from the perspective of numerical evaluation, the integral is highly oscillatory as we take $\epsilon \to 0^{+}$. To make the integral numerically convergent, we are led to perform a Wick rotation similar to the one used for the toy model:
\begin{equation}\label{eq:wickrotation}
\eta \to -i \tau \textrm{ for }\tau\in (-\infty,0) \ .
\end{equation}
We then set $\epsilon=0$ and evaluate the now-convergent $\tau$ integral. Due to the simplicity of the above integral, it is easy to check that the Wick-rotated integral is equivalent to the original, $i\epsilon$-defined integral. This Wick rotation in Eq.~\eqref{eq:wickrotation} generalizes to all integrals considered in this paper. For each one-loop diagram and Schwinger--Keldysh assignment considered in this paper, we find a correlated choice of signs in 
\begin{equation}\label{eq:generalwickrot}
\begin{split}
\eta_{v} &\to \pm i\tau_{v} \\
\kappa_{e} &\to \pm \bar{\kappa}_{e} \\
\beta_{e} &\to \pm i\overline{\beta}_{e} \ , 
\end{split}
\end{equation}
that yields an admissible integration cycle that is numerically convergent. These rotations make the polynomial governing the exponential damping positive. Other factors need not be sign-definite, so their branches and sector decompositions are fixed in the application sections. Direct numerical evaluation nevertheless remains difficult for sufficiently heavy masses, where the scaling dimension becomes complex and the integrand oscillates. We mitigate this with the reduced Schwinger method in Sec.~\ref{sec:redschwinger}. 

For general Witten diagrams beyond one loop, we do not naively expect complex rotations of the $\eta_{v}$, $\kappa_{e}$ and $\beta_e$ variables as in Eq.~\eqref{eq:generalwickrot} to be universally appropriate, since the integrand can depend on the $\eta_{v}$, $\kappa_{e}$ and $\beta_e$ variables in a highly nontrivial way. Indeed, even in flat space, identifying suitable contour deformations for higher-loop Feynman-parameterized integrals is often more heuristic than systematic~\cite{Soper:1999xk,Binoth:2005ff,Pittau:2021jbs,Mizera:2021icv,Borinsky:2023jdv} and is consequently a significant bottleneck for the explicit evaluation of Feynman integrals in Lorentzian kinematics. However, it is interesting that such simple complex rotations are viable even at one loop for in-in correlators.

\section{The reduced Schwinger method}\label{sec:redschwinger}

As reviewed in Sec.~\ref{sec:wittrop}, the WF parameterization converts
Witten diagrams into generalized Euler--Mellin integrals. For sufficiently
heavy masses, the scaling dimension is
\begin{equation}
\Delta_+=\frac{d}{2}+i\widetilde{\nu}\,,
\qquad \widetilde{\nu}>0\,.
\end{equation}
The resulting Witten--Feynman integrals have complex exponents and hence oscillatory integrands. To evaluate them efficiently, we use the reduced Schwinger method, which performs a sub-integral analytically and expresses the result as an explicit hypergeometric kernel. This particular sub-integral contains a large number of oscillations for fairly generic kinematics and is therefore difficult to evaluate directly by numerical quadrature.
The remaining directions are integrated numerically. Sec.~\ref{sec:redschsimple} illustrates the mechanism in a two-dimensional example, Sec.~\ref{sec:redschgeneral} describes its
application to the generalized Euler integrals considered here, and Sec.~\ref{sec:redschdiv} explains how ultraviolet divergences are treated.

\subsection{A toy model}
\label{sec:redschsimple}

Consider the toy integral
\begin{equation}
\begin{aligned}
\mathcal I(Q)
&=
\int_0^\infty dx\int_0^\infty dy\,
(x+y)^{\alpha-2}U(x,y;Q)^{-N},
\\
U(x,y;Q)
&=
Q+\frac{1}{x+y}+(x+y)+\frac{xy}{x+y},
\end{aligned}
\end{equation}
with
\begin{equation}
\alpha=\frac12+i,
\qquad
N=\frac52+i.
\end{equation}
Here the common imaginary part of $\alpha$ and $N$ plays the role of $\widetilde{\nu}$ and $Q$ models a dimensionless hierarchy of external momenta, and the real observable is $I(Q)=\operatorname{Re}\mathcal I(Q)$. This integral
is a simplified version of the one relevant for the bubble contribution in
Sec.~\ref{sec:bubble}.

The numerical difficulty is exposed by separating the overall size of
$x,y$ from their ratio. Defining
\begin{equation}
x=st,\qquad y=s(1-t),
\qquad
s\in(0,\infty),\quad t\in(0,1),
\end{equation}
the integral becomes
\begin{equation}
\mathcal I(Q)
=
\int_0^1dt\int_0^\infty ds\,
s^{\alpha-1}
\left[
Q+\frac1s+\bigl(1+t(1-t)\bigr)s
\right]^{-N}.
\label{eq:toy-st-form}
\end{equation}
At fixed $t$, its radial monomial contains the logarithmic phase
\begin{equation}
s^{\alpha-1}
=
s^{\operatorname{Re}\alpha-1}
e^{i\,\operatorname{Im}\alpha\log s}
\end{equation}
and therefore oscillates in $\log s$. For $Q\gg1$, the bracket is dominated by $Q$ and slowly varying across the parametrically large region $Q^{-1}\ll s\ll Q$. At large $Q$, direct quadrature must
therefore integrate a large range of $\log s$ over which the phase winds $\sim (\operatorname{Im}\alpha/\pi)\log Q$ number of times.

To avoid this, the reduced Schwinger method performs the scale integral
analytically. With 
\begin{equation}
U=Q+\frac1s+C(t)s,
\qquad
C(t)=1+t(1-t),
\end{equation}
the integral becomes
\begin{equation}
\mathcal I(Q)
=
\int_0^1dt\,
\mathcal K_{\alpha,N}\bigl(Q,1,C(t)\bigr),
\end{equation}
where
\begin{equation}
\label{eq:redschkernel}
\mathcal K_{\alpha,N}(A,B,C)
=
\int_0^\infty ds\,
s^{\alpha-1}
\left(A+\frac Bs+Cs\right)^{-N}.
\end{equation}
Changing to $u = C s/A$, which in our model is $C(t)s/Q$, gives
\begin{equation}
\mathcal K_{\alpha,N}(A,B,C)
=
C^{-\alpha}A^{\alpha-N}
h\left(\frac{BC}{A^2}\right),
\end{equation}
with
\begin{equation}
h(\xi)
=
\int_0^\infty du\,
u^{\alpha-1}
\left(1+u+\frac{\xi}{u}\right)^{-N}.
\end{equation}
Since $N-\alpha=2$,
\begin{equation}
\label{eq:toy-reduced-form}
Q^2\mathcal I(Q)
=
\int_0^1dt\,
C(t)^{-\alpha}
h\left(\frac{C(t)}{Q^2}\right).
\end{equation}
The unbounded oscillatory scale integral has thus been absorbed into the
universal kernel $h$, which is evaluated analytically below, leaving a numerical integral over the compact
interval $t\in[0,1]$.

The exact scale integration offers a numerical advantage in hierarchical
kinematics because it isolates the nonanalytic asymptotic behavior in the
squeezed limit. In this limit,
\begin{equation}
\xi=\frac{C(t)}{Q^2}\ll1,
\end{equation}
the kernel receives contributions from two parametrically distinct regions.
The generic region $u\sim1\;(s\sim Q)$ produces an ordinary power-series
expansion in $\xi$. Near the endpoint $u\sim\xi\;(s\sim Q^{-1})$, however,
the ratio $\xi/u$ remains of order one and cannot be expanded. This endpoint
region instead generates the nonanalytic term $\xi^\alpha$:
\begin{equation}
h(\xi)
=
\mathrm{B}(\alpha,N-\alpha)
+
\mathrm{B}(\alpha+N,-\alpha)\,\xi^\alpha
+
O(\xi),
\end{equation}
where $\mathrm{B}(a,b)\equiv
\Gamma(a)\Gamma(b)/\Gamma(a+b)$, and
$\mathrm{B}(\alpha+N,-\alpha)$ is defined by analytic continuation.
Substituting this expansion into Eq.~\eqref{eq:toy-reduced-form} gives
\begin{equation}
Q^2\mathcal I(Q)
=
c_0+c_{\mathrm{na}}Q^{-2\alpha}
+O(Q^{-2}),
\end{equation}
where
\begin{equation}
c_0
=
\mathrm{B}(\alpha,N-\alpha)
\int_0^1dt\,C(t)^{-\alpha},
\qquad
c_{\mathrm{na}}
=
\mathrm{B}(\alpha+N,-\alpha).
\end{equation}
Consequently,
\begin{equation}
Q^2 I(Q)
=
\operatorname{Re}c_0
+
\frac{1}{Q}
\operatorname{Re}\left[
c_{\mathrm{na}}e^{-2i\log Q}
\right]
+
O(Q^{-2}).
\end{equation}
Thus, the nonanalytic oscillatory contribution is suppressed by $Q^{-1}$
relative to the smooth analytic background. A direct numerical integration
over $s$ would return both contributions together and would therefore require
relative accuracy of order $Q^{-1}$ to resolve the subleading nonanalytic
behavior. The exact scale integration separates the two analytically: the
nonanalytic dependence appears explicitly as the $\xi^\alpha$ term in
$h(\xi)$, rather than having to be extracted by subtracting the dominant
analytic contribution.

More generally, for finite $\xi$, the kernel admits a closed form suitable
for numerical evaluation. Defining
\begin{equation}
a=\frac{1-\sqrt{1-4\xi}}{2}\,,
\qquad
b=\frac{1+\sqrt{1-4\xi}}{2}\,,
\qquad
w=\frac{a}{b}\,,
\end{equation}
the analytically continued kernel is
\begin{equation}\label{eq:redschtwobranch}
\begin{split}
h(\xi)={}&
b^{\alpha-N}B(\alpha,N-\alpha)
{}_{2}F_{1}\!\left(N,N-\alpha;1-\alpha;w\right)
\\
&+
a^\alpha b^{-N}
\frac{\Gamma(\alpha+N)\Gamma(-\alpha)}{\Gamma(N)}
{}_{2}F_{1}\!\left(N,\alpha+N;1+\alpha;w\right).
\end{split}
\end{equation}
The branches are fixed by matching to the defining integral for
$0<\xi<1/4$, where $a,b>0$ and the kernel is unambiguous. In particular,
we choose $\sqrt{1-4\xi}\to1$ as $\xi\to0^+$, so that $a\sim\xi$ and
$b\sim1$, and analytically continue this choice to other values of $\xi$.
\subsection{Generalized Euler integrals}
\label{sec:redschgeneral}

The above reduction applies to the WF integrals derived in
Sec.~\ref{sec:wittendiagramgeul}. Their numerical evaluation proceeds in two
stages. We reduce a common Schwinger scale exactly and numerically integrate
over the remaining variables.

A change of variables separates a common scale $s$ from angular
variables $\Omega$. The transformed integrand then factors exactly as
\begin{equation}\label{eq:redschfactor}
d^n x\,[\text{integrand}]
=
d\Omega\,W(\Omega)\,ds\,s^{\alpha-1}
\left(
A(\Omega)+\frac{B(\Omega)}{s}+C(\Omega)s
\right)^{-N}.
\end{equation}
Here $A$, $B$, and $C$ are positive functions of $\Omega$ and the
external kinematics, while $W(\Omega)$ collects the transformed monomial
and polynomial factors. Equation~\eqref{eq:redschfactor} is exact. The change of variables realizing it is constructed explicitly for each integral in Secs.~\ref{sec:treeanalyt}, \ref{sec:bubble}, and \ref{sec:triangle}. For the tree-level exchange, $\alpha=\Delta_+-\tfrac{d}{2}=i\widetilde{\nu}$ and $N=\Delta_++\tfrac{d}{2}$.

For the bubble, the exponents are
\begin{equation}
\alpha=\frac{3}{2}+2i\widetilde{\nu}-\varepsilon,
\qquad
N=\frac{9}{2}+2i\widetilde{\nu}-3\varepsilon,
\qquad
N-\alpha=3-2\varepsilon.
\end{equation}
For the triangle, they are
\begin{equation}
\alpha=3\Delta_+-d=\frac{3}{2}+3i\widetilde{\nu},
\qquad
N=\frac{9}{2}+3i\widetilde{\nu},
\qquad
N-\alpha=3.
\end{equation}
The precise form of $B$ depends on the integral. Only its positivity is
needed for the subsequent scale integration.\footnote{
The function $A$ collects terms linear in the external energies. The
inverse-scale term $B/s$ contains the momentum-dependent contributions
generated by the Gaussian integrals, each proportional to a
momentum-squared invariant divided by four times the scale, while $Cs$
contains terms that grow with the common scale. The limit $C\to0$
identifies a UV corner.}

The remaining $s$ integral is the kernel
$\mathcal{K}_{\alpha,N}$ defined in Eq.~\eqref{eq:redschkernel}. Its
convergence requires
\begin{equation}
\operatorname{Re}(N-\alpha)>0,
\qquad
\operatorname{Re}(\alpha+N)>0.
\end{equation}
At small $\xi$, we use the exact two-branch continuation in
Eq.~\eqref{eq:redschtwobranch}. The $a^\alpha$ branch must be retained
because it carries the non-analytic behavior. At moderate $\xi$, we
evaluate the one-dimensional scale integral directly in a logarithmic
coordinate, centering the quadrature at the stationary point of the
non-oscillatory envelope obtained by replacing the exponents with their
real parts. Alternatively, we interpolate a table generated using the same
prescription. At large $\xi$, we use a power-law tail expansion. This
expansion is non-oscillatory whenever $N-\alpha$ is real, as it is for
both the bubble and the triangle.

Finally, we numerically integrate over the remaining angular variables. The
integral is two-dimensional for tree-level exchange, four-dimensional for the bubble and seven-dimensional
for the triangle.\footnote{
We use deterministic tensor-product tanh--sinh quadrature, monitor
convergence under successive refinement, and compare with the independent
tropical Monte Carlo calculation of Appendix~\ref{sec:tropdecomp} at
moderate squeezing.}

\subsection{Divergent integrals}
\label{sec:redschdiv}

After the scale reduction, the UV divergence of the bubble appears at an
endpoint of the outer integral. Near $x\to0$, the density behaves as
\begin{equation}
x^{a(\varepsilon)-1}\,,
\qquad
a(\varepsilon)=c\varepsilon+O(\varepsilon^2)\,,
\qquad
c\neq0.
\end{equation}
Suppose the remaining factor satisfies
\begin{equation}
g(x,\Omega,\varepsilon)-g(0,\Omega,\varepsilon)=O(x^p),
\end{equation}
for some $p>0$. Adding and subtracting its finite boundary value then gives
the exact identity
\begin{equation}\label{eq:redschsub}
\begin{split}
I(\varepsilon)
&=
\int_{\Omega}d\Omega\int_0^1 dx\,
x^{a(\varepsilon)-1}g(x,\Omega,\varepsilon)
\\
&=
\int_{\Omega}d\Omega\int_0^1 dx\,
x^{a(\varepsilon)-1}
\left[g(x,\Omega,\varepsilon)-g(0,\Omega,\varepsilon)\right]
+
\frac{D(\varepsilon)}{a(\varepsilon)},
\\
D(\varepsilon)
&=
\int_{\Omega}d\Omega\,g(0,\Omega,\varepsilon).
\end{split}
\end{equation}
Expanding $D$ and $a$ in $\varepsilon$ produces a pole of the form
\begin{equation}
\frac{D_0}{c\varepsilon},
\end{equation}
together with a finite boundary contribution. The subtracted outer integral
is finite at $\varepsilon=0$, so the pole and finite part can be extracted
without evaluating the full integral at a nonzero value of the regulator. We
give the bubble residue in Sec.~\ref{sec:bubble}. Appendix~\ref{sec:regdiv} describes the alternative integration-by-parts treatment used for the tropical cross-check.

\section{Illustrative example: tree-level contribution to the bispectrum}\label{sec:treeanalyt}

In this section, we examine the tree-level contribution to non-Gaussianity at leading order in the slow-roll parameter. The closed-form expressions for the relevant Witten diagrams in terms of ${}_{3}F_{2}$ functions are known in the literature~\cite{Qin:2023ejc} and provide a cross-check of our expressions. 

We consider the graphs generated by the interaction
\begin{equation}
\mathcal{L}_{\mathrm{int}} \supset (\partial^{\eta}\phi)(\partial_{\eta}\phi)\,\sigma \, ,
\end{equation}
where $\sigma$ is a scalar field of mass $m$. We set the coupling of this operator to one, since it only multiplies the result. We expand the (canonically normalized) inflaton about its background value,
\begin{equation}
\phi = \phi_0 + \delta\phi \, ,
\end{equation}
and organize perturbation theory in the interaction strength together with the slow-roll suppression carried by the background time dependence,
\begin{equation}
\epsilon_{\rm sr} \equiv \eta\,\partial_{\eta}\phi_0 = -\dot{\phi}_{0}/H \, .
\end{equation}
Interactions linear in $\sigma$ are well-motivated and, for a neutral scalar, are often the generic leading portal allowed by symmetries.
Absent an internal symmetry that forbids odd powers of $\sigma$ (e.g.\ a $\mathbb{Z}_2:\sigma\to-\sigma$ or a shift symmetry), such a coupling can be generated via loop corrections in the low-energy effective action after integrating out heavy degrees of freedom. From a phenomenological standpoint, the linear portal is important because the on-shell production of $\sigma$ is known to imprint a non-analytic signature in the bispectrum, independent of the details of the portal, that cannot be mimicked by a local self-interaction of the inflaton. It is therefore a conceptually clean and analytically tractable benchmark for the numerical methods developed below.

With this motivation in mind, we turn to explicit computations. The relevant Witten diagrams are
\begin{equation}\label{eq:tree1}
\pgfmathsetmacro{\r}{1.9}
W_{++}=\begin{tikzpicture}[baseline={([yshift=-.5ex]current bounding box.center)},every node/.style={font=\scriptsize}]

\pgfmathsetmacro{\yTop}{0.8*\r}
\draw (-2*\r,\yTop) -- (2*\r,\yTop);

\filldraw (\r/2,0)  circle (1pt) node[above=0pt]{};
\filldraw (-\r/2,0) circle (1pt) node[above=0pt]{};

\draw [heavy scalar] (0:\r/2) -- (180:\r/2);

\filldraw (-1.5*\r,\yTop) circle (1pt) node[above=0pt]{$1$};
\filldraw (-0.8*\r,\yTop) circle (1pt) node[above=0pt]{$2$};
\filldraw ( 0.8*\r,\yTop) circle (1pt) node[above=0pt]{$3$};

\draw [thick] (-\r/2,0) -- (-1.5*\r,\yTop);
\draw [thick] (-\r/2,0) -- (-0.8*\r,\yTop);
\draw [thick] (\r/2,0)  -- ( 0.8*\r,\yTop);
\draw [thick] (\r/2,0) -- (-10:\r);

\draw plot[mark=x,mark size=3pt] coordinates {(-10:\r)};

\coordinate (SE) at (3,0);
\draw[->,thick] (SE) -- (3,0.8) node[above] {$\eta_{+}$};

\coordinate (NE) at (3,2*\yTop);
\draw[->,thick] (NE) -- (3,2*\yTop-0.8) node[below] {$\eta_{-}$};

\end{tikzpicture}
\end{equation}

\begin{equation}
\pgfmathsetmacro{\r}{1.9}
W_{+-}=\begin{tikzpicture}[baseline={([yshift=-.5ex]current bounding box.center)},every node/.style={font=\scriptsize}]

\pgfmathsetmacro{\yTop}{0.8*\r}
\draw (-2*\r,\yTop) -- (2*\r,\yTop);

\filldraw (\r/2,2*\yTop)  circle (1pt) node[above=0pt]{};
\filldraw (-\r/2,0) circle (1pt) node[above=0pt]{};

\draw [heavy scalar] (-\r/2,0) -- (\r/2,2*\yTop);

\filldraw (-1.5*\r,\yTop) circle (1pt) node[above=0pt]{$1$};
\filldraw (-0.8*\r,\yTop) circle (1pt) node[above=0pt]{$2$};
\filldraw ( 0.8*\r,\yTop) circle (1pt) node[below=0pt]{$3$};

\draw [thick] (-\r/2,0) -- (-1.5*\r,\yTop);
\draw [thick] (-\r/2,0) -- (-0.8*\r,\yTop);
\draw [thick] (\r/2,2*\yTop)  -- ( 0.8*\r,\yTop);
\draw [thick] (\r/2,2*\yTop) -- (\r/2+1,2.2*\yTop);

\draw plot[mark=x,mark size=3pt] coordinates {(\r/2+1,2.2*\yTop)};

\coordinate (SE) at (3,0);
\draw[->,thick] (SE) -- (3,0.8) node[above] {$\eta_{+}$};

\coordinate (NE) at (3,2*\yTop);
\draw[->,thick] (NE) -- (3,2*\yTop-0.8) node[below] {$\eta_{-}$};

\end{tikzpicture}
\end{equation}
\begin{equation}
\pgfmathsetmacro{\r}{1.9}
W_{--}=\begin{tikzpicture}[baseline={([yshift=-.5ex]current bounding box.center)},every node/.style={font=\scriptsize}]

\pgfmathsetmacro{\yTop}{0.8*\r}
\draw (-2*\r,\yTop) -- (2*\r,\yTop);

\filldraw (\r/2,2*\yTop)  circle (1pt) node[above=0pt]{};
\filldraw (-\r/2,2*\yTop) circle (1pt) node[above=0pt]{};

\draw [heavy scalar] (-\r/2,2*\yTop) -- (\r/2,2*\yTop);

\filldraw (-1.5*\r,\yTop) circle (1pt) node[below=0pt]{$1$};
\filldraw (-0.8*\r,\yTop) circle (1pt) node[below=0pt]{$2$};
\filldraw ( 0.8*\r,\yTop) circle (1pt) node[below=0pt]{$3$};

\draw [thick] (-\r/2,2*\yTop) -- (-1.5*\r,\yTop);
\draw [thick] (-\r/2,2*\yTop) -- (-0.8*\r,\yTop);
\draw [thick] (\r/2,2*\yTop)  -- ( 0.8*\r,\yTop);
\draw [thick] (\r/2,2*\yTop) -- (\r/2+1,2.2*\yTop);

\draw plot[mark=x,mark size=3pt] coordinates {(\r/2+1,2.2*\yTop)};

\coordinate (SE) at (3,0);
\draw[->,thick] (SE) -- (3,0.8) node[above] {$\eta_{+}$};

\coordinate (NE) at (3,2*\yTop);
\draw[->,thick] (NE) -- (3,2*\yTop-0.8) node[below] {$\eta_{-}$};

\end{tikzpicture}
\end{equation}
\begin{equation}
\pgfmathsetmacro{\r}{1.9}
W_{-+}=\begin{tikzpicture}[baseline={([yshift=-.5ex]current bounding box.center)},every node/.style={font=\scriptsize}]

\pgfmathsetmacro{\yTop}{0.8*\r}
\draw (-2*\r,\yTop) -- (2*\r,\yTop);

\filldraw (\r/2,0)  circle (1pt) node[above=0pt]{};
\filldraw (-\r/2,2*\yTop) circle (1pt) node[above=0pt]{};

\draw [heavy scalar] (-\r/2,2*\yTop) -- (\r/2,0);

\filldraw (-1.5*\r,\yTop) circle (1pt) node[below=0pt]{$1$};
\filldraw (-0.8*\r,\yTop) circle (1pt) node[below=0pt]{$2$};
\filldraw ( 0.8*\r,\yTop) circle (1pt) node[above=0pt]{$3$};

\draw [thick] (-\r/2,2*\yTop) -- (-1.5*\r,\yTop);
\draw [thick] (-\r/2,2*\yTop) -- (-0.8*\r,\yTop);
\draw [thick] (\r/2,0)  -- ( 0.8*\r,\yTop);
\draw [thick] (\r/2,0) -- (-10:\r);

\draw plot[mark=x,mark size=3pt] coordinates {(-10:\r)};

\coordinate (SE) at (3,0);
\draw[->,thick] (SE) -- (3,0.8) node[above] {$\eta_{+}$};

\coordinate (NE) at (3,2*\yTop);
\draw[->,thick] (NE) -- (3,2*\yTop-0.8) node[below] {$\eta_{-}$};

\end{tikzpicture}\label{eq:tree4}
\end{equation}
where the cross denotes the insertion of the background field. Since the tree diagram is ultraviolet finite, we keep the vertex time measures at $d=3$ and continue only the massive propagator to $d$ dimensions. This matches the treatment of the bubble in Sec.~\ref{sec:bubble}, and we set $d$ to $3$ before numerical evaluation. The integrals are
\begin{equation}\label{wittendiagram}
\begin{split}
W_{\pm_{1}\pm_{2}}=&\int_{-\infty}^{0}\frac{d\eta_{1}}{(-\eta_{1})^{2}}\frac{d\eta_{2}}{(-\eta_{2})^{2}} [\partial_{\eta_{1}}G^{\pm_{1}}(\vec{k}_{1},\eta_{1})][\partial_{\eta_{1}}G^{\pm_{1}}(\vec{k}_{2},\eta_{1})] \\
&\quad \times G^{\pm_{1}\pm_{2}}(\vec{k}_{3},\eta_{1},\eta_{2})[\partial_{\eta_{2}}G^{\pm_{2}}(\vec{k}_{3},\eta_{2})][\partial_{\eta_{2}}\phi_{0}(\eta_{2})]
\end{split}
\end{equation}
where $\pm_{1}$ and $\pm_{2}$ label the contours containing $\eta_{1}$ and $\eta_{2}$. Following the notation of~\cite{Qin:2023ejc}, the relevant integrals can be written as 
\begin{equation}
\begin{split}
\tilde{\mathcal{I}}_{\pm_{1}\pm_{2}}&=-\pm_{1}\pm_{2}|\vec{k}_{3}|^{3}\int_{-\infty}^{0}d\eta_{1}d\eta_{2}(-\eta_{2})^{-2}e^{\pm_{1}i(|\vec{k}_{1}|+|\vec{k}_{2}|)\eta_1\pm_{2}i|\vec{k}_{3}|\eta_2}G^{\pm_{1}\pm_{2}}(\vec{k}_{3},\eta_{1},\eta_{2})
\end{split}
\end{equation}
Using Eq.~\eqref{ex:extprop} and $\partial_{\eta}\phi_{0}=\epsilon_{\rm sr}/\eta$, the Witten diagrams and these integrals are related by
\begin{equation}
W_{\pm_{1}\pm_{2}}=-\pm_{1}\pm_{2}\,\frac{\epsilon_{\rm sr}}{8|\vec{k}_{1}||\vec{k}_{2}||\vec{k}_{3}|^{4}}\,\tilde{\mathcal{I}}_{\pm_{1}\pm_{2}}\,.
\end{equation}
Complex conjugation gives $(\tilde{\mathcal{I}}_{\pm_{1}\pm_{2}})^{*}=\tilde{\mathcal{I}}_{\mp_{1}\mp_{2}}$, so it is enough to evaluate $\tilde{\mathcal{I}}_{++}$ and $\tilde{\mathcal{I}}_{+-}$. Including the factor of four from the contractions of the unit-normalized interaction above, the $\vec{k}_3$ exchange channel is
\begin{equation}\label{eq:treechannel}
\left\langle
\delta\phi_{\vec{k}_1}\delta\phi_{\vec{k}_2}\delta\phi_{\vec{k}_3}
\right\rangle'_{\sigma_{\vec{k}_3}}
=
\frac{\epsilon_{\rm sr}}
{|\vec{k}_1||\vec{k}_2||\vec{k}_3|^4}
\operatorname{Re}\!\left(
\tilde{\mathcal I}_{++}+\tilde{\mathcal I}_{+-}
\right) \, .
\end{equation}
Here the prime means that the momentum-conserving delta function has been stripped, and the subscript denotes the channel in which $\sigma$ carries $\vec{k}_{3}$. This channel is symmetric under $\vec{k}_1\leftrightarrow\vec{k}_2$. The full bispectrum includes the two other exchange channels. We now evaluate the two independent integrals using the WF parameterization.

\subsection{Evaluating \texorpdfstring{$\tilde{\mathcal{I}}_{++}$}{I++}}

We first consider $\tilde{\mathcal{I}}_{++}$. We derive the Schwinger form of the WF representation by replacing the bulk propagator with its Schwinger form because there are no loops to integrate over. For convenience, we also make the change of variables $\eta_{i}\to -\eta_{i}$. This is not a contour deformation.
\begin{equation}
\begin{split}
\tilde{\mathcal{I}}_{++}&=-|\vec{k}_{3}|^{3}N_{\Delta_{+}}\int_{0}^{\infty} d\eta_{1}d\eta_{2}d\beta d\kappa \ \frac{\eta _1^{\Delta _+} \eta _2^{\Delta _+-2} \kappa ^{-2 \Delta _++d-1} (\beta -i \kappa )^{\Delta _+-\frac{d}{2}-\frac{1}{2}}}{\sqrt{\beta}} \\
&\times \exp\left (-\frac{1}{4} i \beta  |\vec{k}_{3}|^2-i \eta _2 |\vec{k}_{3}|-\frac{i(\eta_{1}-\eta_{2})^{2}}{\beta}-i \eta _1 |\vec{k}_{1}|-i \eta _1 |\vec{k}_{2}|-\epsilon \beta-\epsilon(2\eta_1+\eta_2) -\frac{4 \eta _1 \eta _2}{\kappa } \right )
\end{split}
\end{equation}
As emphasized in Sec.~\ref{sec:WF-ieps}, this form is still unsuitable for integration due to the explicit $i\epsilon$ dependence. To remedy the situation, we perform an explicit contour deformation.
\begin{equation}\label{eq:contour_deform_pp}
\begin{split}
\eta_{i}&\rightarrow e^{-(1/2) i \theta}\tau_{i} \\
\kappa&\rightarrow e^{ -i  \theta}\bar{\kappa} \\
\beta&\rightarrow e^{-(1/2) i \theta}\overline{\beta}
\end{split}
\end{equation}
where we ultimately set $\theta = \pi$ and then $\epsilon=0$. One can explicitly check that the integral is convergent for $0\leq \theta\leq \pi$, so it is clear we are allowed to deform the original contour to this new contour, which is more suitable for numerical integration. The integral then becomes 
\begin{equation}\label{eq:contourdeformedIpp}
\begin{split}
\tilde{\mathcal{I}}_{++}=&-|\vec{k}_{3}|^{3}e^{-\frac{1}{4} i \pi  \left(3 d-2 \Delta _+\right)}N_{\Delta_{+}}\int_{0}^{\infty}d\tau_{1}d\tau_{2}d\overline{\beta}d\bar{\kappa}\tau _1^{\Delta _+} \tau _2^{\Delta _+-2} \bar{\kappa }^{-2 \Delta _++d-1} \left(\overline{\beta }-\bar{\kappa }\right)^{\Delta _+-\frac{d}{2}-\frac{1}{2}} \\
&\times \frac{1}{\sqrt{\overline{\beta}}}\exp \left(-\frac{1}{4} \overline{\beta } |\vec{k}_{3}|^2-\frac{\tau _2 \bar{\kappa } |\vec{k}_{3}|+(|\vec{k}_{1}|+|\vec{k}_{2}|) \tau _1 \bar{\kappa }+4 \tau _2 \tau _1}{\bar{\kappa }}-\frac{\left(\tau _1-\tau _2\right){}^2}{\overline{\beta }}\right)
\end{split}
\end{equation}
where the exponent is now negative definite, and the integral is trivially convergent. The next step is to gauge fix the remaining $GL(1)$ symmetry, as described at the end of Sec.~\ref{sec:wittendiagramgeul}, thereby replacing the exponential factor with its exponent raised to a power.
\begin{equation}\label{sector1}
\begin{split}
&\tilde{\mathcal{I}}_{++}=-e^{-\frac{1}{4} i \pi  \left(3 d-2 \Delta _+\right)}|\vec{k}_{3}|^{3}N_{\Delta_{+}}\Gamma \left(\Delta _++\frac{d}{2}\right)\int_{0}^{\infty}d\tau_{1} d\tau_{2}d \overline{\beta}d\bar{\kappa} \frac{\tau _1^{\Delta _+} \tau _2^{\Delta _+-2}}{\sqrt{\overline{\beta }}} \\
&\quad \times \bar{\kappa }^{-2 \Delta _++d-1} \left(\overline{\beta }-\bar{\kappa }\right)^{\Delta _+-\frac{d}{2}-\frac{1}{2}} \times \delta(1-\mathbf{m}\  \overline{\beta}-\mathbf{n}\ \bar{\kappa}-\mathbf{q}_{1}\ \tau_{1}-\mathbf{q}_{2}\ \tau_{2})\\
&\quad \times \left(\frac{1}{4} \overline{\beta } |\vec{k}_{3}|^2+\frac{\left(\tau _1-\tau _2\right){}^2}{\overline{\beta }}+\frac{4 \tau _1 \tau _2}{\bar{\kappa }}+\tau _2 |\vec{k}_{3}|+|\vec{k}_{1}| \tau _1+|\vec{k}_{2}| \tau _1\right){}^{-\Delta _+-\frac{d}{2}}
\end{split}
\end{equation}
While this integral looks quite complicated, it is actually very straightforward to numerically integrate. We evaluate Eq.~\eqref{sector1} with the reduced Schwinger method of Sec.~\ref{sec:redschwinger}.

We first decompose Eq.~\eqref{sector1} into sign-definite regions. Since the polynomial contains terms with negative coefficients, the form of the integral in Eq.~\eqref{eq:contourdeformedIpp} is not quite optimal for numerical evaluation. However, this problem is easy to deal with. We use the regions listed below.
\begin{equation}\label{treedec}
\begin{split}
\textrm{Region 1}:& \quad \overline{\beta}>\bar{\kappa}, \quad \tau_{1}>\tau_{2} \\
\textrm{Region 2}:& \quad \overline{\beta}<\bar{\kappa}, \quad \tau_{1}>\tau_{2} \\
\textrm{Region 3}:& \quad \overline{\beta}>\bar{\kappa}, \quad \tau_{1}<\tau_{2} \\
\textrm{Region 4}:& \quad \overline{\beta}<\bar{\kappa}, \quad \tau_{1}<\tau_{2} \\
\end{split}
\end{equation}
In each region, we change coordinates so the integration variables range from 0 to infinity. For example, in Region 1, we make the replacement 
\begin{equation}
\overline{\beta}\to \bar{\kappa}+\Delta\overline{\beta}, \quad \tau_{1}\to \tau_{2}+\Delta\tau 
\end{equation}
where $\bar{\kappa}$, $\tau_{2}$, $\Delta\overline{\beta}$, and $\Delta\tau$ are all integrated from zero to infinity. In these coordinates, all coefficients in the polynomials are positive. We consider the integral from Region 1 as an illustrative example. The relevant polynomials are
\begin{equation}
\begin{split}
P_{1}=&2 \bar{\kappa }^2 \left(|\vec{k}_{3}| \left(\Delta \overline{\beta } |\vec{k}_{3}|+2 \tau _2\right)+2 (|\vec{k}_{1}|+|\vec{k}_{2}|) \left(\Delta \tau+\tau _2\right)\right)\\
&+\bar{\kappa } \left(4 \Delta \overline{\beta } \left(\tau _2 (|\vec{k}_{3}|+|\vec{k}_{1}|+|\vec{k}_{2}|)+\Delta \tau (|\vec{k}_{1}|+|\vec{k}_{2}|)\right)+\Delta \overline{\beta }^2 |\vec{k}_{3}|^2+4 \left(\Delta \tau+2 \tau _2\right){}^2\right)\\
&+\bar{\kappa }^3 |\vec{k}_{3}|^2+16 \tau _2 \Delta \overline{\beta } \left(\Delta \tau+\tau _2\right) \ , \\
P_{2}=&\bar{\kappa}+\Delta \overline{\beta} \ , \\
P_{3}=&\tau_{2}+\Delta \tau \ .
\end{split}
\end{equation}
Here $P_{1}$ is $4\overline{\beta}\bar{\kappa}$ times the bracket in Eq.~\eqref{sector1}, while $P_{2}=\overline{\beta}$ and $P_{3}=\tau_{1}$ are the other two polynomial factors of the integrand, all written in the shifted coordinates. 

In Region 1, the positive coordinate $\Delta\tau=\tau_1-\tau_2$ can be used to fix the common scale. We choose the sector-local gauge $\delta(1-\Delta\tau)$. Equivalently, this sets $\mathbf m=\mathbf n=0$, $\mathbf q_1=1$, and $\mathbf q_2=-1$ in Eq.~\eqref{sector1}. This gauge is valid in Region 1 because $\tau_1-\tau_2>0$ throughout the sector. Writing $r$ for $\tau_2$ gives
\begin{equation}
\tau_1=1+r\,,\qquad \tau_2=r\,.
\end{equation}
With this choice, the delta function is saturated. Its constraint is satisfied, and all remaining variables are integrated from zero to infinity with no constraint left.
We then separate the common scale. Since Region 1 has $\overline{\beta}>\bar{\kappa}$, we can write
\begin{equation}
\frac{1}{\overline{\beta}}=s\,,\qquad \frac{1}{\bar{\kappa}}=s(1+z)\,,\qquad \text{so that}\quad d\!\left(\tfrac{1}{\overline{\beta}}\right)d\!\left(\tfrac{1}{\bar{\kappa}}\right)=s\,ds\,dz\,.
\end{equation}
The integrand then takes the factorized form of Eq.~\eqref{eq:redschfactor}, with
\begin{equation}
\begin{split}
A &= (|\vec{k}_1|+|\vec{k}_2|)(1+r)+|\vec{k}_3| r \,,\\
B &= \frac{|\vec{k}_3|^2}{4}\,,\\
C &= 1+4r+4r^2+4rz+4r^2z\,.
\end{split}
\end{equation}
When we set $d$ to $3$, the exponents $\alpha$ and $N$ take the values $i\widetilde{\nu}$ and $3+i\widetilde{\nu}$, respectively. The $s$ integral is exactly the kernel $\mathcal{K}_{\alpha,N}(A,B,C)$, and the remaining factors are $W=z^{i\widetilde{\nu}-1/2}(1+z)^{i\widetilde{\nu}-1/2}r^{i\widetilde{\nu}-1/2}(1+r)^{3/2+i\widetilde{\nu}}$. The remaining numerical integral is the two-dimensional $(z,r)$ quadrature. The other three regions are treated in the same way. In Regions 2 and 4, the continuous contour fixes the branch
\begin{equation}
(\overline{\beta}-\bar{\kappa})^{q}
=e^{-i\pi q}(\bar{\kappa}-\overline{\beta})^{q}\,,
\qquad
q=\Delta_+-\frac{d}{2}-\frac{1}{2}\,.
\end{equation}

Summing the contributions from the four regions, we find agreement with the analytic expression in Ref.~\cite{Qin:2023ejc} for a collection of random kinematic points. Given the simplicity of the integrals, we note that Mathematica's NIntegrate was also suitable.

\subsection{Evaluating \texorpdfstring{$\tilde{\mathcal{I}}_{+-}$}{I+-}}

We now consider the $\tilde{\mathcal{I}}_{+-}$ integral. The Schwinger form of the WF-parameterized integral is 
\begin{equation}\label{eq:Ipm}
\begin{split}
\tilde{\mathcal{I}}_{+-}&=|\vec{k}_{3}|^{3}N_{\Delta_{+}} \int_{0}^{\infty} d\eta_{1}d\eta_{2} d\kappa \int_{0}^{\operatorname{sgn}(\eta_{1}-\eta_{2})\infty} d\beta \ \frac{\eta _1^{\Delta _+} \eta _2^{\Delta _+-2} \kappa ^{-2 \Delta _++d-1} (\beta -i \kappa )^{\Delta _+-\frac{d}{2}-\frac{1}{2}}}{\sqrt{\beta}} \\
&\times \exp\left (-\frac{1}{4} i \beta  |\vec{k}_{3}|^2+i \eta _2 |\vec{k}_{3}|-i \eta _1 |\vec{k}_{1}|-i \eta _1 |\vec{k}_{2}|-\frac{i(\eta_{1}-\eta_{2})^{2}}{\beta} -\frac{4 \eta _1 \eta _2}{\kappa } -\epsilon \beta\operatorname{sgn}(\eta_{1}-\eta_{2})-\epsilon(2\eta_1+\eta_2)\right )
\end{split}
\end{equation}
where the convergent contour for $\beta$ depends on $\operatorname{sgn}(\eta_{1}-\eta_{2})$. On the negative ray, we take $\beta=|\beta|e^{-i\pi}$ as fixed by the prescription $\beta\to\beta-i0$, and we continue the fractional powers $\sqrt{\beta}$ and $(\beta-i\kappa)^{\Delta_+-d/2-1/2}$ on this branch. We find that the contour deformation
\begin{equation}\label{eq:contour_deform_pm}
\begin{split}
\eta_{1}&\rightarrow-i\tau_{1} \\
\eta_{2}&\rightarrow i \tau_{2} \\
\kappa&\rightarrow\bar{\kappa} \\
\beta&\rightarrow -i \overline{\beta}
\end{split}
\end{equation}
is related via a smooth contour deformation to the contour in Eq.~\eqref{eq:Ipm}. After fixing the $GL(1)$ redundancy, the final integral is 
\begin{equation}\label{eq:Ipmwf}
\begin{split}
&\tilde{\mathcal{I}}_{+-}=-e^{-\frac{1}{4} i \pi  \left(2 \Delta _+-d\right)}|\vec{k}_{3}|^{3}N_{\Delta_{+}}\Gamma \left(\Delta _++\frac{d}{2}\right)\int_{0}^{\infty}d\tau_{1} d\tau_{2}d \overline{\beta}d\bar{\kappa} \frac{\tau _1^{\Delta _+} \tau _2^{\Delta _+-2}}{\sqrt{\overline{\beta }}} \\
&\quad \times \bar{\kappa }^{-2 \Delta _++d-1} \left(\overline{\beta }+\bar{\kappa }\right)^{\Delta _+-\frac{d}{2}-\frac{1}{2}} \times \delta(1-\mathbf{m}\  \overline{\beta}-\mathbf{n}\ \bar{\kappa}-\mathbf{q}_{1}\ \tau_{1}-\mathbf{q}_{2}\ \tau_{2})\\
&\quad \times \left(\frac{1}{4} \overline{\beta } |\vec{k}_{3}|^2+\frac{\left(\tau _1+\tau _2\right){}^2}{\overline{\beta }}+\frac{4 \tau _1 \tau _2}{\bar{\kappa }}+\tau _2 |\vec{k}_{3}|+|\vec{k}_{1}| \tau _1+|\vec{k}_{2}| \tau _1\right){}^{-\Delta _+-\frac{d}{2}}
\end{split}
\end{equation}
We apply the same procedure as for $\tilde{\mathcal{I}}_{++}$. Using the delta function in Eq.~\eqref{eq:Ipmwf}, we choose the gauge $\tau_1=1$ and write $r$ for $\tau_2$. We then set $1/\overline{\beta}=s$ and $1/\bar{\kappa}=sz$, for which $d(1/\overline{\beta})\,d(1/\bar{\kappa})=s\,ds\,dz$. This gives
\begin{equation}
\begin{split}
A &= |\vec{k}_1|+|\vec{k}_2|+|\vec{k}_3| r \,,\\
B &= \frac{|\vec{k}_3|^2}{4}\,,\\
C &= 1+2r+r^2+4rz\,,\\
W &= z^{i\widetilde{\nu}-1/2}(1+z)^{i\widetilde{\nu}-1/2} r^{i\widetilde{\nu}-1/2}\,,\\
\tilde{\mathcal{I}}_{+-} &\propto \int_0^\infty dz\, dr\, W \,\mathcal{K}_{\alpha,N}(A,B,C)\,,
\end{split}
\end{equation}
up to the prefactors in Eq.~\eqref{eq:Ipmwf}. We verify agreement with the closed-form expression in Ref.~\cite{Qin:2023ejc} for a collection of random kinematic points.

\section{Bubble contribution to the bispectrum}\label{sec:bubble}

With explicit numerical evaluations for the tree-level contribution, we now turn to the bubble. Analytic results for the bubble exist only in special kinematic limits~\cite{Chen:2016uwp,Chen:2016hrz, Chen:2016nrs} or, for principal-series masses $\widetilde{\nu}=\sqrt{m^{2}-d^{2}/4}>0$, as series expansions in the squeezed ratio~\cite{Xianyu:2022jwk,Zhang:2025nzd,Liu:2026jzn}. This section gives a direct numerical computation of the bubble for completely general kinematics and masses.

We consider the graphs generated by the interaction term
\begin{equation}\label{desirabelcontri}
\mathcal{L}_{\textrm{int}}\ni \frac{c}{\Lambda^{2}}(\partial^{\eta}\phi)(\partial_{\eta}\phi)\sigma^{2}\,,
\end{equation}
where $\sigma$ is again a real scalar of mass $m$ at leading order in slow-roll. 
Couplings that are quadratic in $\sigma$ arise very naturally in well-motivated UV completions, as emphasized in the introduction. Before doing any calculations, we briefly summarize the general motivation below.
\begin{itemize}
\item First, if $\sigma$ is charged under a gauge symmetry (or carries any
conserved quantum number), then a coupling linear in a single~$\sigma$ is
forbidden by charge conservation~\cite{Bodas:2025vpb}, whereas the neutral
bilinear~$\sigma\sigma^{\star}$ is invariant.  The leading portal from the
inflaton sector to the charged sector is therefore generically bilinear in
the heavy field.
\item Second, this derivative structure naturally arises from the K\"ahler potential in supersymmetric completions. Because $K$ determines the
field-space metric for the scalar kinetic terms, the higher-dimensional
operator
\begin{equation}
  K \;\supset\; \frac{c}{\Lambda^{2}}\,
  \Phi^{\dagger}\Phi\;\Sigma^{\dagger}\Sigma \, ,
  \label{eq:KahlerPortal}
\end{equation}
modifies the inflaton kinetic term through the K\"ahler
metric $K_{\Phi\bar\Phi}=1+c\,|\sigma|^{2}/\Lambda^{2}$, yielding in
components a field-dependent kinetic term
\begin{equation}
  \mathcal{L}\;\supset\;
  \frac{c}{\Lambda^{2}}\,|\sigma|^{2}\,
  \partial_{\mu}\phi\,\partial^{\mu}\phi^{\ast}\,,
\end{equation}
which for a real inflaton leads to the contribution as Eq.~\eqref{desirabelcontri}.  Note that gauge invariance forces
matter to enter $K$ in the neutral combination $\Sigma^{\dagger}\Sigma$.  Such
K\"ahler-induced kinetic portals are well motivated in SUSY
grand-unified scenarios whose characteristic mass scale lies near the
inflationary scale, since heavy GUT-sector multiplets then couple
naturally to the inflaton through operators of the form in Eq.~\eqref{eq:KahlerPortal}~\cite{Kumar:2017ecc,Kumar:2018jxz}.
\end{itemize}
Having motivated the computation, we now explain the object we will compute.

The relevant diagrams are the same as at tree level in Eqs.~\eqref{eq:tree1}--\eqref{eq:tree4}, except that the internal propagator is replaced by a bubble. In contrast to the tree-level exchange, the loop integral over the spatial momentum $\vec{\ell}$ is ultraviolet divergent in $D=d+1=4$, so we must specify a regulator before writing down the integrals. We employ dimensional regularization, continuing the number of spatial dimensions to
\begin{equation}\label{eq:dimregd}
d=3-2\varepsilon
\end{equation}
and expanding the results in powers of $\varepsilon$. In dimensional regularization the masses of the fields become $d$-dependent~\cite{Senatore:2009cf,Chen:2016hrz}. For the exchanged field $\sigma$, we continue the scaling dimensions as
\begin{equation}\label{eq:sigmadimreg}
\Delta_{\pm}=\frac{d}{2}\pm i\widetilde{\nu}\,,
\end{equation}
holding $\widetilde{\nu}$ fixed, so that the Schwinger representation of Sec.~\ref{sec:wittrop} applies verbatim in $d$ dimensions. The background field $\phi_{0}(\eta)$ is a classical solution rather than a mode function, and its derivative $\partial_{\eta}\phi_{0}=\epsilon_{\rm sr}/\eta$ is not continued. With these choices the Witten diagrams are
\begin{equation}\label{wittendiagram2}
\begin{split}
W_{\pm_{1}\pm_{2}}=&\int_{-\infty}^{0}\frac{d\eta_{1}}{(-\eta_{1})^{d-1}}\frac{d\eta_{2}}{(-\eta_{2})^{d-1}} [\partial_{\eta_{1}}G_{d}^{\pm_{1}}(\vec{k}_{1},\eta_{1})][\partial_{\eta_{1}}G_{d}^{\pm_{1}}(\vec{k}_{2},\eta_{1})][\partial_{\eta_{2}}G_{d}^{\pm_{2}}(\vec{k}_{3},\eta_{2})][\partial_{\eta_{2}}\phi_{0}(\eta_{2})] \\
&\quad \times \int_{-\infty}^{\infty} \frac{d^{d}\vec{\ell}}{(2\pi)^{d}} \ G_{d}^{\pm_{1}\pm_{2}}(\vec{\ell},\eta_{1},\eta_{2})G_{d}^{\pm_{1}\pm_{2}}(\vec{k}_{3}-\vec{\ell},\eta_{1},\eta_{2}) \ .
\end{split}
\end{equation}
However, this integral is UV divergent as the dimensional regulator $\varepsilon$ goes to zero, so we must add a counterterm to the action to cancel the UV divergence. This counterterm gives a tree contribution proportional to
\begin{equation}\label{counterterm1}
\begin{split}
C_{\pm_{1}\pm_{2}}=&\delta_{\pm_{1},\pm_{2}}\int_{-\infty}^{0}\frac{d\eta_{1}}{(-\eta_{1})^{d-3}} [\partial_{\eta_{1}}G_{d}^{\pm_{1}}(\vec{k}_{1},\eta_{1})][\partial_{\eta_{1}}G_{d}^{\pm_{1}}(\vec{k}_{2},\eta_{1})][\partial_{\eta_{1}}G_{d}^{\pm_{2}}(\vec{k}_{3},\eta_{1})][\partial_{\eta_{1}}\phi_{0}(\eta_{1})] \ , 
\end{split}
\end{equation}
corresponding to the Witten diagrams
\begin{equation}\label{eq:ctree1}
\pgfmathsetmacro{\r}{1.9}
C_{++}=
\begin{tikzpicture}[
    baseline={([yshift=-.5ex]current bounding box.center)},
    every node/.style={font=\scriptsize}
]

\pgfmathsetmacro{\yTop}{0.8*\r}
\draw (-2*\r,\yTop) -- (2*\r,\yTop);

\filldraw (0,0) circle (1pt);

\filldraw (-1.5*\r,\yTop) circle (1pt) node[above=0pt]{$1$};
\filldraw (-0.8*\r,\yTop) circle (1pt) node[above=0pt]{$2$};
\filldraw ( 0.8*\r,\yTop) circle (1pt) node[above=0pt]{$3$};

\draw[thick] (0,0) -- (-1.5*\r,\yTop);
\draw[thick] (0,0) -- (-0.8*\r,\yTop);
\draw[thick] (0,0) -- ( 0.8*\r,\yTop);
\draw[thick] (0,0) -- (-10:\r);

\draw plot[mark=x,mark size=3pt] coordinates {(-10:\r)};

\coordinate (SE) at (3,0);
\draw[->,thick] (SE) -- (3,0.8) node[above] {$\eta_{+}$};

\coordinate (NE) at (3,2*\yTop);
\draw[->,thick] (NE) -- (3,2*\yTop-0.8) node[below] {$\eta_{-}$};

\end{tikzpicture}
\end{equation}
\begin{equation}\label{eq:ctree1minus}
\pgfmathsetmacro{\r}{1.9}
C_{--}=
\begin{tikzpicture}[
    baseline={([yshift=-.5ex]current bounding box.center)},
    every node/.style={font=\scriptsize}
]

\pgfmathsetmacro{\yTop}{0.8*\r}
\draw (-2*\r,\yTop) -- (2*\r,\yTop);

\filldraw (0,2*\yTop) circle (1pt);

\filldraw (-1.5*\r,\yTop) circle (1pt) node[above=0pt]{$1$};
\filldraw (-0.8*\r,\yTop) circle (1pt) node[above=0pt]{$2$};
\filldraw ( 0.8*\r,\yTop) circle (1pt) node[above=0pt]{$3$};

\draw[thick] (0,2*\yTop) -- (-1.5*\r,\yTop);
\draw[thick] (0,2*\yTop) -- (-0.8*\r,\yTop);
\draw[thick] (0,2*\yTop) -- ( 0.8*\r,\yTop);

\begin{scope}[shift={(0,2*\yTop)}]
    \draw[thick] (0,0) -- (10:\r);
    \draw plot[mark=x,mark size=3pt] coordinates {(10:\r)};
\end{scope}

\coordinate (SE) at (3,0);
\draw[->,thick] (SE) -- (3,0.8) node[above] {$\eta_{+}$};

\coordinate (NE) at (3,2*\yTop);
\draw[->,thick] (NE) -- (3,2*\yTop-0.8) node[below] {$\eta_{-}$};

\end{tikzpicture} \ .
\end{equation}
The sum of the relevant Witten diagrams becomes
\begin{equation}\label{wittendiagram3}
\begin{split}
&W_{\pm_{1}\pm_{2}}+ c_{\rm ct} \ C_{\pm_{1}\pm_{2}}=\int_{-\infty}^{0}\frac{d\eta_{1}}{(-\eta_{1})^{d-1}}\frac{d\eta_{2}}{(-\eta_{2})^{d-1}} [\partial_{\eta_{1}}G_{d}^{\pm_{1}}(\vec{k}_{1},\eta_{1})][\partial_{\eta_{1}}G_{d}^{\pm_{1}}(\vec{k}_{2},\eta_{1})][\partial_{\eta_{2}}G_{d}^{\pm_{2}}(\vec{k}_{3},\eta_{2})]\\
&\quad \times [\partial_{\eta_{2}}\phi_{0}(\eta_{2})]  \left [ \int_{-\infty}^{\infty} \frac{d^{d}\vec{\ell}}{(2\pi)^{d}} \ G_{d}^{\pm_{1}\pm_{2}}(\vec{\ell},\eta_{1},\eta_{2})G_{d}^{\pm_{1}\pm_{2}}(\vec{k}_{3}-\vec{\ell},\eta_{1},\eta_{2})+c_{\rm ct} \ \delta_{\pm_{1},\pm_{2}} (-\eta_{1})^{d+1}\delta (\eta_{1}-\eta_{2})\right ] \ .
\end{split}
\end{equation}
where $c_{\rm ct}\propto \varepsilon^{-1}$ to cancel the divergence from the second line of Eq.~\eqref{wittendiagram2}. In this representation, it is clear that we do not need to keep the external propagators or the vertex time measures in general $d$ because the factor in brackets in Eq.~\eqref{wittendiagram3} is manifestly finite. Therefore, we consider the integral 
\begin{equation}\label{wittendiagram4}
\begin{split}
&W_{\pm_{1}\pm_{2}}+ c_{\rm ct} \ C_{\pm_{1}\pm_{2}} = \int_{-\infty}^{0}\frac{d\eta_{1}}{(-\eta_{1})^{2}}\frac{d\eta_{2}}{(-\eta_{2})^{2}} [\partial_{\eta_{1}}G_{d=3}^{\pm_{1}}(\vec{k}_{1},\eta_{1})][\partial_{\eta_{1}}G_{d=3}^{\pm_{1}}(\vec{k}_{2},\eta_{1})][\partial_{\eta_{2}}G_{d=3}^{\pm_{2}}(\vec{k}_{3},\eta_{2})]\\
&\left. \quad \times [\partial_{\eta_{2}}\phi_{0}(\eta_{2})]  \left [ \int_{-\infty}^{\infty} \frac{d^{d}\vec{\ell}}{(2\pi)^{d}} \ G_{d}^{\pm_{1}\pm_{2}}(\vec{\ell},\eta_{1},\eta_{2})G_{d}^{\pm_{1}\pm_{2}}(\vec{k}_{3}-\vec{\ell},\eta_{1},\eta_{2})\right ] \right  |_{\mathcal{O}(\varepsilon^{0})}\;+\;\Delta^{\rm ct}_{\pm_{1}\pm_{2}}\ ,
\end{split}
\end{equation}
where the $|_{\mathcal{O}(\varepsilon^{0})}$ means we are projecting the integral onto the $\mathcal{O}(\varepsilon^{0})$ term in the $\varepsilon$ Laurent expansion, and $\Delta^{\rm ct}_{\pm_1\pm_2}$ is the finite remainder of the counterterm, coming from the $c_{\rm ct}\propto \varepsilon^{-1}$ pole multiplying the $\mathcal{O}(\varepsilon)$ term from expanding $(-\eta_1)^{d+1}$. Explicitly,
\begin{equation}\label{eq:ctremainder}
\Delta^{\rm ct}_{++}=-2P\left[\log\frac{E}{H}-\psi(3)\right],
\qquad
\Delta^{\rm ct}_{+-}=0\,,
\end{equation}
where $E=|\vec{k}_{1}|+|\vec{k}_{2}|+|\vec{k}_{3}|$, $\psi(3)=\tfrac{3}{2}-\gamma_{E}$, and $P$ is the ultraviolet pole coefficient of the basis integral, given in Eq.~\eqref{eq:bubblepole} below. The bracket also carries an $i\pi$, which drops out
of the real part that enters the bispectrum, and the $(--)$ branch is the complex conjugate. We note that this regularization scheme amounts to imposing that $c_{\rm ct}$ is proportional to $\varepsilon^{-1}$ and does not have a sub-leading term that is $\mathcal{O}(\varepsilon^{0})$. We return to this choice in Sec. \ref{sec:CMBdata}.

The integral in Eq.~\eqref{wittendiagram4} is proportional to the basis integral
\begin{equation}\label{eq:bubblebasis}
\begin{split}
\tilde{\mathcal{B}}_{\pm_{1}\pm_{2}}&=-|\vec{k}_{3}|^{3}\int_{-\infty}^{0}d\eta_{1}d\eta_{2}(-\eta_{1})^{p_{1}}(-\eta_{2})^{p_{2}}e^{\pm_{1}i(|\vec{k}_{1}|+|\vec{k}_{2}|)\eta_{1}\pm_{2}i|\vec{k}_{3}|\eta_{2}} \\
&\quad \times \int \frac{d^{d}\vec{\ell}}{(2\pi)^d}\ G_{d}^{\pm_{1}\pm_{2}}(\vec{\ell},\eta_{1},\eta_{2})G_{d}^{\pm_{1}\pm_{2}}(\vec{k}_{3}-\vec{\ell},\eta_{1},\eta_{2}) \, ,
\end{split}
\end{equation}
with
\begin{equation}\label{eq:bubblepowers}
p_{1}=0\,,\qquad p_{2}=-2\,.
\end{equation}
This is the object we evaluate numerically. We call this the mixed dimensional-regularization scheme, since only the loop bracket is continued to $d$ dimensions. With this normalization, the constant relating Eq.~\eqref{eq:bubblebasis} to Eq.~\eqref{wittendiagram4} is independent of $d$. The ultraviolet pole of $\tilde{\mathcal{B}}_{++}$,
\begin{equation}\label{eq:bubblepole}
\tilde{\mathcal{B}}_{++}\ni\frac{P}{\varepsilon}\,,
\qquad
P\equiv-\frac{\pi|\vec{k}_{3}|^{3}}{(2\pi)^3E^{3}}\,,
\end{equation}
is independent of the mass, while $\tilde{\mathcal{B}}_{+-}$ is finite. This pole is the divergence canceled by the counterterm in Eq.~\eqref{wittendiagram3}. Since the full factor $(2\pi)^{-d}$ is continued in $d$, its expansion contributes $2P\log(2\pi)$ to the finite part relative to freezing it at $(2\pi)^{-3}$. We now turn to the evaluation of $\tilde{\mathcal{B}}_{\pm_{1}\pm_{2}}$ using the WF parameterization, following the approach outlined in the tree-level calculation.

\subsection{Evaluating \texorpdfstring{$\tilde{\mathcal{B}}_{++}$}{B++}}

We begin by computing the WF form of $\tilde{\mathcal{B}}_{++}$. Relative to the tree-level case, the new ingredient is the momentum bubble, whose $\vec\ell$-integration has already been carried out in Eq.~\eqref{eq:bubble_loop}. Thus, after inserting the Schwinger/WF form of the bulk propagators and making the innocuous change of variables $\eta_i\to -\eta_i$ (again a reparametrization, not a contour deformation), we arrive at the expression below. The external legs also carry the regulator $e^{-\epsilon(2\eta_{1}+\eta_{2})}$ of Sec.~\ref{sec:treeanalyt}, which we suppress here and in Sec.~\ref{sec:triangle}.
\begin{equation}
\begin{split}
&\tilde{\mathcal{B}}_{++}=-e^{-\frac{1}{4} i \pi  d}|\vec{k}_{3}|^{3}N_{\Delta_{+}}^{2}\int_{0}^{\infty} d\eta_{1}d\eta_{2} \ \eta _1^{2\Delta _+} \eta _2^{2\Delta _+-2} \prod_{i=1}^{2}d\beta_{i}d\kappa_{i}\frac{ \kappa_{i} ^{d-2 \Delta _+-1} (\beta_{i} -i \kappa_{i} )^{\Delta _+-\frac{d}{2}-\frac{1}{2}}}{\sqrt{\beta_{i}}} \\
&\times \frac{1}{(2\pi)^d}\left ( \frac{4\pi}{\beta_{1}+\beta_{2}}\right )^{d/2}\exp\bigg (-i \frac{ \beta_{1}\beta_{2} |\vec{k}_{3}|^2}{4(\beta_{1}+\beta_{2})}-i \eta _2 |\vec{k}_{3}|-i \eta _1 |\vec{k}_{1}|-i \eta _1 |\vec{k}_{2}|-i(\eta_{1}-\eta_{2})^{2}(\frac{1}{\beta_{1}}+\frac{1}{\beta_{2}})\\
&\quad \quad \quad \quad \quad \quad \quad \quad \quad  -4 \eta _1 \eta _2(\frac{1}{\kappa_{1}}+\frac{1}{\kappa_{2}})-\epsilon \beta_{1}-\epsilon \beta_{2} \bigg )
\end{split}
\end{equation}
The remaining integrals over $\eta$ and the Schwinger parameters inherit the same oscillatory Lorentzian phase and explicit $i\epsilon$ regulator as in the tree-level example, so at this stage the representation is still poorly suited to direct numerics. We again consider the contour deformations
\begin{equation}\label{contourdef3}
\begin{split}
\eta_{i}&\rightarrow e^{-(1/2) i \theta}\tau_{i}\,, \\
\kappa_{i}&\rightarrow e^{- i \theta}\bar{\kappa}_{i}\,, \\
\beta_{i}&\rightarrow e^{-(1/2) i \theta}\overline{\beta}_{i}\,,
\end{split}
\end{equation}
and then take $\theta=\pi$ followed by $\epsilon\to0$. As in the tree-level case, the deformation is along a domain where the integral remains convergent for $0\le \theta\le\pi$. The resulting representation is therefore absolutely convergent and numerically stable. We then fix the $GL(1)$ rescaling symmetry as in the tree-level case. The final result is 
\begin{align}\label{eq:bubblefinalWF}
&\tilde{\mathcal{B}}_{++}=e^{i \pi  \left(\Delta _+-\frac{d}{2}\right)}e^{i\pi(3-d)}|\vec{k}_{3}|^{3}\Gamma \!\left(2\Delta_{+}+\frac{d}{2}\right)N_{\Delta_{+}}^{2}\int_{0}^{\infty} d\tau_{1}d\tau_{2} \ \tau _1^{2\Delta _+} \tau _2^{2\Delta _+-2}  \\
&\times \left (\prod_{i=1}^{2}d\overline{\beta}_{i}d\bar{\kappa}_{i}\frac{ \bar{\kappa}_{i} ^{d-2 \Delta _+-1} (\overline{\beta}_{i} -\bar{\kappa}_{i} )^{\Delta _+-\frac{d}{2}-\frac{1}{2}}}{\sqrt{\overline{\beta}_{i}}} \right )\frac{1}{(2\pi)^d}\left ( \frac{4\pi}{\overline{\beta}_{1}+\overline{\beta}_{2}}\right )^{d/2} \delta\!\left(1-\sum_{i}(\mathbf{m}_{i}\  \overline{\beta}_{i}+\mathbf{n}_{i}\ \bar{\kappa}_{i})-\sum_{i}\mathbf{q}_{i}\ \tau_{i}\right)\nonumber \\
&\times \mathcal{P}_{++}^{\,-(2\Delta_{+}+d/2)}\nonumber
\end{align}
where the polynomial $\mathcal{P}_{++}$ is
\begin{equation}\label{eq:Ppp}
\mathcal{P}_{++}=\frac{\overline{\beta }_1 \overline{\beta }_2 |\vec{k}_{3}|^2}{4 \left(\overline{\beta }_1+\overline{\beta }_2\right)}+\left(\tau _1-\tau _2\right){}^2\!\left (\frac{1}{\overline{\beta}_{1}}+\frac{1}{\overline{\beta}_{2}}\right )+4 \tau _1 \tau _2\!\left (\frac{1}{\bar{\kappa}_{1}}+\frac{1}{\bar{\kappa}_{2}}\right )+\tau _2 |\vec{k}_{3}|+\tau _1 |\vec{k}_{1}| +\tau _1|\vec{k}_{2}| \ .
\end{equation}
The overall exponent
\begin{equation}\label{eq:bubbleNtot}
N=2\Delta_{+}+p_{1}+p_{2}+2+\tfrac{d}{2}=2\Delta_{+}+\tfrac{d}{2}
\end{equation}
is the total scaling weight of the integrand. In contrast to the tree-level case, Eq.~\eqref{eq:bubblefinalWF} suffers from a UV divergence in $D=d+1=4$, so we use the reduced Schwinger method of Sec.~\ref{sec:redschwinger} together with the subtraction at the level of the reduced integrand described in Sec.~\ref{sec:redschdiv}, evaluating the resulting integrals as Laurent series in $\varepsilon$ with $\widetilde{\nu}$ held fixed, as specified in Eqs.~\eqref{eq:dimregd} and \eqref{eq:sigmadimreg}.

We now turn to numerical integration. The polynomial coefficients are again not all positive, so we break up the integration region into subregions as we did at tree level. There are now eight regions.
\begin{equation}\label{bubbledec}
\begin{split}
\textrm{Region 1}:& \quad \overline{\beta}_{1}>\bar{\kappa}_{1}, \quad \overline{\beta}_{2}>\bar{\kappa}_{2}, \quad \tau_{1}>\tau_{2} \\
\textrm{Region 2}:& \quad \overline{\beta}_{1}>\bar{\kappa}_{1}, \quad \overline{\beta}_{2}<\bar{\kappa}_{2}, \quad \tau_{1}>\tau_{2} \\
\textrm{Region 3}:& \quad \overline{\beta}_{1}>\bar{\kappa}_{1},\quad \overline{\beta}_{2}>\bar{\kappa}_{2}, \quad \tau_{1}<\tau_{2} \\
\textrm{Region 4}:&\quad \overline{\beta}_{1}>\bar{\kappa}_{1}, \quad \overline{\beta}_{2}<\bar{\kappa}_{2}, \quad \tau_{1}<\tau_{2} \\
\textrm{Region 5}:& \quad \overline{\beta}_{1}<\bar{\kappa}_{1}, \quad \overline{\beta}_{2}>\bar{\kappa}_{2}, \quad \tau_{1}>\tau_{2} \\
\textrm{Region 6}:& \quad \overline{\beta}_{1}<\bar{\kappa}_{1}, \quad \overline{\beta}_{2}<\bar{\kappa}_{2}, \quad \tau_{1}>\tau_{2} \\
\textrm{Region 7}:& \quad \overline{\beta}_{1}<\bar{\kappa}_{1},\quad \overline{\beta}_{2}>\bar{\kappa}_{2}, \quad \tau_{1}<\tau_{2} \\
\textrm{Region 8}:&\quad \overline{\beta}_{1}<\bar{\kappa}_{1}, \quad \overline{\beta}_{2}<\bar{\kappa}_{2}, \quad \tau_{1}<\tau_{2}  \\
\end{split}
\end{equation}
In each region, we change coordinates so the integration variables range from zero to infinity. For example, in Region 1, we make the replacement 
\begin{equation}
\overline{\beta}_{1}\to \bar{\kappa}_{1}+\Delta\overline{\beta}_{1},\quad \overline{\beta}_{2}\to \bar{\kappa}_{2}+\Delta\overline{\beta}_{2}, \quad \tau_{1}\to \tau_{2}+\Delta\tau 
\end{equation}
where $\bar{\kappa}_{i}$, $\tau_{2}$, $\Delta\overline{\beta}_{i}$, and $\Delta\tau$ are all integrated from zero to infinity. In regions with $\overline{\beta}_{i}<\bar{\kappa}_{i}$, the factor $(\overline{\beta}_{i}-\bar{\kappa}_{i})^{\Delta _+-\frac{d}{2}-\frac{1}{2}}$ in Eq.~\eqref{eq:bubblefinalWF} follows the branch dictated by the continuous contour deformation in Eq.~\eqref{contourdef3}. On this branch, $(\overline{\beta}_{i}-\bar{\kappa}_{i})^{\Delta _+-\frac{d}{2}-\frac{1}{2}}=e^{-i \pi  \left(\Delta _+-\frac{d}{2}-\frac{1}{2}\right)}(\bar{\kappa}_{i}-\overline{\beta}_{i})^{\Delta _+-\frac{d}{2}-\frac{1}{2}}$. There is one such factor for each edge with $\overline{\beta}_{i}<\bar{\kappa}_{i}$. This choice differs from the principal branch by $e^{-2 i \pi  \left(\Delta _+-\frac{d}{2}-\frac{1}{2}\right)}$ per flipped ordering.

After the region shifts, we use the delta function in Eq.~\eqref{eq:bubblefinalWF} to set the gauge variable $\Delta\tau$ to unity. In Region 1, this sets $\tau_1=1+r$ and $\tau_2=r$. We then apply to the shifted variables the change of variables
\begin{equation}
s=\frac{1}{\overline{\beta}_1}+\frac{1}{\overline{\beta}_2},
\qquad
t=\frac{1/\overline{\beta}_1}{s}.
\end{equation}
For the second pair, we define
\begin{equation}
sz=
\left(\frac{1}{\bar{\kappa}_1}-\frac{1}{\overline{\beta}_1}\right)
+
\left(\frac{1}{\bar{\kappa}_2}-\frac{1}{\overline{\beta}_2}\right),
\qquad
w=
\frac{1}{sz}\left (
\frac{1}{\bar{\kappa}_1}-\frac{1}{\overline{\beta}_1}
\right ).
\end{equation}
The exponents $\alpha$ and $N$ are defined by
\begin{equation}
\alpha=\frac{3}{2}+2i\widetilde{\nu}-\varepsilon,
\qquad
N=\frac{9}{2}+2i\widetilde{\nu}-3\varepsilon,
\end{equation}
as quoted in Sec.~\ref{sec:redschgeneral}. In Region 1, the functions $A$, $B$, and $C$ are
\begin{equation}
\begin{split}
A &= (|\vec{k}_1|+|\vec{k}_2|)(1+r)+|\vec{k}_3| r\,,\\
B &= \frac{|\vec{k}_3|^2}{4}\,,\\
C &= 1+4r+4r^2+4rz+4r^2z\,.
\end{split}
\end{equation}
The remaining regions yield analogous functions. Only Regions 6 and 8 contain the UV divergence. In these regions, we refine the change of variables by replacing $1/\bar{\kappa}_1$ with $st\,p^2$ and $1/\bar{\kappa}_2$ with $s(1-t)q^2$, where $0<p,q<1$. In the $\Delta\tau=1$ gauge, the ultraviolet endpoint is approached as $p$ and $q$ tend to zero and $r$ tends to infinity. We handle this endpoint with the subtraction of Sec.~\ref{sec:redschdiv}. The outer integral is the four-dimensional quadrature anticipated in Sec.~\ref{sec:redschgeneral}. Its variables are $(t,p,q,r)$ in Regions 6 and 8 and $(t,z,w,r)$ in the other regions.

\subsection{Evaluating \texorpdfstring{$\tilde{\mathcal{B}}_{+-}$}{B+-}}

Running through the same procedure for $\tilde{\mathcal{B}}_{+-}$, we find that, as in the tree-level case, the contour deformation replaces each $(\overline{\beta}_{i}-\bar{\kappa}_{i})$ factor with $(\overline{\beta}_{i}+\bar{\kappa}_{i})$.
\begin{align}\label{eq:bubblefinalWFpm}
&\tilde{\mathcal{B}}_{+-}=e^{i \pi  \left(\frac{d}{2}-\Delta _+\right)}|\vec{k}_{3}|^{3}\Gamma \!\left(2\Delta_{+}+\frac{d}{2}\right)N_{\Delta_{+}}^{2}\int_{0}^{\infty} d\tau_{1}d\tau_{2} \ \tau _1^{2\Delta _+} \tau _2^{2\Delta _+-2}  \\
&\times \left (\prod_{i=1}^{2}d\overline{\beta}_{i}d\bar{\kappa}_{i}\frac{ \bar{\kappa}_{i} ^{d-2 \Delta _+-1} (\overline{\beta}_{i} +\bar{\kappa}_{i} )^{\Delta _+-\frac{d}{2}-\frac{1}{2}}}{\sqrt{\overline{\beta}_{i}}} \right )\frac{1}{(2\pi)^d}\left ( \frac{4\pi}{\overline{\beta}_{1}+\overline{\beta}_{2}}\right )^{d/2} \delta\!\left(1-\sum_{i}(\mathbf{m}_{i}\  \overline{\beta}_{i}+\mathbf{n}_{i}\ \bar{\kappa}_{i})-\sum_{i}\mathbf{q}_{i}\ \tau_{i}\right)\nonumber \\
&\times \mathcal{P}_{+-}^{\,-(2\Delta_{+}+d/2)}\nonumber
\end{align}
where
\begin{equation}\label{eq:Ppm}
\mathcal{P}_{+-}=\frac{\overline{\beta }_1 \overline{\beta }_2 |\vec{k}_{3}|^2}{4 \left(\overline{\beta }_1+\overline{\beta }_2\right)}+\left(\tau _1+\tau _2\right){}^2\!\left (\frac{1}{\overline{\beta}_{1}}+\frac{1}{\overline{\beta}_{2}}\right )+4 \tau _1 \tau _2\!\left (\frac{1}{\bar{\kappa}_{1}}+\frac{1}{\bar{\kappa}_{2}}\right )+\tau _2 |\vec{k}_{3}|+\tau _1 |\vec{k}_{1}| +\tau _1|\vec{k}_{2}| \ .
\end{equation}
As in the tree-level $\tilde{\mathcal{I}}_{+-}$, the key structural difference from $\tilde{\mathcal{B}}_{++}$ is that each $(\overline{\beta}_{i}-\bar{\kappa}_{i})$ is replaced by $(\overline{\beta}_{i}+\bar{\kappa}_{i})$, and the overall phase is modified. Since all polynomial factors are positive, we evaluate the integral numerically using the reduced Schwinger method of Sec.~\ref{sec:redschwinger}.
We apply the same procedure as in the tree-level example. Using the delta function in Eq.~\eqref{eq:bubblefinalWFpm}, we choose the gauge $\tau_1=1$ and write $r$ for $\tau_2$. The analogous change of variables, now with $sz=1/\bar{\kappa}_1+1/\bar{\kappa}_2$ and $w=(1/\bar{\kappa}_1)/(sz)$, then yields
\begin{equation}
\begin{split}
A &= |\vec{k}_1|+|\vec{k}_2|+|\vec{k}_3| r\,,\\
B &= \frac{|\vec{k}_3|^2}{4}\,,\\
C &= 1+2r+r^2+4rz\,,\\
W &= t^{1/2-\varepsilon-i\widetilde{\nu}}(1-t)^{1/2-\varepsilon-i\widetilde{\nu}} [w(1-w)]^{-1/2+i\widetilde{\nu}} z^{2i\widetilde{\nu}} r^{1-2\varepsilon+2i\widetilde{\nu}} [(t+wz)((1-t)+(1-w)z)]^{-1/2+i\widetilde{\nu}}\,,
\end{split}
\end{equation}
up to the prefactors already displayed in Eq.~\eqref{eq:bubblefinalWFpm}.

\section{Triangle contribution to the bispectrum}\label{sec:triangle}

We now consider the triangle contributions to the toy theory in Eq.~\eqref{desirabelcontri}. Unlike the bubble, the triangle topology involves three bulk vertices, each carrying one external inflaton leg and one background field insertion $\dot{\phi}_0$, connected by three internal $\sigma$ propagators forming a closed loop. Representative diagrams, for the all-plus assignment and for one mixed assignment, are
\begin{equation}
\label{trianglepicture1}
\begin{tikzpicture}[baseline={([yshift=-.5ex]current bounding box.center)},every node/.style={font=\scriptsize}]
\pgfmathsetmacro{\r}{1.9}

\pgfmathsetmacro{\yTop}{0.8*\r}

\draw (-2*\r,\yTop) -- (2*\r,\yTop);

\tikzset{decoration={snake,amplitude=.4mm,segment length=1.5mm,post length=0mm,pre length=0mm}}

\filldraw (0*\r,\yTop) circle (1pt) node[above=0pt]{$2$};
\filldraw ( 1.4*\r,\yTop) circle (1pt) node[above=0pt]{$1$};
\filldraw ( -1.4*\r,\yTop) circle (1pt) node[above=0pt]{$3$};

\filldraw (-30:\r/1.7)   circle (1pt) node[right=0pt]{};
\filldraw (90:\r/1.7) circle (1pt) node[above=0pt]{};
\filldraw (210:\r/1.7) circle (1pt) node[below=0pt]{};

\draw [thick] (-30:\r/1.7) -- (-50:\r/1.5);
\draw plot[mark=x,mark size=3pt] coordinates {(-50:\r/1.5)};

\draw [thick] (90:\r/1.7) -- (70:\r/1.5);
\draw plot[mark=x,mark size=3pt] coordinates {(70:\r/1.5)};

\draw [thick] (210:\r/1.7) -- (230:\r/1.5);
\draw plot[mark=x,mark size=3pt] coordinates {(230:\r/1.5)};

\draw [thick] (90:\r/1.7) -- (-0*\r,\yTop);
\draw [thick] (-30:\r/1.7)   -- ( 1.4*\r,\yTop);
\draw [thick] (210:\r/1.7) -- ( -1.4*\r,\yTop);

\draw [heavy scalar] (90:\r/1.7) -- (-30:\r/1.7);
\draw [heavy scalar] (210:\r/1.7) -- (90:\r/1.7);
\draw [heavy scalar] (210:\r/1.7) -- (-30:\r/1.7);

\filldraw (30:0.4*\r)  circle (0pt) node{$\beta_{1}$};
\filldraw (150:0.4*\r) circle (0pt) node{$\beta_{2}$};
\filldraw (270:0.4*\r) circle (0pt) node{$\beta_{3}$};

\coordinate (SE) at ([xshift=-2mm,yshift=2mm]current bounding box.south east);
\draw[->,thick] (SE) -- ([yshift=8mm]SE) node[above] {$\eta_{+}$};

\end{tikzpicture}
\end{equation}
and
\begin{equation}
\label{trianglepicture2}
\begin{tikzpicture}[baseline={([yshift=-.5ex]current bounding box.center)},every node/.style={font=\scriptsize}]
\pgfmathsetmacro{\r}{1.9}

\pgfmathsetmacro{\yTop}{0.8*\r}

\draw (-2*\r,\yTop) -- (2*\r,\yTop);

\tikzset{decoration={snake,amplitude=.4mm,segment length=1.5mm,post length=0mm,pre length=0mm}}

\filldraw (0*\r,\yTop) circle (1pt) node[below=0pt]{$2$};
\filldraw ( 1.4*\r,\yTop) circle (1pt) node[above=0pt]{$1$};
\filldraw ( -1.4*\r,\yTop) circle (1pt) node[above=0pt]{$3$};

\filldraw (-30:\r/1.7)   circle (1pt) node[right=0pt]{};
\filldraw (90:1.5*\r) circle (1pt) node[above=0pt]{};
\filldraw (210:\r/1.7) circle (1pt) node[below=0pt]{};

\draw [thick] (-30:\r/1.7) -- (-50:\r/1.5);
\draw plot[mark=x,mark size=3pt] coordinates {(-50:\r/1.5)};

\draw [thick] (90:1.5*\r) -- (70:1.5*\r);
\draw plot[mark=x,mark size=3pt] coordinates {(70:1.5*\r)};

\draw [thick] (210:\r/1.7) -- (230:\r/1.5);
\draw plot[mark=x,mark size=3pt] coordinates {(230:\r/1.5)};

\draw [thick] (90:1.5*\r) -- (-0*\r,\yTop);
\draw [thick] (-30:\r/1.7)   -- ( 1.4*\r,\yTop);
\draw [thick] (210:\r/1.7) -- ( -1.4*\r,\yTop);

\draw [heavy scalar] (90:1.5*\r) -- (-30:\r/1.7);
\draw [heavy scalar] (210:\r/1.7) -- (90:1.5*\r);
\draw [heavy scalar] (210:\r/1.7) -- (-30:\r/1.7);

\filldraw (30:0.55*\r)  circle (0pt) node{$\beta_{1}$};
\filldraw (150:0.55*\r) circle (0pt) node{$\beta_{2}$};
\filldraw (270:0.4*\r) circle (0pt) node{$\beta_{3}$};

\coordinate (SE) at ([xshift=-2mm,yshift=2mm]current bounding box.south east);
\draw[->,thick] (SE) -- ([yshift=8mm]SE) node[above] {$\eta_{+}$};

\coordinate (NE) at ([xshift=-2mm,yshift=2mm]current bounding box.north east);
\draw[->,thick] (NE) -- ([yshift=-8mm]NE) node[below] {$\eta_{-}$};

\end{tikzpicture}
\end{equation}
Before turning to the computation, we briefly address the relative size of the triangle contribution compared to the bubble. Restoring the Hubble scale, we work in the scaling regime where $\lambda = c^2\dot{\phi}_0^2/\Lambda^4$ is perturbatively small, as required for the validity of the slow-roll expansion. Concretely, this perturbativity condition arises from requiring that the one-loop correction to the inflaton two-point function is small.
\begin{equation}
\begin{tikzpicture}[
    baseline=(current bounding box.center),
    thick,
    inflaton/.style={solid, thick},
    vertex/.style={circle, fill=black, inner sep=1.5pt},
    bginsert/.style={draw, circle, inner sep=3pt, thick},
]
  \draw[thick] (-3, 0) node[left] {$\delta \phi$} -- (-1.5, 0);
  \draw[thick] ( 1.5, 0) -- ( 3, 0) node[right] {$\delta \phi$};
  \draw[thick] ( 1.5, 0) -- ( 2.5, -0.8);
  \draw plot[mark=x, mark size=3pt] coordinates {( 2.5, -0.8)};
  \draw[thick] ( -1.5, 0) -- ( -2.5, -0.8);
  \draw plot[mark=x, mark size=3pt] coordinates {( -2.5, -0.8)};
  \draw[heavy scalar] (-1.5, 0) arc (180:0:1.5cm and 0.9cm);
  \draw[heavy scalar] ( 1.5, 0) arc (0:-180:1.5cm and 0.9cm);
  \fill (-1.5, 0) circle (2pt);
  \fill ( 1.5, 0) circle (2pt);
  \node at (0,  1.2) {$\sigma$};
  \node at (0, -1.2) {$\sigma$};
\end{tikzpicture}
\;\sim\; \mathcal{O}\!\left(c^2\frac{\dot{\phi}_0^2}{\Lambda^4}\right) \ll 1\,.
\end{equation}
The parametric dependence of the two contributions on $\lambda$ and $\dot{\phi}_0$ is
\begin{equation}
\begin{split}
f_{\mathrm{NL}}^{\textrm{bubble}} &\sim \mathcal{O}\!\left(\frac{c^2\dot{\phi}_0^2}{\Lambda^4}\right) = \mathcal{O}\!\left(\lambda\right), \\
f_{\mathrm{NL}}^{\textrm{triangle}} &\sim \mathcal{O}\!\left(\frac{c^3(\dot{\phi}_0)^4}{\Lambda^6H^2}\right) = \mathcal{O}\!\left(\lambda^{3/2}\,\frac{\dot{\phi}_0}{H^2}\right),
\end{split}
\end{equation}
which gives
\begin{equation}\label{eq:triangle_vs_bubble}
\frac{f_{\mathrm{NL}}^{\textrm{triangle}}}{f_{\mathrm{NL}}^{\textrm{bubble}}} \sim \mathcal{O}\!\left(\lambda^{1/2}\,\frac{\dot{\phi}_0}{H^2}\right).
\end{equation}
The triangle carries an extra factor of $\sqrt{\lambda}$ relative to the bubble. However, the power spectrum normalization implies $\dot{\phi}_0/H^2 \simeq 3.5 \times 10^3$~\cite{Planck:2018vyg}, so the ratio~\eqref{eq:triangle_vs_bubble} can be $\mathcal{O}(1)$ even for moderately small $\lambda$. Therefore, the triangle contribution cannot be neglected a priori in phenomenologically relevant regions of parameter space.

Given the importance of the triangle diagram, we now turn to its explicit computation. The relevant Witten diagrams are
\begin{equation}\label{eq:triangle_witten}
\begin{split}
W_{\pm_1\pm_2\pm_3} &= \int_{-\infty}^{0} \left(\prod_{i=1}^{3} \frac{d\eta_i}{(-\eta_i)^{2}}\, [\partial_{\eta_i} G^{\pm_i}(\vec{k}_i, \eta_i)]\,[\partial_{\eta_i}\phi_0(\eta_i)]\right) \\
&\quad \times \int_{-\infty}^{\infty} \frac{d^d \vec{\ell}}{(2\pi)^d}\, G^{\pm_1\pm_2}(\vec{\ell},\eta_1,\eta_2)\, G^{\pm_2\pm_3}(\vec{\ell}-\vec{k}_2,\eta_2,\eta_3)\, G^{\pm_3\pm_1}(\vec{\ell}-\vec{k}_2-\vec{k}_3,\eta_3,\eta_1)\,,
\end{split}
\end{equation}
which are spanned by the basis functions
\begin{equation}\label{eq:triangle_basis}
\begin{split}
\mathcal{T}_{\pm_1\pm_2\pm_3} &= \int_{-\infty}^{0} d\eta_1\,d\eta_2\,d\eta_3 \prod_{i=1}^{3} (-\eta_i)^{q_i}\, \exp\!\left(\pm_1 i|\vec{k}_1|\eta_1 \pm_2 i|\vec{k}_2|\eta_2 \pm_3 i|\vec{k}_3|\eta_3\right) \\
&\quad \times \int \frac{d^d\vec{\ell}}{(2\pi)^d}\, G^{\pm_1\pm_2}(\vec{\ell},\eta_1,\eta_2)\, G^{\pm_2\pm_3}(\vec{\ell}-\vec{k}_2,\eta_2,\eta_3)\, G^{\pm_3\pm_1}(\vec{\ell}-\vec{k}_2-\vec{k}_3,\eta_3,\eta_1)\,,
\end{split}
\end{equation}
where $q_i=-2$ at each vertex, since the factor $\eta$ from the derivative leg cancels against the factor $\eta^{-1}$ from the background insertion. Using Eq.~\eqref{ex:extprop} and $\partial_{\eta}\phi_{0}=\epsilon_{\rm sr}/\eta$, the two are related by $W_{\pm_1\pm_2\pm_3}=\epsilon_{\rm sr}^{3}\,\mathcal{T}_{\pm_1\pm_2\pm_3}/(8|\vec{k}_1||\vec{k}_2||\vec{k}_3|)$. Due to the relation $(\mathcal{T}_{\pm_1\pm_2\pm_3})^* = \mathcal{T}_{\mp_1\mp_2\mp_3}$ and the permutation symmetry of the three vertices, a complete basis is provided by $\mathcal{T}_{+++}$ and $\mathcal{T}_{++-}$. The remaining branch assignments are obtained by complex conjugation and permutation. We evaluate both integrals below. Together with the momentum permutations of $\mathcal{T}_{++-}$ and the vertex factors in Appendix~\ref{sec:reviewinin}, they determine the full in-in result. Writing $v_a=-ia$ for a vertex on branch $a=\pm1$, which is the rule of Appendix~\ref{sec:reviewinin}, the branch sum is
\begin{equation}
\begin{split}
\sum_{a,b,c=\pm1}v_av_bv_cW_{abc}
={}&\frac{\epsilon_{\rm sr}^{3}}
{8|\vec{k}_1||\vec{k}_2||\vec{k}_3|}
\Big[
i\mathcal T_{+++}(k_1,k_2,k_3)
-i\mathcal T_{++-}(k_1,k_2,k_3)\\
&\hspace{4.3em}
-i\mathcal T_{++-}(k_2,k_3,k_1)
-i\mathcal T_{++-}(k_3,k_1,k_2)
+\mathrm{c.c.}
\Big] \, .
\end{split}
\end{equation}
The three mixed terms place the minus-branch vertex on each external leg in turn. An important simplification relative to the bubble is that the triangle integral is UV finite in $D = d+1 = 4$.\footnote{To see this, simply remember that the UV behavior is the same in flat and de Sitter and the triangle is UV finite in flat space.} We therefore keep the vertex time measures at $d=3$, as for the tree and the bubble, and retain $d$ only in the propagators and the loop measure so that the formulas below can be compared with those of Sec.~\ref{sec:bubble}.

To compute the WF parameterization, we substitute the Schwinger form of the propagator and evaluate the Gaussian integral over loop momentum.
\begin{equation}\label{eq:triangle_gaussian}
\int \frac{d^d\vec{\ell}}{(2\pi)^d}\;\exp\!\left(-\frac{i\beta_1\vec{\ell}^{\,2}}{4} - \frac{i\beta_2(\vec{\ell}-\vec{k}_2)^2}{4} - \frac{i\beta_3(\vec{\ell}-\vec{k}_2-\vec{k}_3)^2}{4}\right) = \frac{1}{(2\pi)^{d}}\left(\frac{4\pi}{i\,\mathcal{U}_{\mathrm{tri}}}\right)^{\!d/2}\exp\!\left(-\frac{i\,\mathcal{F}_{\mathrm{tri}}}{4\,\mathcal{U}_{\mathrm{tri}}}\right),
\end{equation}
where the Symanzik polynomials for the triangle graph are
\begin{equation}\label{eq:symanzik_triangle}
\begin{split}
\mathcal{F}_{\mathrm{tri}} &= \beta_1\beta_2\,|\vec{k}_2|^2 + \beta_2\beta_3\,|\vec{k}_3|^2 + \beta_1\beta_3\,|\vec{k}_1|^2\,, \\
\mathcal{U}_{\mathrm{tri}} &= \beta_1 + \beta_2 + \beta_3\,.
\end{split}
\end{equation}

\subsection{Evaluating \texorpdfstring{$\mathcal{T}_{+++}$}{T+++}}

We begin with the all-$+$ assignment. After substituting the Schwinger representation for each internal propagator, performing the change of variables $\eta_i \to -\eta_i$, and carrying out the loop momentum integral via Eq.~\eqref{eq:triangle_gaussian}, we perform the contour deformation
\begin{equation}\label{eq:contour_triangle_ppp}
\begin{split}
\eta_i &\to e^{-i\pi/2}\,\tau_i\,, \\
\kappa_e &\to e^{-i\pi}\,\bar{\kappa}_e\,, \\
\beta_e &\to e^{-i\pi/2}\,\overline{\beta}_e\,,
\end{split}
\end{equation}
which is the same rotation used for the $++$ bubble in Eq.~\eqref{contourdef3} with $\theta = \pi$. One can verify that the integral remains convergent throughout $0 \leq \theta \leq \pi$, so the deformation is admissible. After setting $\epsilon = 0$ and fixing the $GL(1)$ rescaling symmetry, the final WF representation is
\begin{equation}\label{eq:Tppp_final}
\begin{split}
\mathcal{T}_{+++} &= \tilde{C}_{+++}\,\Gamma\!\left(\gamma_{+++}\right) \int_0^\infty \prod_{i=1}^{3}\left(d\tau_i\,d\overline{\beta}_i\,d\bar{\kappa}_i\right)\; \tau_1^{a_1}\,\tau_2^{a_2}\,\tau_3^{a_3} \\
&\quad \times \prod_{e=1}^{3} \frac{\bar{\kappa}_e^{d-2\Delta_+-1}\,(\overline{\beta}_e - \bar{\kappa}_e)^{\Delta_+-\frac{d}{2}-\frac{1}{2}}}{\sqrt{\overline{\beta}_e}}\;\frac{1}{\mathcal{U}_{\mathrm{tri}}(\overline{\beta}_e)^{d/2}} \\
&\quad \times \delta\!\left(1 - \sum_{e}(\mathbf{m}_e\,\overline{\beta}_e + \mathbf{n}_e\,\bar{\kappa}_e) - \sum_{i}\mathbf{q}_i\,\tau_i\right)\;\mathcal{P}_{+++}^{-\gamma_{+++}}\,,
\end{split}
\end{equation}
where $a_{i}=2\Delta_{+}-2$ collects the vertex measure at $d=3$ and the two incident propagators, $\gamma_{+++}=3\Delta_{+}+d-3$ is the overall scaling dimension of the integrand, $\tilde{C}_{+++}$ collects the overall phase and normalization constants, and
\begin{equation}\label{eq:Pppp}
\begin{split}
\mathcal{P}_{+++} &= \frac{\mathcal{F}_{\mathrm{tri}}(\overline{\beta}_e)}{4\ \mathcal{U}_{\mathrm{tri}}(\overline{\beta}_e)} + \frac{(\tau_1 - \tau_2)^2}{\overline{\beta}_1} + \frac{(\tau_2 - \tau_3)^2}{\overline{\beta}_2} + \frac{(\tau_3 - \tau_1)^2}{\overline{\beta}_3} \\
&\quad + \frac{4\tau_1\tau_2}{\bar{\kappa}_1} + \frac{4\tau_2\tau_3}{\bar{\kappa}_2} + \frac{4\tau_3\tau_1}{\bar{\kappa}_3} + |\vec{k}_1|\tau_1 + |\vec{k}_2|\tau_2 + |\vec{k}_3|\tau_3\,.
\end{split}
\end{equation}
At $d=3$ the constant is
\begin{equation}\label{eq:Cppp}
\tilde{C}_{+++}=\frac{(4\pi)^{3/2}}{(2\pi)^{3}}\,N_{\Delta_{+}}^{3}\,\exp\!\left[-i\pi\left(\frac{21}{4}-\frac{3\Delta_{+}}{2}\right)\right],
\end{equation}
which collects the three propagator normalizations, the loop measure, the Gaussian factor, and the phases produced by the rotation in Eq.~\eqref{eq:contour_triangle_ppp}.

As in the bubble case, expanding the conformal-time differences $(\tau_v - \tau_{v'})^2$ in the polynomial~\eqref{eq:Pppp} produces negative coefficients. The factors $(\overline{\beta}_e - \bar{\kappa}_e)$ are also not sign-definite. We therefore decompose the integration domain into regions with positive polynomial coefficients. The three $\overline{\beta}_e \gtrless \bar{\kappa}_e$ conditions give $2^3 = 8$ choices, and the orderings of $\tau_1$, $\tau_2$, and $\tau_3$ give $3! = 6$ choices. Together, they give $8 \times 6 = 48$ regions. In each region, we introduce shifted variables. For example, we use $\overline{\beta}_e = \bar{\kappa}_e + \Delta\overline{\beta}_e$ when $\overline{\beta}_e > \bar{\kappa}_e$, and we use $\tau_1 = \tau_2 + \Delta\tau_{12}$ when $\tau_1 > \tau_2$. These shifts make all polynomial coefficients positive. The triangle is UV finite, so no subtraction is needed.

To apply the reduced Schwinger method of Sec.~\ref{sec:redschwinger}, we use the delta function in Eq.~\eqref{eq:Tppp_final} to set $\tau_1$ to unity. We then define the common scale $s$ through the change of variables associated with the $(\tau_3,\tau_1)$ edge.
\begin{equation}
\begin{split}
s &= 1/\overline{\beta}_3+1/\bar{\kappa}_3\,,  \quad t_a = (1/\overline{\beta}_3)/s\,,\\
s z_b &= 1/\overline{\beta}_2+1/\bar{\kappa}_2\,,  \quad t_b = (1/\overline{\beta}_2)/(s z_b)\,,\\
s z_c &= 1/\overline{\beta}_1+1/\bar{\kappa}_1\,,  \quad t_c = (1/\overline{\beta}_1)/(s z_c)\,.
\end{split}
\end{equation}
This change of variables yields the factorization
\begin{equation}
\begin{split}
A &= |\vec{k}_1| + |\vec{k}_2|\tau_2 + |\vec{k}_3|\tau_3\,,\\
B &= \frac{t_b z_b |\vec{k}_1|^2 + t_a |\vec{k}_2|^2 + t_c z_c |\vec{k}_3|^2}{4(t_b z_b t_c z_c + t_a(t_b z_b + t_c z_c))}\,,\\
C &= t_a(1-\tau_3)^2 + 4(1-t_a)\tau_3 + z_b t_b(\tau_3-\tau_2)^2 + 4 z_b(1-t_b)\tau_2\tau_3\\
&\quad + z_c t_c(\tau_2-1)^2 + 4 z_c(1-t_c)\tau_2\,.
\end{split}
\end{equation}
The exponents $\alpha$ and $N$ are, respectively, $\tfrac32+3i\widetilde{\nu}$ and $\tfrac92+3i\widetilde{\nu}$. Writing $\mathcal{D}=t_bz_bt_cz_c+t_a(t_bz_b+t_cz_c)$ for the combination in the denominator of $B$, the remaining factor at $d=3$ is
\begin{equation}\label{eq:Wppp}
W_{+++}=(\tau_2\tau_3)^{1+2i\widetilde{\nu}}\,(z_bz_c)^{\frac12+i\widetilde{\nu}}\,\mathcal{D}^{-\frac32}\prod_{x=t_a,t_b,t_c}x^{\frac12-i\widetilde{\nu}}(1-x)^{-\frac12+i\widetilde{\nu}}(1-2x)^{-\frac12+i\widetilde{\nu}}\,.
\end{equation}
In a region with $\overline{\beta}_e<\bar{\kappa}_e$, that is $x>\tfrac12$ for the corresponding edge, the continuous contour fixes $(1-2x)^{q}=e^{-i\pi q}(2x-1)^{q}$ with $q=\Delta_{+}-2$, exactly as for the tree in Sec.~\ref{sec:treeanalyt}. The region shifts make the squared-difference terms sign-definite in each region. The outer integral is a seven-dimensional quadrature.

\subsection{Evaluating \texorpdfstring{$\mathcal{T}_{++-}$}{T++-}}

Running through the same procedure for $\mathcal{T}_{++-}$, we find that, as in the tree-level and bubble cases, the contour deformation for the mixed-branch integral is
\begin{equation}\label{eq:contour_triangle_ppm}
\begin{split}
\eta_1 &\to -i\,\tau_1\,, \qquad \eta_2 \to -i\,\tau_2\,, \qquad \eta_3 \to +i\,\tau_3\,, \\
\kappa_1 &\to e^{-i\pi}\,\bar{\kappa}_1\,, \qquad \kappa_2 \to \bar{\kappa}_2\,, \qquad \kappa_3 \to \bar{\kappa}_3\,, \\
\beta_e &\to -i\,\overline{\beta}_e\,,
\end{split}
\end{equation}
where the sign of the $\tau_3$ rotation is flipped because vertex~3 sits on the opposite branch. On the two mixed edges, we continue $(\beta_e-i\kappa_e)^{\Delta_{+}-\frac{d}{2}-\frac{1}{2}}$ from the positive $\beta_e$ axis before performing the contour rotation. After fixing the gauge, the result takes the same structural form as Eq.~\eqref{eq:Tppp_final}, with three modifications. The overall prefactor changes to 
\begin{equation}
\tilde{C}_{++-}=e^{2\pi\widetilde{\nu}}\,\tilde{C}_{+++}
\end{equation}
at $d=3$, which follows from the phases of the rotation in Eq.~\eqref{eq:contour_triangle_ppm}. The edge-1 factor remains $(\overline{\beta}_1-\bar{\kappa}_1)$, while the two cross-branch factors become $(\overline{\beta}_2+\bar{\kappa}_2)$ and $(\overline{\beta}_3+\bar{\kappa}_3)$, and the polynomial becomes
\begin{equation}\label{eq:Pppm}
\begin{split}
\mathcal{P}_{++-} &= \frac{\overline{\beta}_1\overline{\beta}_2\,|\vec{k}_2|^2 + \overline{\beta}_2\overline{\beta}_3\,|\vec{k}_3|^2 + \overline{\beta}_1\overline{\beta}_3\,|\vec{k}_1|^2}{4(\overline{\beta}_1 + \overline{\beta}_2 + \overline{\beta}_3)} + \frac{(\tau_1 - \tau_2)^2}{\overline{\beta}_1} + \frac{(\tau_2 + \tau_3)^2}{\overline{\beta}_2} + \frac{(\tau_3 + \tau_1)^2}{\overline{\beta}_3} \\
&\quad + \frac{4\tau_1\tau_2}{\bar{\kappa}_1} + \frac{4\tau_2\tau_3}{\bar{\kappa}_2} + \frac{4\tau_3\tau_1}{\bar{\kappa}_3} + |\vec{k}_1|\tau_1 + |\vec{k}_2|\tau_2 + |\vec{k}_3|\tau_3\,.
\end{split}
\end{equation}
On the two edges whose endpoints lie on opposite branches, the conformal-time differences $(\tau_v - \tau_{v'})^2$ are replaced by $(\tau_v + \tau_{v'})^2$. The same replacement occurs for the bubble, where 
\begin{equation}
(\tau_1 - \tau_2)^2 \to (\tau_1 + \tau_2)^2
\end{equation}
when passing from $\mathcal{P}_{++}$~\eqref{eq:Ppp} to $\mathcal{P}_{+-}$~\eqref{eq:Ppm}. The polynomial $\mathcal{P}_{++-}$ is positive throughout the integration domain. Only the expansion of the single $(\tau_1-\tau_2)^2$ term on the edge connecting the two $+$ vertices produces a negative coefficient. We use $2 \times 6 = 12$ regions, compared with 48 for $\mathcal{T}_{+++}$, so that the six time-ordering sectors can be reused, although four regions suffice for sign-definite polynomial coefficients.

To apply the reduced Schwinger method of Sec.~\ref{sec:redschwinger}, we use the same change of variables as for $\mathcal{T}_{+++}$ above, now applied to Eq.~\eqref{eq:Pppm}. The functions $A$ and $B$ remain unchanged, while $C$ becomes
\begin{equation}
\begin{split}
C_{++-} &= t_a(1+\tau_3)^2 + 4(1-t_a)\tau_3 + z_b t_b(\tau_3+\tau_2)^2 + 4 z_b(1-t_b)\tau_2\tau_3\\
&\quad + z_c t_c(1-\tau_2)^2 + 4 z_c(1-t_c)\tau_2\,.
\end{split}
\end{equation}

The factor $W$ also changes. On the two cross-branch edges the combination $\overline{\beta}_e+\bar{\kappa}_e$ is $1/(s\,t_a(1-t_a))$ for edge 3 and $1/(s z_b t_b(1-t_b))$ for edge 2, and neither produces a factor of $(1-2t_e)$, so $W_{++-}$ is $W_{+++}$ in Eq.~\eqref{eq:Wppp} with the factors $(1-2t_a)^{-\frac12+i\widetilde{\nu}}$ and $(1-2t_b)^{-\frac12+i\widetilde{\nu}}$ removed. The exponents $\alpha$ and $N$ are unchanged.

\section{Numerical results and observational constraints}\label{sec:CMBdata}

In this section, we present the bispectrum shapes from the interaction Lagrangian
\begin{equation}
\mathcal{L}_{\mathrm{int}}\supset \frac{c}{\Lambda^2}(\partial^{\eta}\phi)(\partial_{\eta}\phi)\,\sigma^2
\end{equation}
computed with the reduced Schwinger method described above. We also present constraints on $\fnl$ from a comparison with \textit{Planck} data~\cite{Planck:2019kim} using the \texttt{CMB-BEST} package~\cite{Sohn:2023fte}. The shape function of the bispectrum $\cs(k_1, k_2, k_3)$ read by \texttt{CMB-BEST} is defined as
\begin{equation}
    \langle \zeta(\vec{k}_1) \zeta(\vec{k}_2) \zeta(\vec{k}_3)\rangle \equiv (2\pi)^3\delta^{(3)}(\vec{k}_1+\vec{k}_2+\vec{k}_3)\frac{18}{5} \fnl (2\pi^2 A_s)^2 \frac{\cs(k_1, k_2, k_3)}{k_1^2 k_2^2 k_3^2},
\end{equation}
where $A_s \equiv (k^3/(2\pi^2)) (k_*/k)^{n_s-1}\langle\zeta(\vec{k})\zeta(-\vec{k})\rangle'$ is the dimensionless power spectrum, and we use the normalization convention of $\cs(k_*, k_*, k_*) = 1$. Because of exact scale invariance of our theory, any choice of $k_*$ is equivalent, but we will choose $k_* = k_{\rm max} = 0.209\; {\rm Mpc}^{-1}$ for clarity.

\subsection{Parameter choices, numerical convergence, and computational cost}

Using the reduced Schwinger method, we compute the bubble and triangle diagram contributions to the bispectrum throughout the full kinematic space relevant for observational data. In particular, we compute the bispectrum on a discrete three-dimensional grid of $(k_1, k_2, k_3)$ with uniform logarithmic spacing by a factor of $10^{1/20}$ in each direction in the \texttt{CMB-BEST} dynamic range $10^{-3}\leq k_i/k_{\rm max}\leq  1$, $k_{\rm max} = 0.209\; {\rm Mpc}^{-1}$. Because of scale invariance of our theory, it suffices to compute on the two-dimensional grid $10^{-3} \leq k_1/k_3, k_2/k_3 \leq 10^3$ of $121^2$ points. The $k_1\leftrightarrow k_2$ symmetry of the diagrams reduces this to $121\times 122/2=7381$ kinematic points. We choose two mass parameter values for the massive scalar, $\widetilde{\nu} \equiv \sqrt{m_{\sigma}^2/H^2-9/4} = 0.5$ and $0.7$. 

As outlined in Sec.~\ref{sec:redschwinger}, the reduced Schwinger method performs the integration over the overall scale of the Schwinger parameters analytically, yielding a kernel in terms of ${}_2F_1$. The remaining four- and seven-dimensional integrals for the bubble and triangle diagrams, respectively, are evaluated numerically. For the rapidly convergent bubble integral, the numerical uncertainty is estimated from the change between two successive numerical resolutions. For the triangle, we conservatively take the largest discrepancy observed upon increasing the grid resolution, recalculating the integral using an independent Sobol sampling of the integration domain, and testing exact symmetry and internal consistency relations. The \texttt{C++} source files for the numerical integrations and the \texttt{Python} scripts to assemble the resulting bispectra are provided as ancillary files with the arXiv submission of this paper.

The computational cost in CPU-hours and the numerical error for the bispectrum computations are shown in Table~\ref{tab:stats}. For the bubble diagram we combine the $++$ and $+-$ Schwinger--Keldysh branches, while for the triangle diagram we list the $+++$ and $++-$ branches separately. The numerical error listed in the table is the ``normalized average error'', defined as
\begin{equation}
    \bar{\delta}_{\rm num} \equiv \frac{1}{N_{\rm physical }} \sum_{\vec{k}\in {\rm physical}} \frac{|\delta(\vec{k})|}{|\cs(k_*, k_*, k_*)|},
\end{equation}
where $\delta(\vec{k})$ is the numerical convergence error as discussed above, and $N_{\rm physical}$ is the number of $\vec{k}$ points satisfying the triangle inequality. For both the bubble and triangle diagrams, varying each shape function by its numerical error has negligible effect on the observational constraints from \texttt{CMB-BEST}. We therefore consider only the central numerical values in the following discussion.

\begin{table}[ht!]
\renewcommand{\arraystretch}{1.2}
\centering
\begin{tabular}{c | c  c c}
\hline \hline 
$\widetilde{\nu}$ & \hspace*{0.5em}components \hspace*{0.5em}& total CPU-hours &  Numerical errors $\bar{\delta}_{\rm num}$  \\ 
\hline \hline
\multirow{3}{*}{0.5}& bubble  & 2,738 & $<10^{-5}\%$\\
 & triangle $+++$ & 11,950 & $2.8\%$\\
 & triangle $++-$ & 6,463 & $1.3\%$\\
\Cline{0.2pt}{1-4}
\multirow{3}{*}{0.7}& bubble  & 2,689 & $< 10^{-5}\%$ \\
 & triangle $+++$ & 12,019 & $2.1\%$  \\
 & triangle $++-$ & 15,550 & $6.5\%$  \\ 
\hline \hline
\end{tabular}
\caption{Computational cost and numerical convergence of the bubble and triangle computations. Each computation contains 7,381 momentum configurations. See the text for details. The bubble rows include both $++$ and $+-$ Schwinger--Keldysh branches, while for the triangle we list the $+++$ and $++-$ branches separately. The total CPU-hours include the highest-resolution run used for the central value and the lower-resolution runs used to estimate numerical integration errors. The latter contribute roughly 25\% of the total CPU-hours for both the bubble and the triangle. The \texttt{C++} source files for the numerical integrations and the \texttt{Python} scripts to assemble the resulting bispectra are provided as ancillary files with the arXiv submission of this paper.} 
\label{tab:stats}
\end{table}

\subsection{Definition of the template and likelihood}

Before presenting the shape functions calculated by the numerical methods, we need to discuss two obstacles to defining a unique shape function for this theory for use in a data search. The first obstacle is the UV divergence of the bubble diagram. As discussed in Sec.~\ref{sec:bubble}, the UV divergence is renormalized by the counterterm operator
\begin{equation}
    \mathcal{L}_{\rm ct}  = \frac{c_{\rm ct}\left(\partial^{\eta}\phi\,\partial_{\eta}\phi\right)^{2}}{\Lambda^4},
\end{equation}
whose finite piece contributes an equilateral-like shape to the bispectrum. The bispectrum from the bubble diagram is therefore renormalization scheme dependent~\cite{Jain:2025maa}. Given that this paper is a proof of principle for the reduced Schwinger numerical integration method, we simply use the $\overline{\rm MS}$ scheme following Ref.~\cite{Xianyu:2022jwk} with the renormalization scale $\mu_R = \widetilde{\nu} H$, and we also set the renormalized tree-level coefficient of the counterterm to zero. More specifically, the renormalized combination of Eq.~\eqref{wittendiagram4} is defined with a pure-pole coefficient at renormalization scale $H$. It differs from the $\overline{\rm MS}$ convention of
Ref.~\cite{Xianyu:2022jwk} at $\mu_R=\widetilde{\nu}H$ only by the standard $\overline{\rm MS}$ redefinition of the pole subtraction and by the change of renormalization scale, which amount to the constant shift
\begin{equation}\label{eq:XZschemeshift}
\tilde{\mathcal{B}}_{++}\;\longrightarrow\;
\tilde{\mathcal{B}}_{++}
+\frac{\pi|\vec{k}_{3}|^{3}}{(2\pi)^{3}E^{3}}
\Big[\log 4\pi-\gamma_{E}-2\log\widetilde{\nu}\Big],
\qquad \tilde{\mathcal{B}}_{+-}\ \text{unchanged},
\end{equation}
with $E=|\vec{k}_{1}|+|\vec{k}_{2}|+|\vec{k}_{3}|$.

The second obstacle comes from the relative amplitude of the bubble and triangle diagrams. The numerical methods explained above compute the dimensionless bispectrum of the dimensionless inflaton perturbation, $\tilde{\phi}\equiv\delta\phi/H$. Assuming that our interaction Lagrangian does not significantly correct the power spectrum from the inflaton potential (which we will check for consistency later when interpreting the results from \textit{Planck} data), we define
\begin{equation}
P_{\rm sr}\equiv\frac{|\dot{\phi}_0|}{H^2}
=\frac{1}{2\pi\sqrt{A_s}}\simeq(59)^2,
\qquad
s_\phi\equiv-\frac{\dot\phi_0}{|\dot\phi_0|}= \frac{\epsilon_{\rm sr}}{HP_{\rm sr}}\, .
\end{equation}
With the branch vertex factor $v_a$ of Sec.~\ref{sec:triangle}, the coupling-stripped (superscript ``st'') bubble bispectrum is
\begin{equation}\label{eq:bst_bubble}
B_{\rm bub}^{\rm st}(k_1,k_2,k_3)
=2s_\phi
\sum_{(i,j,k)=(1,2,3),(2,3,1),(3,1,2)}
\frac{1}{k_i k_j k_k^4}
\operatorname{Re}\!\left[
\tilde{\mathcal B}_{++}^{\rm ren}(k_i,k_j;k_k)
-\tilde{\mathcal B}_{+-}(k_i,k_j;k_k)
\right] \, .
\end{equation}
Here $\tilde{\mathcal B}_{\pm\pm}(k_i,k_j;k_k)$ is the basis integral of Sec.~\ref{sec:bubble} for the channel in which the background vertex carries $k_k$. The superscript ${\rm ren}$ denotes the value after the renormalization in Eqs.~\eqref{wittendiagram4} and~\eqref{eq:ctremainder} and the conversion in Eq.~\eqref{eq:XZschemeshift}, evaluated at $|\vec{k}_{3}|\to k_{k}$ and applied point by point before the templates are normalized. The coupling-stripped triangle bispectrum is
\begin{equation}\label{eq:bst_triangle}
\begin{split}
B_{\rm tri}^{\rm st}(k_1,k_2,k_3)
={}&\frac{8s_\phi}{k_1k_2k_3}
\Big[
i\mathcal T_{+++}(k_1,k_2,k_3)
-i\mathcal T_{++-}(k_1,k_2,k_3)\\
&\hspace{7.0em}
-i\mathcal T_{++-}(k_2,k_3,k_1)
-i\mathcal T_{++-}(k_3,k_1,k_2)
+\mathrm{c.c.}
\Big] \, .
\end{split}
\end{equation}
The constants $2$ and $8$ collect the contraction factors of the interaction, the relation between the Witten diagrams and the basis integrals, and the branch vertex factors. The interaction Lagrangian then fixes the relative normalization,
\begin{equation}
\begin{split}
\langle\tilde{\phi}(\vec{k}_1)\tilde{\phi}(\vec{k}_2)\tilde{\phi}(\vec{k}_3)\rangle'
&=\frac{c^2|\dot\phi_0|H^2}{\Lambda^4}B_{\rm bub}^{\rm st}
+\left(\frac{c|\dot\phi_0|}{\Lambda^2}\right)^3B_{\rm tri}^{\rm st}\\
&=g^2P_{\rm sr}\left(B_{\rm bub}^{\rm st}
+gP_{\rm sr}^2B_{\rm tri}^{\rm st}\right) \, ,
\end{split}
\end{equation}
where we have defined the dimensionless combination $g\equiv cH^2/\Lambda^2$. The two diagrams thus share a common amplitude of $g^2P_{\rm sr}$ with a relative coupling of $gP_{\rm sr}^2$. We therefore cannot do a conventional single-template search with $ B_{\rm theory}\equiv B_{\rm bub}^{\rm st}+gP_{\rm sr}^2B_{\rm tri}^{\rm st}$ without knowing the observational constraint on $\fnl$ and thus $g$ in the first place.

Instead, we jointly search for $B^{\rm st}_{\rm bub}$ and $B^{\rm st}_{\rm tri}$ as independent shape functions with independent amplitudes $f_B$ and $f_T$ in \texttt{CMB-BEST}. We then maximize the joint likelihood along the theoretical curve that fixes the relation between $f_B$ and $f_T$.\footnote{Due to the linear dependence of the model prediction on $f_B$ and $f_T$ and the assumption of Gaussianity of the likelihood, this is equivalent to the likelihood computed directly for the coupling $g$.} From the definition of $\cs$ used by \texttt{CMB-BEST} and listed above and our convention to normalize the shape function at the equilateral point $k_1=k_2=k_3 = k_*$, the relation between each $\fnl$ value constrained by \texttt{CMB-BEST} and the corresponding product of couplings is 
\begin{equation}
    f_B = s_\phi\, g^2 P_{\rm sr} \left(\frac{(k_1^2 k_2^2 k_3^2 B^{\rm st}_{\rm bub})|_{k_*}}{\frac{9}{5} \pi A_s^{1/2}}\right),\quad
    f_T = s_\phi\, g^3 P_{\rm sr}^3 \left(\frac{(k_1^2 k_2^2 k_3^2 B^{\rm st}_{\rm tri})|_{k_*}}{\frac{9}{5} \pi A_s^{1/2}}\right),\label{eq:fnlcurve}
\end{equation}
where $(k_1^2 k_2^2 k_3^2 B^{\rm st}_i)|_{k_*}$ is the amplitude of the coupling-stripped $\tilde{\phi}$ shape function at the equilateral point that we have divided out to normalize $\cs$, and the factor $s_{\phi}$ comes from applying $\zeta = -H\delta \phi/\dot{\phi}_0 = s_{\phi}\tilde{\phi}/P_{\rm sr}$ three times. The $s_{\phi}$ here cancels with the $s_{\phi}$ in Eq.~\eqref{eq:bst_bubble} and Eq.~\eqref{eq:bst_triangle}, and the observational constraint on the coupling $g$ is $s_{\phi}$-independent, as it should be. The joint $(f_B, f_T)$ likelihood is sliced with this theoretical relation to obtain the constraint on $g$, for each value of the $\sigma$ mass.

\subsection{Shapes and \textit{Planck} constraints}

\begin{figure}
    \centering
    \begin{subfigure}[b]{0.5\textwidth}
        \centering
        \includegraphics[width=\textwidth]{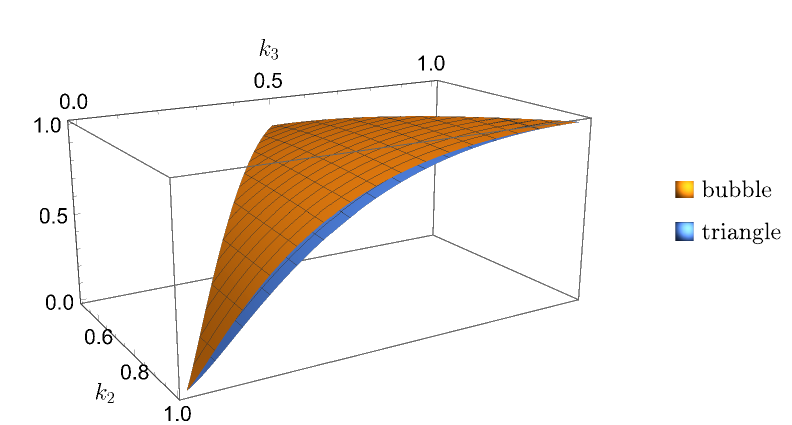}
    \end{subfigure}%
    \begin{subfigure}[b]{0.5\textwidth}
        \centering
        \includegraphics[width=\textwidth]{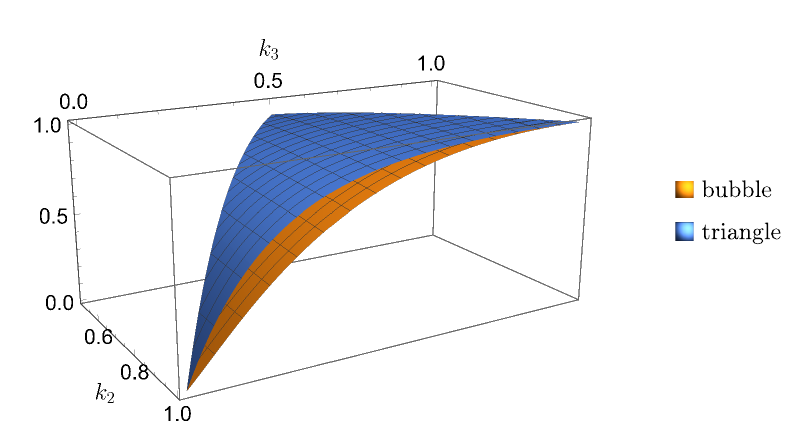}
    \end{subfigure}
    \caption{Normalized bubble and triangle shape functions for $\widetilde{\nu}=0.5$ (left panel) and $0.7$ (right panel), shown on the $k_1=k_*$ slice of one symmetric half of the physical momentum-triangle domain. The bubble and triangle shape functions are strongly correlated with Fisher correlations above 99\% for both mass values. Both shapes are also strongly correlated with the standard equilateral shape, with Fisher correlations above 90\%. See text for more details.}
    \label{fig:btshapes}
\end{figure}

We first compare the momentum dependence of the bispectra from the bubble and triangle diagrams. Fig.~\ref{fig:btshapes} shows the bubble and triangle shape functions with $k_1=k_*$, normalized independently to 1 at the equilateral point. Both contributions are smooth and largest near the equilateral region, and decrease toward the squeezed region. The non-analytic oscillatory behavior from on-shell propagation of the $\sigma$ particle thus has negligible amplitude compared to the analytic background in the full bispectrum.

This similarity is quantified by the \texttt{CMB-BEST} Fisher inner product.  For $\widetilde{\nu}=0.5$, the bubble-triangle correlation is $0.998$, while the correlations of the bubble and triangle with the equilateral template are $0.927$ and $0.936$, respectively. For $\widetilde{\nu}=0.7$, the corresponding correlations are $0.995$, $0.932$, and $0.915$.  Thus, the bubble and triangle are nearly degenerate with each other in the CMB, and each also has a substantial overlap with the equilateral direction. This is expected from the smooth, equilateral-dominated full shapes in Fig.~\ref{fig:btshapes}.

\begin{figure}
    \centering
    \begin{subfigure}[b]{0.4\textwidth}
        \centering
        \includegraphics[width=\textwidth]{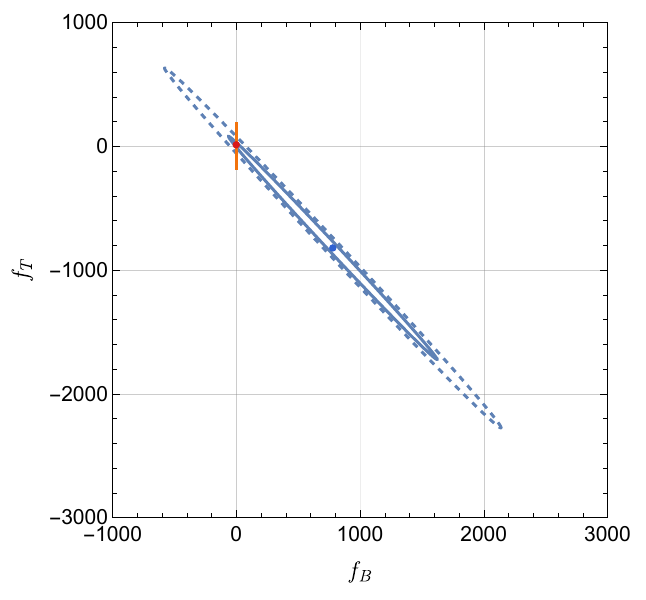}
    \end{subfigure}%
    \hspace*{1cm}
    \begin{subfigure}[b]{0.4\textwidth}
        \centering
        \includegraphics[width=\textwidth]{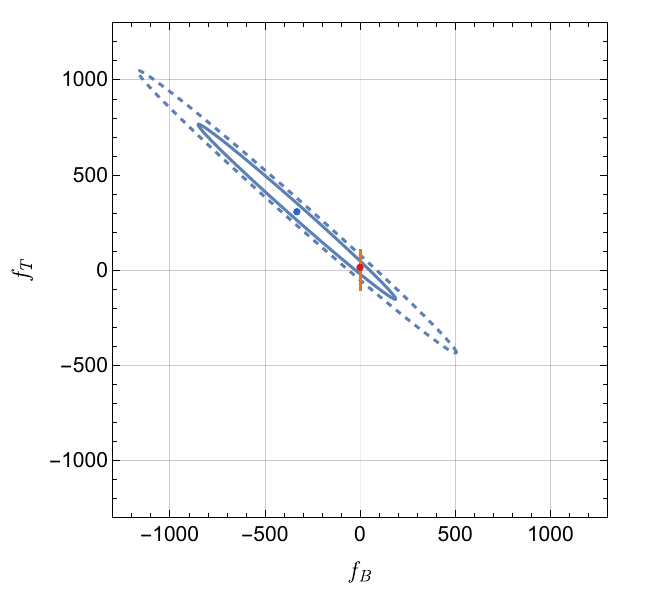}
    \end{subfigure}
    \centering
    \begin{subfigure}[b]{0.46\textwidth}
        \centering
        \includegraphics[width=\textwidth]{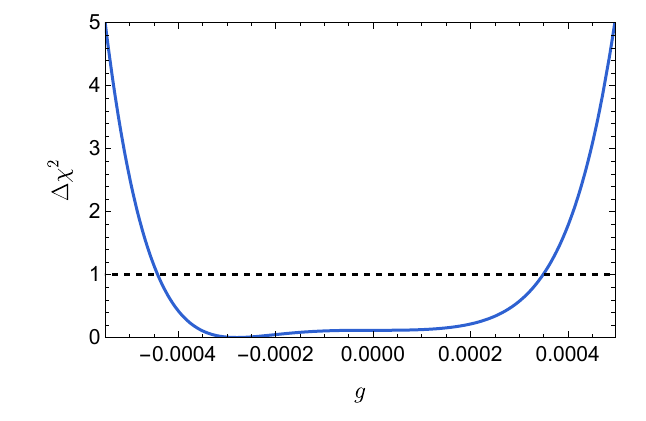}
    \end{subfigure}%
    \hspace*{0.5cm}
    \begin{subfigure}[b]{0.46\textwidth}
        \centering
        \includegraphics[width=\textwidth]{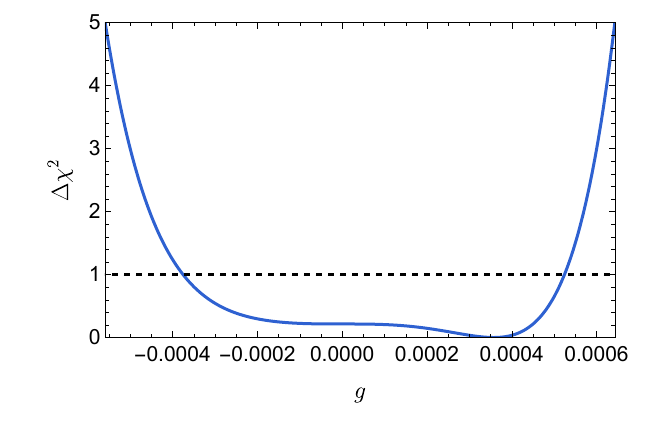}
    \end{subfigure}
    \caption{The top row shows the joint $(f_B,f_T)$ likelihood from \texttt{CMB-BEST} for $\widetilde{\nu}=0.5$ (left panel) and $0.7$ (right panel).  The solid and dashed blue curves are the $1\sigma$ and $2\sigma$ joint confidence contours. The orange curve is the theoretical relation between $f_B$ and $f_T$ from Eq.~\eqref{eq:fnlcurve}. The red point marks the maximum on this curve. The bottom row shows the corresponding one-dimensional profile likelihood for $g$. The horizontal dashed line marks $\Delta\chi^2 = 1$, which defines the $1\sigma$ one-parameter confidence interval.}\label{fig:likelihoods}
\end{figure}

As explained above, we first perform an unconstrained two-template fit, using independent amplitudes $f_B$ and $f_T$ for the bubble and triangle, and then restrict the likelihood to the theory constraint in Eq.~\eqref{eq:fnlcurve}. The joint $(f_B,f_T)$ likelihoods are shown in the top row of Fig.~\ref{fig:likelihoods}. The elongated contours along the negative-slope direction reflect the large positive Fisher correlation between the two templates and the resulting anticorrelation of their fitted amplitudes. The blue points mark the unconstrained maxima, but these points do not correspond to the interaction Lagrangian, for which $f_B$ and $f_T$ are related through the single coupling $g$. The orange curves show this theoretical relation, which passes through $(f_B, f_T) = (0, 0)$. The profile likelihood for $g$ after imposing the theoretical relation is shown in the bottom row of Fig.~\ref{fig:likelihoods}. The $\Delta\chi^2 = 1$ interval contains $g=0$, and we find no evidence for a nonzero coupling.

Finally, we compare the bispectrum limits with an order-of-magnitude estimate of the perturbativity constraint on the correction to the power spectrum and the fine-tuning requirement of the $\sigma$ mass. A complete calculation of the correction to the power spectrum, including its mass dependence and renormalization, is beyond the scope of the present work. For an order-of-magnitude estimate, we retain the loop and combinatorial factor and write
\begin{equation}
 \left|\frac{\Delta\mathcal{P}_{\zeta}}{\mathcal{P}_{\zeta}}\right|
 \simeq C_P\,(59^2 g)^2 F_P(\widetilde{\nu}),
 \qquad C_P=\frac{8}{16\pi^2}.
 \label{eq:powerestimate}
\end{equation}
If we set the unknown mass function $F_P(\widetilde{\nu})$ to 1, Eq.~\eqref{eq:powerestimate} gives $|g|\lesssim 1.3\times10^{-3}$ if the loop correction is required to be smaller than the observed tree-level power spectrum. Because of scale invariance of our theory, a large loop correction to the power spectrum from the interaction Lagrangian does not contradict observation. But it does require resummation of multiple insertions of the $\sigma$-loop correction to the inflaton propagator, which is beyond the scope of our current calculation. The $|g|\lesssim 1.3\times10^{-3}$ constraint is weaker than the $\Delta\chi^2 = 1$ interval from the bispectrum shown in Fig.~\ref{fig:likelihoods}. Our calculation is therefore perturbatively consistent at the level of this order-of-magnitude estimate.

A much stronger constraint is the lack of fine-tuning of the $\sigma$ mass. Due to the classical rolling of the inflaton, the operator $c(\partial_{\eta}\phi\partial^{\eta}\phi)\sigma^2/\Lambda^2$ gives a correction to the $\sigma$ mass by $\delta m_{\sigma}^2\sim c \dot{\phi}_0^2/\Lambda^2$. The mass parameter we have been using throughout the paper is the effective mass after the tree-level mass is corrected by this contribution. For the effective mass to not arise from fine-tuning the EFT operator coefficients, we need $\delta m_{\sigma}^2\lesssim H^2$, or $g \lesssim P_{\rm sr}^{-2} \sim 8\times 10^{-8}$, many orders of magnitude stronger than the constraint from bispectrum search in \textit{Planck} data. For the simple model we have chosen in this paper as a demonstration for the numerical method, observational sensitivity from bispectrum search is currently still far from probing interesting parameter spaces.

\section{Discussion}\label{sec:disc}

In this paper, we developed an error-controlled numerical framework for evaluating one-loop Witten diagrams contributing to the inflationary bispectrum at general external momenta and internal masses. The framework combines the Witten--Feynman parameterization~\cite{Herderschee:2025znl}, the reduced Schwinger method, and analytic ultraviolet subtraction in dimensional regularization. We validated it by reproducing the known ${}_3F_2$ expression for the tree-level bispectrum generated by the $(\partial_\eta\phi)(\partial^\eta\phi)\sigma$ interaction~\cite{Qin:2023ejc}. We then applied it to the one-loop bubble and triangle topology generated by $(\partial_\eta\phi)(\partial^\eta\phi)\sigma^2$ at general kinematics. Finally, we compared the resulting templates with CMB data. Under the CMB weighting considered here, the bubble and triangle templates overlap substantially with one another and with the equilateral template, limiting their observational distinguishability. This conclusion is specific to the models studied here: the same framework applies to a much broader class of interactions and particle content, which may produce more distinctive one-loop shapes.

Previous analytic work has established substantial control over loop-level cosmological correlators in special regimes. Squeezed-limit approximations were studied in Refs.~\cite{Chen:2016uwp,Chen:2016hrz,Chen:2016nrs}, while spectral decompositions of the one-loop bubble and systematic expansions in the squeezed ratio were developed in Refs.~\cite{Xianyu:2022jwk,Qin:2024gtr}. For the triangle diagram, squeezed ratio expansion of the signal component was computed in Refs.~\cite{Qin:2023bjk, XianyuQinTriangle, You:2026xoq}. What remained missing was a systematic and verifiable method for evaluating both bubble and triangle contributions at general masses and kinematics. Because reductions to a small basis of special functions are unlikely at higher loops, we adopt an operational notion of ``solving'' these integrals: a method solves them if it robustly returns a value and a controlled error estimate throughout the relevant parameter range. By this criterion, the reduced Schwinger method solves the class of generalized Euler--Mellin integrals considered here. It analytically integrates the problematic common-scale direction and reduces the remaining integrations to a form suitable for deterministic quadrature. The tropical Monte Carlo method of Appendix~\ref{sec:tropdecomp} provides an independent numerical cross-check.

Looking ahead, the framework admits extensions that would broaden the space of bispectrum templates available for observational searches. Current constraints on primordial non-Gaussianity~\cite{Planck:2019kim} are highly template-dependent. Computing loop-level shapes at general kinematics therefore provides a direct way to enlarge the template library in directions that are not already tightly constrained. The following extensions are feasible within the framework developed here; they require modifying the interaction vertices and propagator structures but not the core integration machinery of Witten-Feynman parameterization and the reduced Schwinger method.
\begin{itemize}
\item \emph{Spinning exchanges.} Replacing the scalar $\sigma$ with fields carrying nonzero spin---massive vectors, spin-$\frac{1}{2}$ fermions, or spin-2 fields---changes the tensor structure of the bulk-to-bulk propagators and introduces angular dependence encoding the exchanged spin. The Schwinger parameterization and Symanzik polynomial structure generalize to these cases, and the resulting shapes have squeezed-limit behavior distinguishable from the scalar case~\cite{Arkani-Hamed:2015bza,Noumi:2012vr,Lee:2016vti}. But similar to the scalar case studied in this paper, more complex interaction structure such as a chemical potential is needed to boost the bispectrum to observable level without fine tuning~\cite{Wang:2019gbi}. Whether our framework could accommodate the more complex Whittaker function propagator in such scenario is left for future work. More exotic matter content includes partially massless fields~\cite{Baumann:2017jvh}, continuous-spin representations~\cite{Bekaert:2017khg}, and unparticle sectors from strongly coupled conformal hidden sectors~\cite{Pimentel:2025rds}, would produce non-standard spectral signatures computable within the same framework.

\item \emph{Extended interaction vertices.} The derivative coupling studied here is a simple representative of a much larger class of interactions. Dimension-5 couplings of the inflaton to heavy-particle currents, generating an effective chemical potential scale $\dot\phi_0^{1/2}\sim 60H$~\cite{Bodas:2020yho,Chen:2018xck}, can overcome Boltzmann suppression and yield $\fnl\sim\mathcal{O}(0.01\text{--}10)$ for masses as high as $\dot\phi_0^{1/2}$. Periodically varying couplings from discrete shift symmetries, such as axion monodromy where the coupling oscillates as $\sim\sin(\phi/f)$, produce resonance-enhanced collider signals when the oscillation frequency exceeds the field mass~\cite{Chen:2022vzh, Wang:2025qww}. These vertex structures can be decomposed into complex conformal-time powers, which change the polynomial integrand but are accommodated by the reduced Schwinger method, and the resulting shapes populate currently unconstrained regions of shape space.

\item \emph{Higher-loop topologies.} The same method extends in principle to sunset diagrams and more complicated higher-loop topologies. The principal challenge is the rapid growth both in the dimensionality of Schwinger-parameter space and in the number of sign-definite regions. This is a computational rather than conceptual limitation and may be mitigated by more efficient enumeration algorithms. Its severity is already apparent in the one-loop triangle, for which the all-plus branch alone required 48 sign-definite regions (cf.~Sec.~\ref{sec:triangle}). The phenomenological value of pursuing such extensions is less clear, however. Higher-loop contributions are generally strongly suppressed, and, absent a symmetry or another mechanism that eliminates lower-order terms, there is no evident reason for them to provide the leading signal.

\item \emph{Four-point function.} Extending the numerical pipeline to the trispectrum would access collapsed-limit singularities and consistency relations complementary to those probed by the bispectrum, providing an additional channel for constraining particle masses and spins. Moreover, for sufficiently singular collapsed configurations, the trispectrum sensitivity can improve with survey resolution faster than the corresponding bispectrum sensitivity, making it a potentially competitive observational probe~\cite{Kalaja:2020mkq}.
\end{itemize}

More broadly, the ability to evaluate loop-level inflationary correlators at general kinematics directly supports the program of using non-Gaussianity to probe particle physics at energy scales far above terrestrial reach~\cite{Chen:2009zp,Arkani-Hamed:2015bza}. The cosmological collider signal encodes the masses and spins of heavy fields through oscillatory and scaling signatures in the squeezed bispectrum~\cite{Chen:2009zp,Arkani-Hamed:2015bza, Noumi:2012vr}, but extracting these signatures from data requires templates accurate across the full triangle domain, not only in asymptotic limits. One-loop contributions are the leading signals when tree-level exchange is forbidden, and otherwise the leading corrections to it. They can shift both the amplitude and the shape of the signal in ways that depend on the full spectrum of the theory. The shapes produced by the extensions discussed above are not necessarily well approximated by any linear combination of existing templates~\cite{Babich:2004gb}, providing new directions in shape space that next-generation measurements can probe. Robust numerical control over these contributions is therefore necessary for quantitative interpretation of non-Gaussianity measurements from the CMB~\cite{Planck:2019kim}, galaxy surveys~\cite{Alvarez:2014vva,Meerburg:2019qqi,Schlegel:2019eqc}, and 21-cm observations~\cite{Munoz:2015eqa}.

\paragraph{Note added.}
We thank Zhehan Qin and Zhong-Zhi Xianyu for coordinating the submission of their concurrent work~\cite{XianyuQinTriangle}, which analytically computes the cosmological collider signal generated by a massive scalar triangle loop in the squeezed bispectrum (and soft-limit trispectrum). The present work numerically evaluates the complete triangle contribution to the bispectrum over full momentum space. A direct comparison would require isolating the corresponding signal from our full numerical result, which we do not attempt in this work.

\section*{Acknowledgments}
In addition to his collaboration on Appendix \ref{sec:tropdecomp}, AH and QL thank Michael Borinsky for valuable discussions of the integrals, their contour deformations, and the associated integration algorithms, as well as for providing preliminary numerical code. We thank Xingang Chen, Soubhik Kumar, and Zhong-Zhi Xianyu for helpful comments on a previous draft. We are grateful to Nima Arkani-Hamed, Daniel Baumann, Daniel Harlow, Guilherme L. Pimentel, Giulio Salvatori, Nathan Seiberg and Zimo Sun for valuable discussion. This material is based upon work supported by the U.S. Department of Energy, Office of Science, Office of High Energy Physics of U.S. Department of Energy under grant Contract Number  DE-SC0012567. In addition, AH is grateful to the Simons Foundation as well as the Edward and Kiyomi Baird Founders’ Circle Member Recognition for their support while working at the IAS. QL is supported in part by the NSF grants PHY-2210498 and PHY-2514611 and by the Simons Foundation. The work of QL was performed in part at the Aspen Center for Physics, which is supported by NSF grant PHY-2210451. Research at Perimeter Institute is supported in part by the Government of Canada through the Department of Innovation, Science and Economic Development Canada and by the Province of Ontario through the Ministry of Colleges and Universities. Fable (Anthropic) and ChatGPT 5.6 (OpenAI) were used extensively for coding. In addition, as mentioned in footnote~\ref{fableusedec}, Fable developed the reduced Schwinger method. Finally, Fable and ChatGPT 5.6 were used to search for typos and logical inconsistencies in the draft, along with minor editing. 

\appendix

\section{Review: the in-in formalism in quantum mechanics}\label{sec:reviewinin}

In this appendix, we illustrate the in-in (Schwinger--Keldysh) formalism with a quantum-mechanical example. We consider the first-order correction to $\langle x^{2}\rangle$ in a harmonic oscillator perturbed by $\lambda x^{4}$. This is the quantum-mechanical analog of computing the one-loop correction to the power spectrum from a $\lambda\phi^{4}$ self-interaction during inflation. The harmonic oscillator ground state $|0\rangle$ plays the role of the Bunch--Davies vacuum~\cite{Bunch:1978yq}.

Consider a harmonic oscillator perturbed by a quartic interaction,
\begin{equation}\label{eq:qm_hamiltonian}
H=H_{0}+H_{I}, \qquad H_{0}=\frac{p^{2}}{2m}+\frac{1}{2}m\omega^{2}x^{2}, \qquad H_{I}=\lambda x^{4} \ .
\end{equation}
We set $\hbar = 1$ throughout. Writing $x=\sqrt{1/(2m\omega)}\,(a+a^{\dagger})$, the elementary interaction-picture propagators are the positive- and negative-frequency Wightman functions,
\begin{equation}\label{eq:wightman}
G^{+}(t-t') \equiv \langle 0|x_{I}(t)\,x_{I}(t')|0\rangle = \frac{1}{2m\omega}\,e^{-i\omega(t-t')} \ , \qquad G^{-}(t-t') \equiv \langle 0|x_{I}(t')\,x_{I}(t)|0\rangle = \frac{1}{2m\omega}\,e^{+i\omega(t-t')} \ .
\end{equation}
Note that $G^{-}(\tau)=[G^{+}(\tau)]^{*}$ and $G^{+}(0)=G^{-}(0)=1/(2m\omega)$. The four contour propagators are
\begin{equation}
\begin{aligned}
G^{++}(t,t')&=\theta(t-t')G^+(t-t')+\theta(t'-t)G^-(t-t'),\\
G^{--}(t,t')&=\theta(t'-t)G^+(t-t')+\theta(t-t')G^-(t-t'),\\
G^{-+}(t,t')&=G^+(t-t'),\qquad G^{+-}(t,t')=G^-(t-t').
\end{aligned}
\end{equation}

\begin{figure}[t]
\centering
\begin{tikzpicture}[every node/.style={font=\footnotesize}]
\draw [thick,-stealth] (-3.5,0.2) -- (2.8,0.2);
\node [above] at (2.2,0.2) {\footnotesize $T$ (forward, $+$)};
\draw [thick,-stealth] (2.8,-0.2) -- (-3.5,-0.2);
\node [left] at (-3.65,-0.2) {$\bar{T}$ (backward, $-$)};
\draw [thick] (2.8,0.2) to[out=0,in=0] (2.8,-0.2);
\filldraw [white] (2.8,0) circle (3pt);
\draw (2.8,0) circle (3pt);
\node [right=6pt] at (3.05,0) {$x^{2}(t)$};
\filldraw (-0.5,0.2) circle (2pt) node[above=3pt]{$-i\lambda$};
\filldraw (-2,-0.2) circle (2pt) node[below=3pt]{$+i\lambda$};
\draw [gray,-stealth] (-3.5,-0.7) -- (2.8,-0.7);
\node [gray,right] at (2.8,-0.7) {$t$};
\end{tikzpicture}
\caption{The Schwinger--Keldysh contour for the first-order in-in computation. The operator insertion $x^{2}(t)$ sits at the tip of the contour. The interaction vertex at $t'$ can sit on the forward ($+$) branch, contributing $-i\lambda$, or the backward ($-$) branch, contributing $+i\lambda$. The physical result is obtained by summing both branch contributions, Eq.~\eqref{eq:inin_twobranch}.}
\label{fig:sk_contour}
\end{figure}
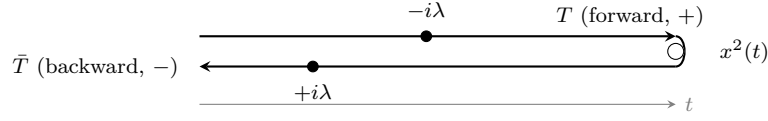

The in-in formalism computes expectation values at a fixed time rather than transition amplitudes between asymptotic states. The expectation value of a Heisenberg-picture operator $Q(t)$ in the interacting vacuum is
\begin{equation}\label{eq:inin_master}
\langle \Omega|Q(t)|\Omega\rangle = \langle 0|\,\bar{T}\exp\!\left (i\int_{-\infty(1+i\epsilon)}^{t}H_{I}\,dt_{-}\right )Q_{I}(t)\, T\exp\!\left (-i\int_{-\infty(1-i\epsilon)}^{t}H_{I}\,dt_{+}\right )|0\rangle  \ ,
\end{equation}
where $T$ and $\bar{T}$ are the time-ordering and anti-time-ordering operators and the $i\epsilon$ prescription projects the free vacuum onto the interacting vacuum~\cite{Weinberg:2005vy,Chen:2017ryl}. The operator $Q_{I}(t)$ sits at the tip of the Schwinger--Keldysh contour. The $+$ branch is time ordered and approaches the past along $t_+\to-\infty(1-i\epsilon)$. The $-$ branch is anti-time ordered and approaches the past along $t_-\to-\infty(1+i\epsilon)$, as shown in Fig.~\ref{fig:sk_contour}. Interaction vertices on the $+$ branch carry a factor $(-i)$ and vertices on the $-$ branch carry $(+i)$. In inflation, the Bunch--Davies vacuum replaces $|0\rangle$ and $Q$ is replaced by products of $\zeta(\vec{k})$ to compute $n$-point correlators~\cite{Maldacena:2002vr}.

At zeroth order in $\lambda$, the expectation value $\langle x^{2}\rangle$ is a single self-contraction.
\begin{equation}\label{eq:x2_LO}
\langle x^{2}\rangle^{(0)} =\quad
\begin{tikzpicture}[baseline={([yshift=-.5ex]current bounding box.center)},every node/.style={font=\scriptsize}]
\filldraw [white] (0,0) circle (2pt);
\draw (0,0) circle (2pt);
\draw [thick] (0,0) .. controls (-0.4,0.7) and (0.4,0.7) .. (0,0);
\end{tikzpicture}
\quad = G^{+}(0) \ .
\end{equation}
Wick-rotating to Euclidean signature, the propagator in frequency space is $\tilde{G}_{E}(p)=1/[m(p^{2}+\omega^{2})]$, so the equal-time contraction evaluates to
\begin{equation}\label{eq:bubbleqm}
G^{+}(0) = \int\frac{dp}{2\pi}\frac{1}{m(p^{2}+\omega^{2})}=\frac{1}{m}\cdot\frac{1}{2\omega} = \frac{1}{2m\omega} \ ,
\end{equation}
where we closed the contour in the upper half-plane and picked up the pole at $p=i\omega$. Unlike in higher dimensions, this integral is UV finite and requires no regularization.

We now compute the $\mathcal{O}(\lambda)$ correction. Expanding Eq.~\eqref{eq:inin_master} to first order, the interaction vertex at time $t'$ can sit on either branch of the Schwinger--Keldysh contour.
\begin{equation}\label{eq:inin_twobranch}
\langle x^{2}(t)\rangle^{(1)} 
= \underbrace{(-i\lambda)\int_{-\infty}^{t}dt_{+}\,\langle 0|\,x_{I}^{2}(t)\,x_{I}^{4}(t_{+})\,|0\rangle}_{\displaystyle \langle x^{2}\rangle^{(1)}_{+}}
\;+\; \underbrace{(+i\lambda)\int_{-\infty}^{t}dt_{-}\,\langle 0|\,x_{I}^{4}(t_{-})\,x_{I}^{2}(t)\,|0\rangle}_{\displaystyle \langle x^{2}\rangle^{(1)}_{-}} \ .
\end{equation}
The first term is the $+$ (forward) branch contribution, in which all operators are already time-ordered since $t'<t$. The second is the $-$ (backward) branch contribution, which is anti-time-ordered. The two terms are complex conjugates of each other, but we evaluate them separately since this structure generalizes directly to the cosmological computation with distinct $\pm$ vertex rules~\cite{Chen:2017ryl}.

The correlator in each branch involves six operators. Applying Wick's theorem yields connected and disconnected classes of contractions. In the connected class, both operators at $t$ are contracted with operators at $t'$, while the remaining two operators at $t'$ form a tadpole. On the $+$ branch, contractions from $t$ to $t'$ produce $G^{+}(t-t')$.
\begin{equation}\label{eq:connected_minus}
\langle x^{2}\rangle^{(1)}_{+}\Big|_{\textrm{conn}} =\quad
\begin{tikzpicture}[baseline={([yshift=-.5ex]current bounding box.center)},every node/.style={font=\scriptsize}]
\draw [thin, dashed] (-1.0,-0.8) -- (1.0,-0.8) node[right]{$t$};
\draw [thick] (0,-0.8) to[bend left=30] (0,0.8);
\draw [thick] (0,-0.8) to[bend right=30] (0,0.8);
\draw [thick] (0,0.8) .. controls (-0.7,1.1) and (-0.7,0.5) .. (0,0.8);
\filldraw [white] (0,-0.8) circle (2pt);
\draw (0,-0.8) circle (2pt);
\filldraw (0,0.8) circle (2pt) node[right=4pt]{$t_{+}$};
\node [left=6pt] at (-0.7,0.8) {$+$};
\end{tikzpicture}
\quad = (-i\lambda)\cdot 12\int_{-\infty}^{t}dt_{+}\,[G^{+}(t-t_{+})]^{2}\,G^{+}(0) \ ,
\end{equation}
where the combinatorial factor is $\binom{4}{2}\times 2! = 12$. On the $-$ branch, the operator ordering is reversed, so contractions from $t'$ to $t$ produce $G^{+}(t'-t)=G^{-}(t-t')$.
\begin{equation}\label{eq:connected_plus}
\langle x^{2}\rangle^{(1)}_{-}\Big|_{\textrm{conn}} =\quad
\begin{tikzpicture}[baseline={([yshift=-.5ex]current bounding box.center)},every node/.style={font=\scriptsize}]
\draw [thin, dashed] (-1.0,0.8) -- (1.0,0.8) node[right]{$t$};
\draw [thick] (0,0.8) to[bend left=30] (0,-0.8);
\draw [thick] (0,0.8) to[bend right=30] (0,-0.8);
\draw [thick] (0,-0.8) .. controls (-0.7,-0.5) and (-0.7,-1.1) .. (0,-0.8);
\filldraw [white] (0,0.8) circle (2pt);
\draw (0,0.8) circle (2pt);
\filldraw (0,-0.8) circle (2pt) node[right=4pt]{$t_{-}$};
\node [left=6pt] at (-0.7,-0.8) {$-$};
\end{tikzpicture}
\quad = (+i\lambda)\cdot 12\int_{-\infty}^{t}dt_{-}\,[G^{-}(t-t_{-})]^{2}\,G^{+}(0) \ .
\end{equation}
The disconnected contraction, in which $x^{2}(t)$ self-contracts and all four operators at $t_{\pm}$ pair among themselves, gives 
\begin{equation}\label{eq:disconnected1}
\begin{tikzpicture}[baseline={([yshift=-.5ex]current bounding box.center)},every node/.style={font=\scriptsize}]
\draw [thin, dashed] (-1.2,0.8) -- (1.2,0.8) node[right]{$t$};
\draw [thick] (0,0.8) .. controls (-0.3,1.5) and (0.3,1.5) .. (0,0.8);
\draw [thick] (0,-0.8) .. controls (-0.7,-0.5) and (-0.7,-1.1) .. (0,-0.8);
\draw [thick] (0,-0.8) .. controls (0.7,-0.5) and (0.7,-1.1) .. (0,-0.8);
\filldraw [white] (0,0.8) circle (2pt);
\draw (0,0.8) circle (2pt);
\filldraw (0,-0.8) circle (2pt) node[above=4pt]{$t_{-}$};
\node [left=6pt] at (-0.7,-0.8) {$-$};
\end{tikzpicture}
\quad = (i\lambda)\cdot 3\,[G^{+}(0)]^{3}\int_{-\infty}^{t}dt_{-} \ ,
\end{equation}
and 
\begin{equation}\label{eq:disconnected2}
\begin{tikzpicture}[baseline={([yshift=-.5ex]current bounding box.center)},every node/.style={font=\scriptsize}]
\draw [thin, dashed] (-1.2,-0.8) -- (1.2,-0.8) node[right]{$t$};
\draw [thick] (0,-0.8) .. controls (-0.3,-1.5) and (0.3,-1.5) .. (0,-0.8);
\draw [thick] (0,0.8) .. controls (-0.7,1.1) and (-0.7,0.5) .. (0,0.8);
\draw [thick] (0,0.8) .. controls (0.7,1.1) and (0.7,0.5) .. (0,0.8);
\filldraw [white] (0,-0.8) circle (2pt);
\draw (0,-0.8) circle (2pt);
\filldraw (0,0.8) circle (2pt) node[below=4pt]{$t_{+}$};
\node [left=6pt] at (-0.7,0.8) {$+$};
\end{tikzpicture}
\quad = (-i\lambda)\cdot 3\,[G^{-}(0)]^{3}\int_{-\infty}^{t}dt_{+} \ ,
\end{equation}
with combinatorial factor $4!/(2!\cdot 2!\cdot 2)=3$. This contribution is proportional to $[G^{+}(0)]^{3}$ and carries no phase that depends on $t-t'$. Crucially, the $-$ and $+$ branch contributions carry opposite signs ($+i\lambda$ versus $-i\lambda$) and therefore cancel when summed.%
\footnote{This is the in-in analog of the linked-cluster theorem. Disconnected vacuum bubbles cancel between the forward and backward branches. In inflation, the same cancellation removes the parts of a diagram that are disconnected from every external operator. Pieces that factorize into lower-point functions but remain attached to external operators are removed by taking the connected correlator.}

We now evaluate the connected contributions. On the $+$ branch, substituting Eq.~\eqref{eq:wightman} into Eq.~\eqref{eq:connected_minus} and performing the time integral with the $i\epsilon$ prescription gives
\begin{equation}\label{eq:minus_branch_integral}
\langle x^{2}\rangle^{(1)}_{+}\Big|_{\textrm{conn}} = \frac{-12\,i\lambda}{(2m\omega)^{3}}\int_{-\infty}^{t}dt_{+}\,e^{-2i\omega(t-t_{+})-\epsilon(t-t_{+})} = \frac{-12\,i\lambda}{(2m\omega)^{3}}\cdot\frac{1}{2i\omega} = -\frac{3\lambda}{4m^{3}\omega^{4}} \ .
\end{equation}
On the $-$ branch, the reversed operator ordering replaces $G^{+}\to G^{-}$, flipping the sign of the oscillating phase.
\begin{equation}\label{eq:plus_branch_integral}
\langle x^{2}\rangle^{(1)}_{-}\Big|_{\textrm{conn}} = \frac{+12\,i\lambda}{(2m\omega)^{3}}\int_{-\infty}^{t}dt_{-}\,e^{+2i\omega(t-t_{-})-\epsilon(t-t_{-})} = \frac{+12\,i\lambda}{(2m\omega)^{3}}\cdot\frac{1}{-2i\omega} = -\frac{3\lambda}{4m^{3}\omega^{4}} \ .
\end{equation}
The two branches contribute equally, as expected from the fact that they are complex conjugates whose real parts add and whose imaginary parts cancel. Summing the two branches gives
\begin{equation}\label{eq:x2_correction}
\langle x^{2}\rangle^{(1)} = \langle x^{2}\rangle^{(1)}_{-} + \langle x^{2}\rangle^{(1)}_{+} = -\frac{3\lambda}{2m^{3}\omega^{4}} \ ,
\end{equation}
The full result to $\mathcal{O}(\lambda)$ is
\begin{equation}\label{eq:x2_full}
\langle x^{2}\rangle = \frac{1}{2m\omega}-\frac{3\lambda}{2m^{3}\omega^{4}}+\mathcal{O}(\lambda^{2}) \ ,
\end{equation}
which agrees with standard time-independent perturbation theory. The negative sign reflects the stiffening of the potential by the quartic term.

Several structural features of this calculation carry over to the computation of non-Gaussianity during inflation. The master formula in Eq.~\eqref{eq:inin_master} has the same structure as the cosmological in-in formula for $\langle \zeta(\vec{k}_{1})\cdots\zeta(\vec{k}_{n})\rangle$, with the Bunch--Davies vacuum replacing $|0\rangle$~\cite{Maldacena:2002vr}. Each vertex carries a branch label $\pm$ and a factor $\mp i\lambda$, as in the diagrammatic rules of~\cite{Chen:2017ryl}, and the physical correlator is obtained by summing over all branch assignments. The propagators connecting pairs of vertices are the four Schwinger--Keldysh propagators $G^{\pm\pm}$, of which $G^{+}$ and $G^{-}$ above are the off-diagonal components $G^{-+}$ and $G^{+-}$, respectively. The $i\epsilon$ prescription ensures convergence at $t'\to -\infty$ and is the analog of selecting the Bunch--Davies vacuum in the infinite past. Finally, disconnected diagrams cancel between the two branches, as demonstrated in Eqs.~\eqref{eq:disconnected1} and \eqref{eq:disconnected2}.

\section{Tropical Monte Carlo for divergent Euler--Mellin integrals}\label{sec:tropdecomp}
\begin{center}\emph{Appendix with Michael Borinsky}\end{center}

In this appendix, we develop the tropical Monte Carlo method used as an independent cross-check of the reduced Schwinger method in Sec.~\ref{sec:redschwinger}. A secondary purpose of this appendix is to demonstrate the compatibility of the tropical Monte Carlo framework with dimensional regularization. To this end, we combine the tropical Monte Carlo method of~\cite{Borinsky:2020rqs} with the directed integration-by-parts algorithm of~\cite{MR3010271}, allowing for the numerical evaluation of divergent Euler--Mellin integrals, including Feynman integrals, through the computation of the coefficients of their $\varepsilon$-expansion.

We adapt the strategy outlined for Feynman integrals in Ref.~\cite{Borinsky:2020rqs} to the de Sitter integrals considered in this paper, using only part of the full tropical-geometric pipeline.
Our approach is essentially equivalent to the geometric formulation of sector decomposition due to Kaneko and Ueda~\cite{Kaneko:2009qx}. Sector decomposition, introduced by Binoth and Heinrich~\cite{Binoth:2000ps}, automates and generalizes a process due to Hepp~\cite{Hepp:1966eg}.
For the practical handling of divergent integrals, we use the directed integration-by-parts algorithm of Nilsson and Passare~\cite{MR3010271}. Independently, a functionally equivalent version of this algorithm, tailored to Feynman integrals, was put forward in Ref.~\cite{vonManteuffel:2014qoa}.

We consider general integrals of the form
\begin{equation}\label{eq:genEulerint}
\int_{\mathbb{R}_{>0}^n} \left ( \prod_{i} dx_{i}x_{i}^{\nu_{i}-1} \right )\prod_{j}P_{j}(x_1,\ldots,x_n)^{-\delta_{j}} \, ,
\end{equation}
where the $P_{j}(x_1,\ldots,x_n)$ are polynomials of the $x_{i}$ variables. 
To guarantee convergence, we need all $P_j$ to be \emph{completely non-vanishing}~\cite{MR3010271} (see, e.g.,~\cite[Def.~2]{Borinsky:2020rqs} for details relevant in our context). This is a slightly stronger condition than only requiring that $P_j$ evaluates to a positive number if the variables $x_i$ are all positive. For example, the polynomial $x_1^2 - 2x_1x_2 + x_2^2 + x_3^2 = (x_1-x_2)^2+ x_3^2$ evaluates to a positive number whenever $x_1,x_2,x_3 >0$, but it is not completely non-vanishing. 
The polynomial $P_j$ is always completely non-vanishing if all coefficients of $P_j$ are positive. Even though this property is not necessary to guarantee convergence, as there are completely non-vanishing polynomials with mixed-sign or complex coefficients, we will aim for the stronger property of positive coefficients.

The goal is to partition the integration domain into regions in which a single monomial dominates in each polynomial in the product. These regions are then mapped to the unit hypercube, making them particularly well suited for numerical integration.

The appendix is organized as follows. Sec.~\ref{simgen} introduces the general tropical approach to integral evaluation via a simple example with real exponents. Sec.~\ref{sec:tropcomplex} explains how the algorithm accommodates the complex exponents $\nu_i$ and $\delta_j$ that arise in the de Sitter integrals considered in this paper. Sec.~\ref{sec:regdiv} describes how manifestly divergent integrals are put into a form amenable to tropical integration, and Sec.~\ref{sec:tropicalfancompute} reviews the algorithmic construction of the tropical fan for a generic polynomial.

\subsection{Sector decomposition from the tropical fan via a simple example}\label{simgen}

In this section, we illustrate the general algorithm in a simple example. 
We consider the integral
\begin{equation}
\label{eq:exampleint}
I(\delta,\nu_1,\nu_2)=\int_{0}^{\infty}\int_{0}^{\infty} \frac{x_1^{\nu_1} x_2^{\nu_2} }{P(x_1, x_2)^{\delta} } \frac{\dd x_{1}}{x_1} \frac{\dd x_{2}}{x_2} \, ,
\end{equation}
where $\delta>0$ and $\nu_1,\nu_2$ are real parameters, which we later continue to complex values, and 
\begin{equation}\label{eq:examplepoly}
P(x_1, x_2) = 1+ 2x_1^2 + x_2^2 + x_1 x_2^2 + 3x_1^2 x_2\,.
\end{equation}
We want to compute the value $I(\delta,\nu_1,\nu_2)$ for different parameters $\delta,\nu_1,\nu_2$. 
Even with advanced numerical algorithms, this is a highly nontrivial problem, because the integrand, even when integrable, is often not square-integrable. 
Naive Monte Carlo integration requires the integrand to be square-integrable for the estimator to have finite variance. Quartic integrability is additionally required for the sample-variance estimator itself to have finite variance. Without it, finite-sample error estimates can be unstable. Black-box variance-reduction methods such as grid adaptation can mitigate the problem, but their success is unpredictable, especially in cases where the naive approach fails completely. 

Here, we use a systematic approach to tackle the numerical integration of~\eqref{eq:exampleint} that builds on an underlying tropical geometry.
It is useful to introduce the following function.
\[
P^\tr(x_1,x_2) = \max(1,x_1^2, x_2^2, x_1 x_2^2,x_1^2 x_2)\,,
\]
where we included all monomials of $P$, but set all coefficients to $1$ and took the maximum instead of summing.
The function $P^\tr$ is called the \emph{tropical approximation} of $P$, due to the following \emph{approximation property} (see~\cite[Thm.~8]{Borinsky:2020rqs} for details and generalizations). For each pair $P$, $P^\tr$, where $P$ has positive coefficients, there are constants $C_1,C_2>0$ such that 
\[
C_1 \leq \frac{P(x_1,x_2)}{P^\tr(x_1,x_2)} \leq C_2 \quad \text{for all} \quad x_1,x_2 >0\,. 
\]
In our example, we can choose $C_1 = 1$ and $C_2 = 8$.
As a consequence of this property, the functions $P$ and $P^\tr$ behave similarly in the integration domain. 
In particular, $P^\tr$ is large precisely where $P$ is large, and small precisely where $P$ is small.

As it is piecewise monomial, the function $P^\tr$ is much easier to handle under the integral sign than the polynomial $P$. We can insert a $1$ into the integrand~\eqref{eq:exampleint} to effectively replace $P$ with $P^\tr$.
\begin{equation}
\label{eq:exampleint2}
I(\delta,\nu_1,\nu_2)=\int_{0}^{\infty}\int_{0}^{\infty} g(x_1,x_2) \frac{x_1^{\nu_1} x_2^{\nu_2}}{P^\tr(x_1, x_2)^{\delta} } \frac{\dd x_{1}}{x_1} \frac{\dd x_{2}}{x_2}\, ,
\end{equation}
where
\[
g(x_1,x_2) = 
\left( \frac{P^\tr(x_1,x_2)}{P(x_1,x_2)} \right)^\delta\,.
\]
From now on, the integration kernel $g(x_1,x_2)$ can play a spectator role. It is bounded from above and below by the approximation property, 
so it affects the value of the integral but not its convergence.

A (piecewise) monomial function becomes a (piecewise) linear function in logarithmic coordinates. For this reason, changing to logarithmic coordinates is a standard theme in tropical geometry. 
In Eq.~\eqref{eq:exampleint2}, we pass to logarithmic coordinates $z_i$, defined by
$x_i = \exp(z_i)$, and get 
\begin{equation}
\label{eq:exampleint3}
I(\delta,\nu_1,\nu_2)=\int_{\mathbb{R}^2} g(e^{z_1},e^{z_2}) \exp\left( \nu_1 z_1 + \nu_2 z_2 - \delta \cdot \textrm{Trop}[P](z_1,z_2) \right) {\dd z_{1}} {\dd z_{2}}\, ,
\end{equation}
where 
\begin{equation}
\label{eq:troppoly}
\textrm{Trop}[P](z_1,z_2) = \log( P^\tr(e^{z_1},e^{z_2} ) ) = \max\!\left(0,\;2z_{1},\;2z_{2},\;z_{1}+2z_{2},\;2z_{1}+z_{2}\right)\,.
\end{equation}
By construction, the linear forms in the argument of the $\max$ function above correspond to monomials of the polynomial $P$ and capture the precise exponents of the variables.
Here, the benefit of our previous manipulations becomes apparent. We have arrived at an integral which is controlled by an exponential with a piecewise linear function in its argument. 
The integration kernel $g$ is bounded from below and above, so the value of the integral mainly depends on the behavior of the exponential. 
For instance, it is clear that the integral converges if the piecewise linear function $\nu_1 z_1 + \nu_2 z_2 - \delta \cdot \textrm{Trop}[P](z_1,z_2)$ 
is strictly negative on every ray $\{tz \in \mathbb{R}^2 : t > 0\}$, $z \neq 0$.
The precise mathematical statement is that the integral~\eqref{eq:exampleint3} is convergent if the vector $(\nu_1,\nu_2)$ is contained in the interior of $\delta \cdot \mathcal{N}(P)$, 
which is the $\delta$-scaled Newton polytope (the convex hull of all exponent vectors) of $P$. This fundamental property of generalized Euler--Mellin integrals was first proven by Nilsson and Passare as Theorem~1 of~\cite{MR3010271}. A similar statement was used in the context of certain string theory integrals in Ref.~\cite{Arkani-Hamed:2019mrd}.

\begin{figure}[t]
    \centering
\includegraphics[width=0.6\textwidth]{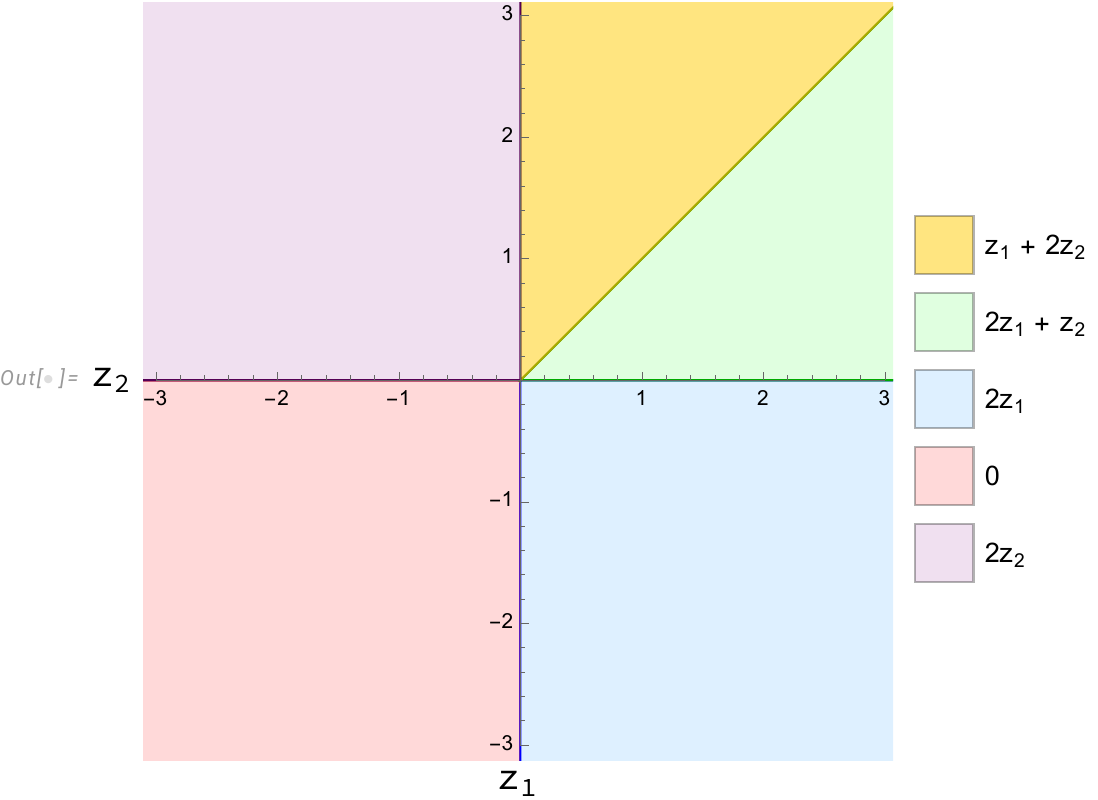}
    \caption{The tropical fan corresponding to Eq.~\eqref{eq:troppoly}. The color of each region corresponds to which term in the polynomial dominates in that region.}
    \label{fig:fignormalfan}
\end{figure}

Fig.~\ref{fig:fignormalfan} shows the different sectors (also known as maximal cones) of $\mathbb{R}^2$ in which the exponent in Eq.~\eqref{eq:exampleint3} is linear. 
 The set of all cones is called the \emph{normal fan} of the Newton polytope $\mathcal{N}(P)$. 
For brevity, we call this normal fan the tropical fan associated with $P$.
Computing the tropical fan and thereby resolving the $\max$ function in the integral above is a nontrivial but algorithmically solvable problem.
We discuss how to compute the tropical fan for a generic polynomial in Sec.~\ref{sec:tropicalfancompute}, noting here that existing algorithms scale comfortably to polynomials in up to approximately 50 variables~\cite{Gawrilow:2000qhs}, well beyond what is needed for the integrals we consider. We provide \textsc{Mathematica} code suitable for polynomials in up to 10 variables, which is again sufficient for our purposes.

Given the tropical fan, we can break up the integration region into maximal cones where a single monomial dominates. Consider the yellow region in Fig.~\ref{fig:fignormalfan}. Since the region is a polyhedral cone, all points inside it can be written as a nonnegative linear combination of the generating vectors $(0,1)^T$ and $(1,1)^T$.
\begin{equation}\label{eq:changeofvariablescones}
\left.\binom{z_{1}}{z_{2}}\right|_{\textrm{yellow region}}\in \{ c_{1}\binom{0}{1}+c_{2}\binom{1}{1}:   c_{1},c_{2}\geq 0 \}\,.
\end{equation}
The contribution to the integral~\eqref{eq:exampleint3} from this cone is
\begin{equation}
\label{eq:exampleint4}
\int_{\textrm{yellow region}} g(e^{z_1},e^{z_2}) \exp\left( \nu_1 z_1 + \nu_2 z_2 - \delta (z_1+2z_2) \right) {\dd z_{1}} {\dd z_{2}}  \, ,
\end{equation}
where we could resolve the $\max$ function because $\textrm{Trop}[P](z_1,z_2)$ is linear in each cone of the tropical fan. Furthermore, changing to the conical, positive $c_1,c_2$ coordinates yields the expression:
\begin{align}
&=
\int_{0}^\infty
\int_{0}^\infty
g(e^{c_2},e^{c_1+c_2}) \exp\left( 
- ( 2\delta - \nu_2) c_1 - (3 \delta - \nu_1 -\nu_2 ) c_2
\right) {\dd c_{1}} {\dd c_{2}}\,.
\intertext{
The convergence conditions arising  
from this specific cone are $2\delta - \nu_2 > 0$ and $3 \delta - \nu_1 -\nu_2 > 0$.
With the further change of coordinates $c_1 = - \log( y_1 ) / ( 2\delta - \nu_2)$ and 
$c_2 = - \log( y_2 ) / ( 3\delta - \nu_1 - \nu_2)$, we pull the integral back to the unit square $[0,1]^2$ and obtain
}
&=
\frac{1}{( 2\delta - \nu_2)( 3\delta - \nu_1 - \nu_2)}
\int_{0}^1
\int_{0}^1
g\left(y_2^{-\frac{1}{3\delta - \nu_1 - \nu_2}},y_1^{-\frac{1}{2\delta - \nu_2}}y_2^{-\frac{1}{3\delta - \nu_1 - \nu_2}}\right)  {\dd y_{1}} {\dd y_{2}}\,.
\end{align}
The resulting integrand is given entirely by evaluating $g$, and is therefore bounded from above and below 
by the approximation property.
Such an integral is straightforward to evaluate numerically, either by naive Monte Carlo or some other algorithm that aims to accelerate the convergence. The important property of the resulting integrand is that it is bounded and therefore belongs to $L^2$, $L^4$, and $L^\infty$. 

Applying this process to every maximal cone (after triangulating non-simplicial cones, see below) in the tropical fan, we can rewrite Eq.~\eqref{eq:exampleint} as a sum over all such cones:
\begin{align}
I(\delta,\nu_1,\nu_2) =
\sum_{\textrm{cones } \sigma} 
I^\tr_\sigma(\delta,\nu_1,\nu_2) \int_{[0,1]^2} g\left( \prod_{i \in \{1,2\} } y_i^{\rho^\sigma_{1,i}}, \prod_{i \in \{1,2\} } y_i^{\rho^\sigma_{2,i} } \right) \dd y_1 \dd y_2\,,
\end{align}
where $I^\tr_\sigma(\delta,\nu_1,\nu_2)$ is the prefactor of the cone integral,
i.e., the tropical contribution, and $\rho^\sigma_{j,i}$ are rational functions in  $\delta, \nu_1,\nu_2$ that can be computed using standard convex geometry algorithms.

The full tropical sampling approach put forward in Refs.~\cite{Borinsky:2020rqs,Borinsky:2023jdv,Borinsky:2025asc} 
not only evaluates the individual integrals in the sum above using Monte Carlo, but 
also evaluates the \emph{sum over all cones} using Monte Carlo. The weighting of the 
cones is given by the tropical prefactors $I^\tr_\sigma(\delta,\nu_1,\nu_2)$. 
In a special case that is highly relevant for Feynman integrals, namely when all Newton polytopes are generalized permutahedra~\cite{MR2487491}, there is a sampling algorithm that avoids enumerating the cones altogether.
 This algorithm is the key to fast integration of Feynman integrals at high loop orders. Here, we are mainly interested in integrals of much smaller dimension, so we will not attempt to apply this more sophisticated sampling approach. 

\begin{figure}[t]
    \centering
    \begin{subfigure}[b]{0.48\textwidth}
        \centering
        \includegraphics[width=\textwidth]{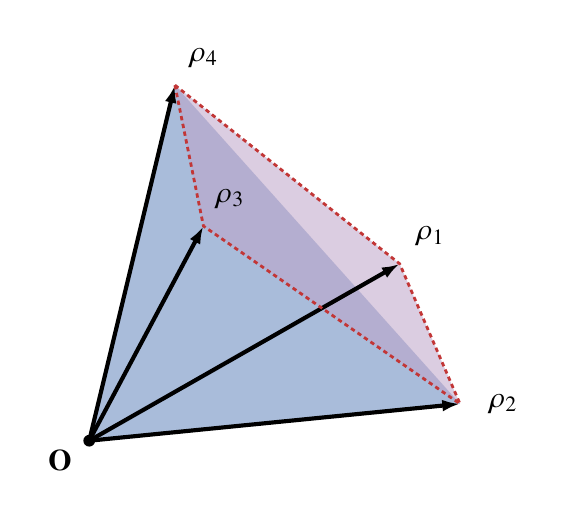}
        \caption{}
        \label{fig:sub_a}
    \end{subfigure}
    \hfill
    \begin{subfigure}[b]{0.5\textwidth}
        \centering
        \includegraphics[width=\textwidth]{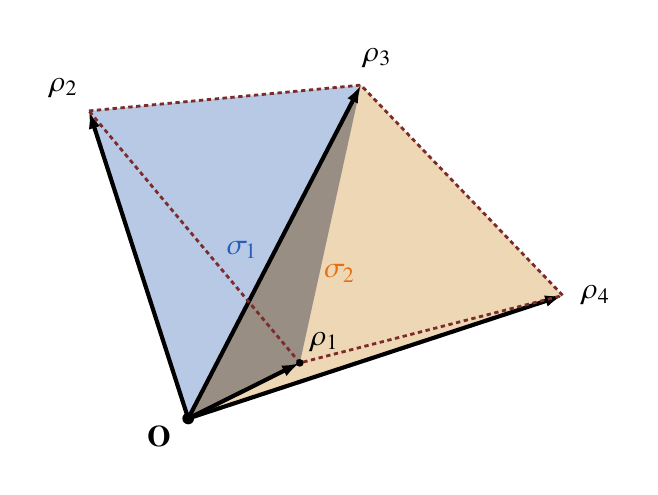}
        \caption{}
        \label{fig:sub_b}
    \end{subfigure}
    \caption{(a) A non-simplicial cone in three dimensions. (b) A non-simplicial cone in three dimensions triangulated into two simplicial cones.}
    \label{fig:nonsimplfans}
\end{figure}

A technical difficulty that arises when generalizing the above process to higher-dimensional integrals 
is that the resulting tropical fan is not necessarily \emph{simplicial}. 
That means that maximal cones are not necessarily generated by $n$ vectors,
where $n$ is the dimension of the integration domain. Fig.~\ref{fig:nonsimplfans} shows a non-simplicial cone in three dimensions
and how to triangulate that cone using simplicial cones. Every non-simplicial fan can be triangulated algorithmically using
standard convex geometry frameworks.

\subsection{Complex exponents}\label{sec:tropcomplex}

The integrals produced by the WF parameterization are of the form~\eqref{eq:genEulerint} and generically have \emph{complex} exponents. As discussed in Sec.~\ref{sec:redschwinger}, the scaling dimensions $\Delta_\pm = \frac{d}{2} \pm i\widetilde{\nu}$ enter the exponents $\nu_i$ and $\delta_j$ directly. The tropical decomposition extends to this case essentially unchanged, with a single modification in the final step of the algorithm.

The key observation is that the imaginary parts of the exponents do not affect the magnitude of the integrand. For $x>0$, we have $|x^{\nu}| = x^{\operatorname{Re}\nu}$. Likewise, for $P(x_1,\ldots,x_n)>0$, we have $|P^{-\delta}| = P^{-\operatorname{Re}\delta}$. The relative dominance of the monomials, and hence the convergence properties of the integral, are therefore controlled entirely by the real parts of the exponents. Accordingly, we run the algorithm of Sec.~\ref{simgen} as before, using only the real parts. The tropical approximation, the tropical fan, and the decomposition into cones are all unchanged when $\operatorname{Re}\delta_j>0$, and the convergence criterion becomes the requirement that the vector $\operatorname{Re}\nu$ lies in the interior of $\sum_j \operatorname{Re}\delta_j \cdot \mathcal{N}(P_j)$. The integration kernel $g$ also remains bounded, since $|g| = (P^{\tr}/P)^{\operatorname{Re}\delta}$ is controlled by the approximation property while the accompanying phase has unit modulus.

The imaginary parts enter only in the final change of variables that maps each cone onto the unit hypercube. Consider the cone integral Eq.~\eqref{eq:exampleint4} with complex parameters. Its exponential factor is $e^{-A_1 c_1 - A_2 c_2}$, where $A_1 = 2\delta - \nu_2$ and $A_2 = 3\delta - \nu_1 - \nu_2$ are now complex. The convergence condition guarantees $\operatorname{Re}A_i > 0$. If the final substitution were performed with the real parts alone, $c_i = -\log(y_i)/\operatorname{Re}A_i$, the pulled-back integrand would retain a residual phase $y_i^{\,i \operatorname{Im}A_i/\operatorname{Re}A_i}$, which oscillates with an ever-shorter period in $y_i$ as $y_i \to 0$. These oscillations are precisely the $\log$-oscillations of the original integrand accumulating when $x_i \to 0$ (and $x_i \to \infty$), and although this factor has unit modulus, its increasingly rapid oscillation near the endpoints can impair numerical integration. Instead, we perform the monomial substitution with the full complex exponents,
\begin{equation}
c_1 = -\frac{\log(y_1)}{A_1}\,,
\qquad
c_2 = -\frac{\log(y_2)}{A_2}\,,
\end{equation}
keeping $y_1, y_2$ real on $[0,1]$. Since $\operatorname{Re}A_i>0$ implies $|\!\arg A_i| < \pi/2$ and the integrand decays exponentially in the closed sector between the real ray and the rotated ray $e^{-i\arg A_i}\,\mathbb{R}_{\geq 0}$, this contour rotation leaves the value of the integral unchanged provided the polynomials $P_j$ remain non-vanishing throughout the deformation. We need to check this condition for each integral before applying the rotation. 

The resulting hypercube representation takes the same form as in the real case, with the (now complex) prefactor $1/(A_1 A_2)$. Its integrand is bounded and free of oscillations near $y_i = 0$. In terms of the original variables, the substitution deforms the integration cycle into the complex domain in the regions $x_i \to 0$ and $x_i \to \infty$, where the oscillations generated by the complex exponents accumulate. The deformation converts them into exponential decay. The kernel $g$ is then evaluated at complex arguments. 

\subsection{Regulating divergences in dimensional regularization}\label{sec:regdiv}

In this section, we describe how to evaluate divergent integrals using dimensional regularization in conjunction with the tropical decomposition algorithm. Many of the integrals we encounter exhibit UV and IR divergences that must be regulated. We employ dimensional regularization, expanding the loop integral as a series in the regulator $\varepsilon = 2 - D/2$ of Eq.~\eqref{eq:dimregd}, which enters through the exponents $\nu_i$ and $\delta_j$ in Eq.~\eqref{eq:genEulerint}. Our integrals are meromorphic functions in $\nu_i$ and $\delta_j$, as proven explicitly in Refs.~\cite{MR3010271,MR3189470}. Therefore, we can analytically continue the integral from the convergent regime in $\nu_i$, $\delta_j$ space to almost any other value of these parameters. However, it is nontrivial to perform this analytic continuation explicitly and obtain a representation of the integral that is amenable to numerical integration.

The strategy is the following. For each divergent integral, one performs a series of integration-by-parts steps that transform the integral into a linear combination of convergent integrals. The prefactors of the convergent integrals will contain $1/\varepsilon$ terms that completely capture the divergence as $\varepsilon \rightarrow 0$.

Again, we will illustrate the general procedure with an example.
We consider the integral 
\begin{equation}
f(\varepsilon)=\int_{0}^{1} \ y^{2\varepsilon-1} \frac{1}{1+y^{2}} \dd y\ .
\end{equation}
For $\operatorname{Re}\varepsilon>0$, this integral converges and develops a pole as $\varepsilon\to 0$. Our goal is to numerically compute $f(\varepsilon)$ as a series expansion in $\varepsilon$. In the $\varepsilon\to 0$ limit, the integrand behaves as $y^{2\varepsilon-1}$. This behavior is responsible for the divergence. We can rewrite the integral as follows:
\begin{equation}
f(\varepsilon)= 
\frac{1}{2\varepsilon} 
\int_{0}^{1} \ (\partial_y y^{2\varepsilon}) \frac{1}{1+y^{2}} \dd y \ \,,
\end{equation}
where we effectively extracted a $1/\varepsilon$. We can now continue via standard integration by parts.
\begin{align}
\begin{aligned}
f(\varepsilon)&= 
\frac{1}{2\varepsilon} 
\left(
\left[
y^{2\varepsilon} \frac{1}{1+y^{2}} \right]_0^1
-
\int_{0}^{1} \ y^{2\varepsilon} \left( \partial_y \frac{1}{1+y^{2}} \right) \dd y 
\right)
\ \,,
\\
&=
\frac{1}{2\varepsilon} 
\left(
\frac12
+
2
\int_{0}^{1} \ y^{2\varepsilon+1}  \frac{1}{(1+y^{2})^2}  \dd y 
\right) \, .
\end{aligned}
\end{align}
The remaining integral is finite for $\varepsilon\to 0$, and we can expand in $\varepsilon$ under the integral sign to get a Laurent expansion in $\varepsilon$. Each of the resulting expansion coefficients is amenable to the tropical decomposition workflow of Sec.~\ref{simgen} and can be computed numerically. 

Our general approach to the systematic regularization of divergent integrals proceeds similarly, via integration by parts. The precise algorithm is due to Nilsson and Passare~\cite{MR3010271}. We will explain the general algorithm using the running example of Eq.~\eqref{eq:exampleint}.

Suppose we wish to evaluate the integral $I(\delta,\nu_1,\nu_2)$ from Eq.~\eqref{eq:exampleint} with the parameters $\delta = 1-\frac12 \varepsilon$, $\nu_1 = \varepsilon$ and $\nu_2 = 1+\varepsilon$. That means
\begin{equation}
\label{eq:exampleinteps}
I(1-\frac12 \varepsilon,\varepsilon,1+\varepsilon)
=\int_{0}^{\infty}\int_{0}^{\infty} \frac{x_1^{\varepsilon} x_2^{1+\varepsilon} }{P(x_1,x_2)^{1-\frac12 \varepsilon} } \frac{\dd x_{1}}{x_1} \frac{\dd x_{2}}{x_2}\, ,
\end{equation}
where $P(x_1,x_2)$ is given in Eq.~\eqref{eq:examplepoly}.
The right-hand side integral is divergent as $\varepsilon \to 0$. 
This divergence occurs because the vector $(\varepsilon, 1+\varepsilon)$ touches 
the boundary of the scaled Newton polytope $(1-\frac12 \varepsilon) \cdot \mathcal{N}(P)$ as
$\varepsilon \to 0$. 
The divergence is associated with the facets of the scaled Newton polytope that are touched by the exponent vector $\nu=(\nu_1,\nu_2)^T$. A facet is a proper face of codimension one.
In the present case, the facet is cut out by the inequality $\nu_1 \ge 0$. 
The associated normal vector is $\mu = (1,0)^T$. Based on this observation, we introduce the rescaling 
$(x_1,x_2) \to (\lambda^{\mu_1} x_1, \lambda^{\mu_2} x_2) = (\lambda x_1, x_2)$ in the integral~\eqref{eq:exampleinteps}, where $\lambda$ is a new auxiliary parameter. We obtain
\begin{equation}
I(1-\frac12 \varepsilon,\varepsilon,1+\varepsilon)
= \lambda^{\varepsilon} \int_{0}^{\infty}\int_{0}^{\infty} \frac{x_1^{\varepsilon} x_2^{1+\varepsilon} }{P(\lambda x_1,x_2)^{1-\frac12 \varepsilon} } \frac{\dd x_{1}}{x_1} \frac{\dd x_{2}}{x_2}\, .
\end{equation}
At fixed $x_1$ and $x_2$, the polynomial satisfies $P(\lambda x_1,x_2)\to1+x_2^2$ as $\lambda\to0$. The prefactor $\lambda^\varepsilon$ isolates the scaling associated with this facet. We now take the derivative with respect to $\lambda$ on both sides.
\begin{align}
\begin{aligned}
0
&= \varepsilon \lambda^{\varepsilon-1} \int_{0}^{\infty}\int_{0}^{\infty} \frac{x_1^{\varepsilon} x_2^{1+\varepsilon} }{P(\lambda x_1,x_2)^{1-\frac12 \varepsilon} } \frac{\dd x_{1}}{x_1} \frac{\dd x_{2}}{x_2}
\\
& -(1-\frac12 \varepsilon)
 \lambda^{\varepsilon} \int_{0}^{\infty}\int_{0}^{\infty} \frac{x_1^{\varepsilon} x_2^{1+\varepsilon} \partial_{\lambda}\!\left[P(\lambda x_1,x_2)\right]  }{P(\lambda x_1,x_2)^{2-\frac12 \varepsilon} } \frac{\dd x_{1}}{x_1} \frac{\dd x_{2}}{x_2}\, .
\end{aligned}
\end{align}
Setting the auxiliary parameter $\lambda$ to $1$ gives an identity that involves the original integral.
\begin{gather}
0=
\varepsilon I(1-\frac12 \varepsilon,\varepsilon,1+\varepsilon)
 -(1-\frac12 \varepsilon)
 \int_{0}^{\infty}\int_{0}^{\infty} \frac{x_1^{\varepsilon} x_2^{1+\varepsilon} \left.\partial_{\lambda}\!\left[P(\lambda x_1,x_2)\right]  \right|_{\lambda=1} }{  P(x_1,x_2)^{2-\frac12 \varepsilon} } \frac{\dd x_{1}}{x_1} \frac{\dd x_{2}}{x_2}\, .
\intertext{Using $\left.\partial_{\lambda}\!\left[P(\lambda x_1,x_2)\right] \right|_{\lambda=1}= 
4 x_1^2 + x_1 x_2^2 + 6 x_1^2 x_2$ gives }
I(1-\frac12 \varepsilon,\varepsilon,1+\varepsilon) 
=
\frac{1-\frac12 \varepsilon}{\varepsilon}
\int_{0}^{\infty}\int_{0}^{\infty} \frac{x_1^{\varepsilon} x_2^{1+\varepsilon} 
\left(4 x_1^2 + x_1 x_2^2 + 6 x_1^2 x_2 \right)
}{P(x_1,x_2)^{2-\frac12 \varepsilon} } \frac{\dd x_{1}}{x_1} \frac{\dd x_{2}}{x_2}\,.
\end{gather}
The integral on the right-hand side is convergent as $\varepsilon \to 0$. The divergence is extracted into the prefactor. The right-hand side integral can be expanded in $\varepsilon$ under the integral sign and is ready to be treated numerically using the workflow of Sec.~\ref{simgen}.

A slightly more complicated example is
\begin{equation}
\label{eq:exampleinteps2}
I(2-\frac12 \varepsilon,3+\varepsilon,3-2\varepsilon)
=\int_{0}^{\infty}\int_{0}^{\infty} \frac{x_1^{3+\varepsilon} x_2^{3-2\varepsilon} }{P(x_1,x_2)^{2-\frac12 \varepsilon} } \frac{\dd x_{1}}{x_1} \frac{\dd x_{2}}{x_2}\, .
\end{equation}
Here, the divergence comes from the exponent vector $(3+\varepsilon, 3-2\varepsilon)$ touching 
the $\nu_1+\nu_2 = 3\delta$ facet as $\varepsilon \rightarrow 0$. To regulate this divergence,  
we rescale along the inward normal of this facet,
$\mu = (-1,-1)^T$. Therefore, rescaling $(x_1,x_2) \rightarrow (\lambda^{-1}x_1,\lambda^{-1} x_2)$ gives
\begin{equation}
I(2-\frac12 \varepsilon,3+\varepsilon,3-2\varepsilon)
= \lambda^{-\frac12\varepsilon} \int_{0}^{\infty}\int_{0}^{\infty} \frac{x_1^{3+\varepsilon} x_2^{3-2\varepsilon} }{ (\lambda^{3} P(\lambda^{-1} x_1,\lambda^{-1} x_2))^{2-\frac12 \varepsilon} } \frac{\dd x_{1}}{x_1} \frac{\dd x_{2}}{x_2}\, .
\end{equation}
Here, we included an auxiliary factor in the denominator such that $\lambda^{6-\frac32 \varepsilon} P(\lambda^{-1} x_1,\lambda^{-1} x_2)^{2-\frac12 \varepsilon}$ is of order $1$ when $\lambda \to 0$.
Taking the $\lambda$ derivative on both sides and then setting $\lambda=1$ yields the identity
\begin{equation}
I(2-\frac12 \varepsilon,3+\varepsilon,3-2\varepsilon)
= -\frac{2(2-\frac12\varepsilon)}{\varepsilon} \int_{0}^{\infty}\int_{0}^{\infty} \frac{x_1^{3+\varepsilon} x_2^{3-2\varepsilon} 
\left( 3+ 2 x_1^2 + x_2^2 \right)
}{  P(x_1,x_2)^{3-\frac12 \varepsilon} } \frac{\dd x_{1}}{x_1} \frac{\dd x_{2}}{x_2}\, ,
\end{equation}
where the form of the numerator follows from $\partial_\lambda \left (\lambda^{3} P(\lambda^{-1} x_1,\lambda^{-1} x_2) \right)|_{\lambda=1} = 3+ 2 x_1^2 + x_2^2$.
Once again, the divergence is resolved, so we can expand in $\varepsilon$ under the integral sign and treat the integral using the numerical workflow of Sec.~\ref{simgen}.

Our implementation of this \emph{directed integration-by-parts} algorithm iterates this process in cases where there are multiple divergences. These divergences might ``overlap'', meaning that the associated exponent vector touches multiple facets. Furthermore, our algorithm generalizes to multiple polynomials in the denominator. To achieve this generalization, all facets of the weighted \emph{Minkowski sum} of the Newton polytopes of the denominator polynomials need to be taken into account. In particular, the integral~\eqref{eq:genEulerint} converges if all $P_j$ have positive coefficients and $\nu$ lies in the interior of $\sum_j \delta_j \mathcal{N}(P_j)$. The generalization of the one-denominator-polynomial algorithm from~\cite{MR3010271} to the multiple-denominator-polynomial case is explained in more detail in Ref.~\cite{MR3189470}. 

Another effective way to handle divergences is the \emph{subtraction} method developed in Refs.~\cite{Arkani-Hamed:2022cqe,Hillman:2023ezp,Salvatori:2024nva,Giroux:2026tgd}. As explained in Ref.~\cite{Salvatori:2024nva}, subtraction behaves well in combination with tropical methods. We also implemented a pipeline that uses a subtraction approach for the evaluation of the generalized Euler--Mellin integrals needed in this work. Both implementations gave consistent results, but the subtraction method combined with tropical sampling resulted in poorer numerical stability than the directed integration-by-parts method, which we attribute to large cancellations between subtraction terms.

\subsection{Computing the tropical fan}\label{sec:tropicalfancompute}

In this section, we review the algorithm for constructing the tropical fan of a generic polynomial with an arbitrary number of variables. 

\begin{figure}[thbp]
    \centering
    \includegraphics[width=0.45\linewidth]{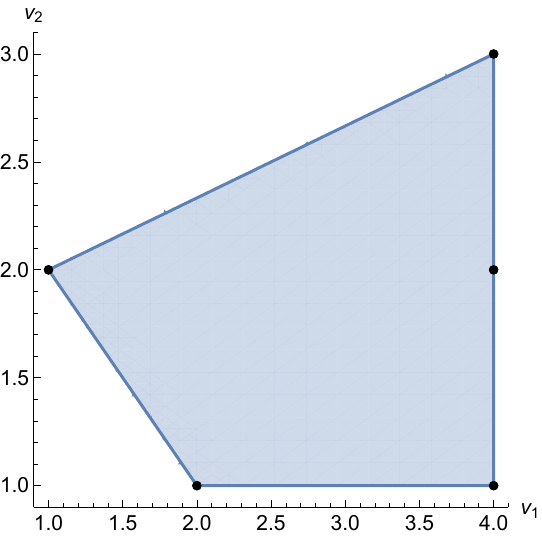}
    \caption{The Newton polytope (blue region) of the 2D polynomial in Eq.~\eqref{eq:appBeg}, defined as the convex hull of the exponent vectors $\vec{v}^{(1)},\ldots,\vec{v}^{(6)}$. Nonvertex monomials, whose exponent vectors lie on an edge or in the interior of the polytope, never uniquely dominates the vertex monomials.}
    \label{fig:convexHull}
\end{figure}

Given a polynomial $P = \sum_i c_i \prod_j x_j^{v_{j}^{(i)}}$ in the variables $x_1, \ldots, x_n$, we first construct the Newton polytope, which is the convex hull of the vertices (i.e.~exponent vectors) $\vec{v}^{(i)} = ( v_1^{(i)}, \ldots, v_n^{(i)} )^T \in \mathbb{R}^n$.
\begin{equation}
\mathcal{N}(P) \;=\; \mathrm{Conv}\!\bigl\{\vec{v}^{\,(i)}\bigr\} \;\subset\; \mathbb{R}^n\,.
\end{equation}
For example, for the 2D polynomial
\begin{equation}
P = x_1 x_2^2 + x_1 x_2^3 + x_1^4 x_2^3 + x_1^4 x_2^2 + x_1^4 x_2 + x_1^2 x_2,
\end{equation}
its exponent vectors are
\begin{equation}
\begin{split}
    &\vec{v}^{(1)} = (1, 2)^T,\quad \vec{v}^{(2)} = (1, 3)^T,\quad \vec{v}^{(3)} = (4, 3)^T\, ,\\
    &\vec{v}^{(4)} = (4, 2)^T,\quad \vec{v}^{(5)} = (4, 1)^T,\quad \vec{v}^{(6)} = (2, 1)^T\, ,
\end{split}\label{eq:appBeg}
\end{equation}
and its Newton polytope is the blue region in Fig.~\ref{fig:convexHull}. The reason we care about the convex hull of the vertices is the following. Notice that in our example, the existence of the monomial $x_1^4 x_2^2$ has no effect on the shape of the Newton polytope because it lies on the line formed by the vertices $x_1^4 x_2^3$ and $x_1^4 x_2$. Since every integration variable satisfies $x_i\in (0,\infty)$, there is no regime where $x_1^4 x_2^2$ strictly dominates both $x_1^4 x_2^3$, whose power of $x_2$ is higher, and $x_1^4 x_2$, whose power of $x_2$ is lower. By the same logic, any monomial whose exponent vector lies strictly inside the Newton polytope would never dominate the existing vertices. Hence, the convex hull encodes all information about the possible dominant monomial behaviors.

In fact, this argument holds in full generality. If an exponent vector $\vec{v}^{\,(k)}$ is not a vertex of $\mathcal{N}(P)$ but lies inside the convex hull, it can always be written as a linear combination of the vertex vectors. Specifically, $\vec{v}^{\,(k)} = \sum_j \lambda_j \vec{v}^{\,(j)}$ with $\lambda_j \geq 0$ and $\sum_j \lambda_j = 1$. Passing to $z_\ell = \log x_\ell$ and using the fact that a weighted average is always bounded by the largest term,
\begin{equation}
\vec{v}^{\,(k)} \cdot \vec{z}
\;=\; \sum_j \lambda_j \bigl(\vec{v}^{\,(j)} \cdot \vec{z}\bigr)
\;\leq\; \max_j \bigl(\vec{v}^{\,(j)} \cdot \vec{z}\bigr)\,,
\end{equation}
which holds for any $\vec{z}\in\mathbb{R}^n$. So the monomial corresponding to $\vec{v}^{\,(k)}$ can never be the uniquely dominant term. There is always at least one vertex monomial that is at least as large.

As discussed in Sec.~\ref{simgen}, the tropical fan lives in $z_i =\log(x_i)$ space, where comparison of monomials such as $x_1^{v_{1}^{(i)}}x_2^{v_{2}^{(i)}}$ in the 2D example becomes a comparison of linear factors $v_{1}^{(i)}z_1 + v_{2}^{(i)}z_2$. In each maximal cone of the tropical fan, one monomial dominates. The boundaries between cones are exactly where some linear factors are equal and greater than all remaining factors. Said differently, the tropical fan is a complete polyhedral fan in $\vec{z}$-space. It tiles all of $\mathbb{R}^n$ into cones, one per vertex of $\mathcal{N}(P)$, according to which linear factor $\vec{v}^{\,(i)}\cdot\vec{z}$ is largest. The maximal cones are the regions
\begin{equation}
\sigma_i \;=\; \bigl\{\vec{z}\in\mathbb{R}^n : \vec{v}^{\,(i)}\cdot\vec{z} \;\geq\; \vec{v}^{\,(k)}\cdot\vec{z}\;\;\forall\, k\bigr\}\,,
\end{equation}
and the boundaries between them are where the maximum is achieved by more than one linear factor simultaneously.

\begin{figure}[thbp]
\centering
\includegraphics[width=0.5\linewidth]{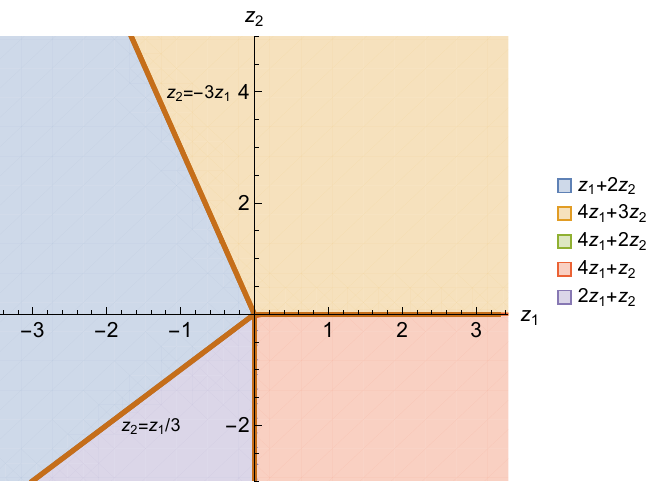}
\includegraphics[width=0.4\linewidth]{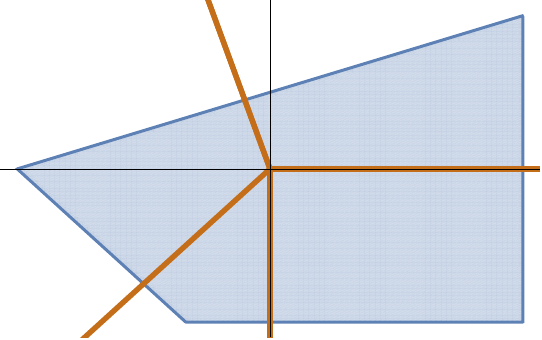}
\caption{The left panel shows the tropical fan of the polynomial in Eq.~\eqref{eq:appBeg}, constructed by solving the equalities and inequalities in Eq.~\eqref{eq:appBeqineq}. Brown lines denote the fan boundaries, and the colored regions indicate which vertex monomial $\vec{v}^{(i)}\cdot\vec{z}$ is largest. The right panel shows the tropical fan boundaries superimposed on the Newton polytope, illustrating the one-to-one correspondence between maximal cones of the fan and vertices of the polytope.}
\label{fig:tropequation}
\end{figure}

Let us restrict to the 2D example again. One obtains a particular boundary between two cones of the tropical fan by solving the equalities and inequalities
\begin{equation}
\vec{z} \in \mathbb{R}^2: \vec{v}^{(i)} \cdot \vec{z} = \vec{v}^{(i')} \cdot \vec{z} \geq \vec{v}^{(k)} \cdot \vec{z},\;\forall \vec{v}^{(i)}, \vec{v}^{(i')} \in S, \; \forall k\label{eq:appBeqineq}
\end{equation}
where $S$ is the subset of vertices that share the boundary\footnote{A nonzero solution exists only when all $\vec{v}^{(i)}\in S$ lie on a common proper face of the Newton polytope. (A face of a polytope is any lower-dimensional flat boundary component. In 2D the faces are edges and vertices. In 3D they are facets, edges, and vertices, and so on in higher dimensions.) A face of dimension $r$ gives a cone of codimension $r$ in the normal fan. Thus an edge gives a codimension-1 boundary, while a two-dimensional face gives a codimension-2 boundary. If the vertices in $S$ do not share a common face, only the trivial solution $\vec{z}=0$ remains.}\textsuperscript{,}\footnote{This construction is invariant under an overall shift of the Newton polytope because Eq.~\eqref{eq:appBeqineq} depends only on differences of exponent vectors.}. The boundaries of the tropical fan of~\eqref{eq:appBeg} constructed this way are shown as brown lines in Fig.~\ref{fig:tropequation}. In Fig.~\ref{fig:tropequation}, we also superimpose a region plot of which linear form attains the maximum
\begin{equation}
    \max\{\vec{v}^{(1)}\cdot\vec{z},\vec{v}^{(2)}\cdot\vec{z},\ldots, \vec{v}^{(6)}\cdot\vec{z}\} = \vec{v}^{(i)}\cdot\vec{z},
\end{equation}
which agrees completely with our boundary construction, as expected.

\begin{figure}[thbp]
    \centering
    \includegraphics[width=0.4\linewidth]{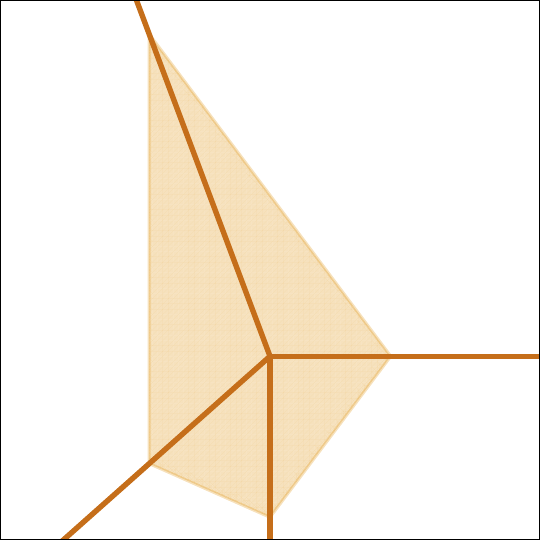}
    \caption{The dual polytope (polar dual) of the shifted Newton polytope, constructed from the hyperplane inequalities in Eq.~\eqref{eq:appBdual}, superimposed with the tropical fan boundaries derived from Eq.~\eqref{eq:appBeqineq}. The facets of the dual polytope correspond to the maximal cones of the tropical fan, confirming that the tropical fan is equivalently the normal fan of the Newton polytope.}
    \label{fig:dualpolytope}
\end{figure}

While the construction based on equalities and inequalities is straightforward to understand since it is the definition of tropical fans, there is another, more geometrical construction. The existence of a geometrical construction can be understood from the observation that there is an obvious one-to-one correspondence between the maximal cones of the tropical fan and the vertices of the Newton polytope. Each cone is defined to be where the monomial corresponding to the vertex dominates. The correspondence can also be seen graphically in our 2D example if we superimpose the tropical fan boundary (that lives in $\vec{z}$ space) with the Newton polytope (that lives in $\vec{v}$ space), making sure that the origin of the tropical fan lies strictly inside the Newton polytope, as shown on the right of Fig.~\ref{fig:tropequation}. The geometrical construction of the tropical fan uses the polar dual Newton polytope: while the Newton polytope itself lives in $\vec v$ space,  the polar dual Newton polytope lives in $\vec z$ space, and is defined by the intersection of the half-spaces 
\begin{equation}
\tilde{\mathcal{N}}^{\circ} = \left\{ \vec{z} \in \mathbb{R}^n : 1 - \tilde{\vec{v}}^{(i)} \cdot \vec{z} \geq 0 \text{ for all } i \right\},\label{eq:appBdual}
\end{equation}
where the $\tilde{\vec{v}}^{(i)}$ are the vertex vectors after an overall shift to ensure the Newton polytope contains the origin in its interior. Again, an overall shift of the Newton polytope has no effect on the relative dominance of the monomials and the structure of the tropical fan, since only the differences $\vec{v}^{\,(i)} - \vec{v}^{\,(k)}$ matter for determining which monomial wins. The containment of the origin is needed to ensure the dual polytope is finite and bounded. If the origin lay on a face of the shifted Newton polytope, the dual would extend to infinity in the direction perpendicular to that face. The form of the hyperplanes that define the dual polytope is familiar from our earlier construction based on equalities and inequalities. The intersection of a pair, a triplet, and so on of the hyperplanes sets the corresponding linear factors equal, and the conical hull of this intersection is the codimension-1, codimension-2, and so on boundary of the tropical fan. But instead of having to solve for the different codimension boundaries with different subsets of vertex vectors and their corresponding equalities-inequalities equations, the hyperplane equations~\eqref{eq:appBdual} contain all of them simultaneously. The conical hull of each facet of the dual polytope is then a maximal cone of the tropical fan. Here, a facet is an $(n{-}1)$-dimensional face of an $n$-dimensional polytope. More generally, the correspondence reverses dimensions. Higher-dimensional faces of $\tilde{\mathcal{N}}^\circ$ map to lower-dimensional faces of the Newton polytope and vice versa. Specifically, $k$-dimensional faces of $\tilde{\mathcal{N}}^\circ$ are in one-to-one correspondence with $(n{-}k{-}1)$-dimensional faces of the shifted Newton polytope $\tilde{\mathcal{N}}$. This is the standard duality between a convex polytope containing the origin and its polar dual. In Fig.~\ref{fig:dualpolytope}, we superimpose the dual polytope constructed from~\eqref{eq:appBdual} and the boundary of the tropical fan derived from Eq.~\eqref{eq:appBeqineq}, and they agree as expected.

\section{Validation against an analytic four-point bubble result}\label{sec:appC}
\label{sec:analyticcomparison}

We validate the reduced Schwinger method against the analytic result for the
one-loop four-point seed integral $\mathcal{J}^{00}$ derived in~\cite{Xianyu:2022jwk} using the de Sitter K\"all\'en--Lehmann representation~\cite{DiPietro:2021sjt,Hogervorst:2021uvp,Loparco:2023rug}. Defining $\vec{k}_s\equiv\vec{k}_1+\vec{k}_2=-(\vec{k}_3+\vec{k}_4)$ and $k_s\equiv|\vec{k}_s|$, and denoting by $k_{12}$ and $k_{34}$ the energy sums entering the two vertices, we consider
\begin{equation}
\begin{aligned}
d &= 3-2\varepsilon,
& \widetilde{\nu} &= 1, \\
k_{12} &= k_1+k_2,
& k_{34} &= k_3+k_4, \\
r_1 &= \frac{k_s}{k_{12}},
& r_2 &= \frac{k_s}{k_{34}}=0.9 .
\end{aligned}
\end{equation}
We vary
\begin{equation}
x\equiv-\log_{10}r_1
\in
\{0.5,1.0,1.5,2.0,2.5,3.0,3.5,4.0\}
\end{equation}
and compare the quantity
\begin{equation}
Y(r_1,r_2)
\equiv
(r_1r_2)^{-4}\operatorname{Re}\mathcal{J}^{00}.
\end{equation}

In the region $0<r_1<r_2<1$,~\cite{Xianyu:2022jwk} expresses
\begin{equation}
\widehat{\mathcal{J}}
\equiv
(r_1r_2)^{-5/2}\mathcal{J}^{00}
\end{equation}
as the sum of the nonlocal signal, local signal, and analytic background
contributions,
\begin{equation}
\widehat{\mathcal{J}}
=
\widehat{\mathcal{J}}_{\mathrm{NS}}
+
\widehat{\mathcal{J}}_{\mathrm{LS}}
+
\widehat{\mathcal{J}}_{\mathrm{BG}}.
\end{equation}
We evaluate these sums with the signal terms truncated at $n\leq 40$ and
the background terms at $\ell_{\max}=20$ and $m_{\max}=6$, using the
Richardson-extrapolated spectral table for the latter. We set the
renormalization scale to $\mu_R/H=\widetilde{\nu}=1$.

The loop measure is $d^{d}\vec{\ell}/(2\pi)^{d}$ in both calculations. The full factor $(2\pi)^{-d}$ is continued in $d$. Our numerical result drops the pole of the continued
integral and then adds $P_{Y}\left[\gamma_{E}-\log 4\pi\right]$, which completes the
$\overline{\rm MS}$ subtraction, where $P_{\mathcal J}$ denotes the coefficient of the simple
$1/\varepsilon$ pole in $\mathcal{J}^{00}$ and $P_{Y}=(r_{1}r_{2})^{-4}P_{\mathcal J}$.  But as discussed at~\eqref{wittendiagram4} and~\eqref{eq:ctremainder}, an analytic shift is required to convert the basis integral (with the pole dropped) computed in our numerical calculation to a renormalized correlation function as computed in Ref.~\cite{Xianyu:2022jwk}. The required analytic conversion is 
\begin{equation}
\mathcal{J}^{00}_{\mathrm{mixed}}
=
\mathcal{J}^{00}_{\mathrm{paper}}
+
P_{\mathcal J}
\left[
2\log\kappa_{\mathrm{tot}}+2\log\frac{k_{s}}{\mu_{R}}-2\psi(5)
\right],
\qquad
\kappa_{\mathrm{tot}}
=
\frac{k_{12}+k_{34}}{k_s},
\label{eq:analytic-scheme-conversion}
\end{equation}
where $\psi$ is the digamma function. In the comparison below we set $k_{s}=\mu_{R}=H$, so the middle term vanishes. All analytic values quoted below include this conversion.

For the same kinematics, the reduced Schwinger representation is a
five-dimensional integral decomposed into nine regions, with
\begin{equation}
\alpha=\frac{3}{2}-\varepsilon+2i,
\qquad
N=\frac{13}{2}-3\varepsilon+2i,
\qquad
N-\alpha=5-2\varepsilon.
\end{equation}
The remaining four-dimensional integral is evaluated using
double-exponential tensor quadrature. We quote the final-rung result and
use the change from the preceding rung, denoted
$\delta_{\mathrm{rung}}$, as a numerical convergence diagnostic.
The table and figure were generated from the same comparison run, in the convention described above. The analytic column is the renormalized series of~\cite{Xianyu:2022jwk} converted by Eq.~\eqref{eq:analytic-scheme-conversion}.

\begin{table}[htbp]
\centering
\begin{tabular}{c r r}
\hline
$x=-\log_{10}r_1$
& Analytic (mixed $\overline{\rm MS}$ scheme)
& Reduced Schwinger $\pm\delta_{\mathrm{rung}}$ \\
\hline
$0.5$ & $+0.009528486$ & $+0.009529284 \pm 6.24\times10^{-9}$ \\
$1.0$ & $+0.012329972$ & $+0.012329972 \pm 2.10\times10^{-8}$ \\
$1.5$ & $+0.012980731$ & $+0.012980730 \pm 7.27\times10^{-8}$ \\
$2.0$ & $+0.027529645$ & $+0.027529623 \pm 3.80\times10^{-6}$ \\
$2.5$ & $-0.007289204$ & $-0.007289870 \pm 8.30\times10^{-6}$ \\
$3.0$ & $+0.004211492$ & $+0.004199399 \pm 2.16\times10^{-4}$ \\
$3.5$ & $+0.011308339$ & $+0.011304186 \pm 9.92\times10^{-4}$ \\
$4.0$ & $-0.015441295$ & $-0.015139988 \pm 1.25\times10^{-3}$ \\
\hline
\end{tabular}
\caption{Comparison of $Y$ obtained from the scheme-converted analytic
series of~\cite{Xianyu:2022jwk} and from the reduced Schwinger
integral. The quoted $\delta_{\mathrm{rung}}$ is the change between the
final two quadrature refinements.}
\label{tab:analyticcomparison}
\end{table}

\begin{figure}[t]
\centering
\includegraphics[width=0.8\textwidth]{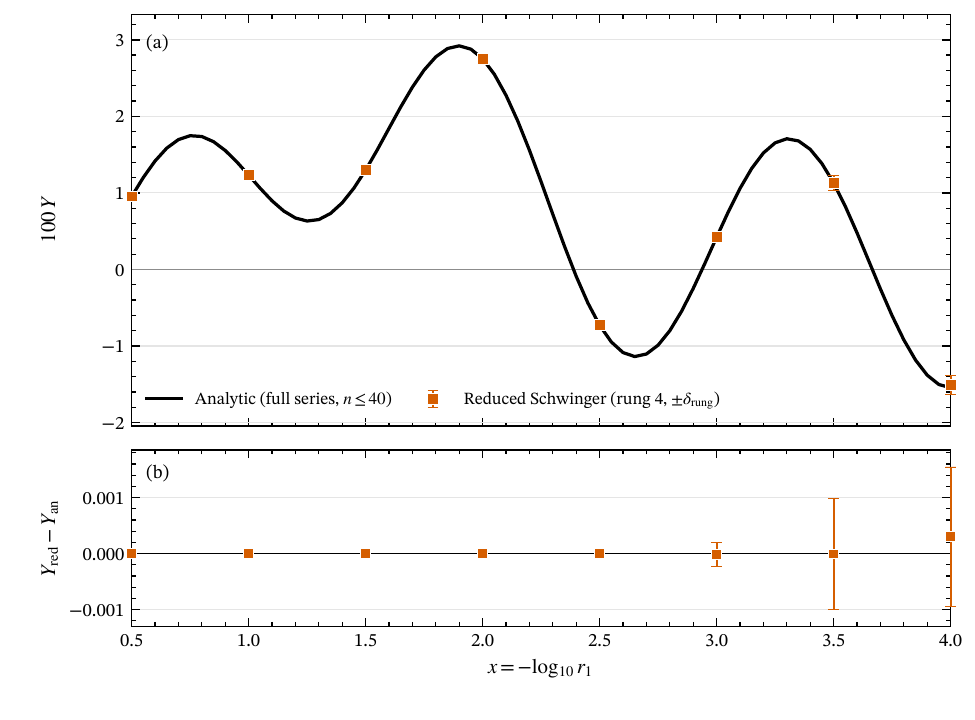}
\caption{Validation of the reduced Schwinger method against the analytic
bubble result. The upper panel shows $100Y$ as a function of
$x=-\log_{10}r_1$, with the scheme-converted analytic series shown as a
solid curve and the reduced Schwinger results shown as squares. The lower
panel shows the difference between the reduced Schwinger and analytic
results. The error bars indicate the movement between the final two
quadrature rungs.}
\label{fig:analytic_vs_reduced}
\end{figure}


\bibliographystyle{apsrev4-1long}
\bibliography{GeneralBibliography.bib}

@article{Chen:2022vzh,
    author = "Chen, Xingang and Ebadi, Reza and Kumar, Soubhik",
    title = "{Classical cosmological collider physics and primordial features}",
    eprint = "2205.01107",
    archivePrefix = "arXiv",
    primaryClass = "hep-ph",
    doi = "10.1088/1475-7516/2022/08/083",
    journal = "JCAP",
    volume = "08",
    pages = "083",
    year = "2022"
}

@article{XianyuQinTriangle,
author = "Qin, Zhehan and Xianyu, Zhong-Zhi",
title = "Cosmological Collider Signals From a Triangle Loop",
eprint = "appearing simultaneously"
}

@article{Xianyu:2025lbk,
    author = "Xianyu, Zhong-Zhi and Zang, Jiaju",
    title = "{Massive inflationary amplitudes: new representations and degenerate limits}",
    eprint = "2511.08677",
    archivePrefix = "arXiv",
    primaryClass = "hep-th",
    doi = "10.1007/JHEP03(2026)122",
    journal = "JHEP",
    volume = "03",
    pages = "122",
    year = "2026"
}

@article{Liu:2024str,
    author = "Liu, Haoyuan and Xianyu, Zhong-Zhi",
    title = "{Massive inflationary amplitudes: differential equations and complete solutions for general trees}",
    eprint = "2412.07843",
    archivePrefix = "arXiv",
    primaryClass = "hep-th",
    reportNumber = "USTC-ICTS/PCFT-24-56",
    doi = "10.1007/JHEP09(2025)183",
    journal = "JHEP",
    volume = "09",
    pages = "183",
    year = "2025"
}

@article{You:2026xoq,
    author = "You, Jingtao and Song, Linghao and Han, Chengcheng and He, Hong-Jian and Chen, Xingang and Xianyu, Zhong-Zhi",
    title = "{Cosmological Collider Signatures from Right-Handed Neutrino Loop}",
    eprint = "2605.21419",
    archivePrefix = "arXiv",
    primaryClass = "hep-ph",
    month = "5",
    year = "2026"
}

@article{Qin:2023bjk,
    author = "Qin, Zhehan and Xianyu, Zhong-Zhi",
    title = "{Inflation correlators at the one-loop order: nonanalyticity, factorization, cutting rule, and OPE}",
    eprint = "2304.13295",
    archivePrefix = "arXiv",
    primaryClass = "hep-th",
    doi = "10.1007/JHEP09(2023)116",
    journal = "JHEP",
    volume = "09",
    pages = "116",
    year = "2023"
}

@article{An:2017hlx,
    author = "An, Haipeng and McAneny, Michael and Ridgway, Alexander K. and Wise, Mark B.",
    title = "{Quasi Single Field Inflation in the non-perturbative regime}",
    eprint = "1706.09971",
    archivePrefix = "arXiv",
    primaryClass = "hep-ph",
    doi = "10.1007/JHEP06(2018)105",
    journal = "JHEP",
    volume = "06",
    pages = "105",
    year = "2018"
}

@article{Xianyu:2023ytd,
    author = "Xianyu, Zhong-Zhi and Zang, Jiaju",
    title = "{Inflation correlators with multiple massive exchanges}",
    eprint = "2309.10849",
    archivePrefix = "arXiv",
    primaryClass = "hep-th",
    doi = "10.1007/JHEP03(2024)070",
    journal = "JHEP",
    volume = "03",
    pages = "070",
    year = "2024"
}

@article{Aoki:2024uyi,
    author = "Aoki, Shuntaro and Pinol, Lucas and Sano, Fumiya and Yamaguchi, Masahide and Zhu, Yuhang",
    title = "{Cosmological correlators with double massive exchanges: bootstrap equation and phenomenology}",
    eprint = "2404.09547",
    archivePrefix = "arXiv",
    primaryClass = "hep-th",
    doi = "10.1007/JHEP09(2024)176",
    journal = "JHEP",
    volume = "09",
    pages = "176",
    year = "2024"
}

@article{Aoki:2023wdc,
    author = "Aoki, Shuntaro and Noumi, Toshifumi and Sano, Fumiya and Yamaguchi, Masahide",
    title = "{Analytic formulae for inflationary correlators with dynamical mass}",
    eprint = "2312.09642",
    archivePrefix = "arXiv",
    primaryClass = "hep-th",
    reportNumber = "CTPU-PTC-23-47, UT-Komaba/23-13",
    doi = "10.1007/JHEP03(2024)073",
    journal = "JHEP",
    volume = "03",
    pages = "073",
    year = "2024"
}

@article{Jazayeri:2022kjy,
    author = "Jazayeri, Sadra and Renaux-Petel, S{\'e}bastien",
    title = "{Cosmological bootstrap in slow motion}",
    eprint = "2205.10340",
    archivePrefix = "arXiv",
    primaryClass = "hep-th",
    doi = "10.1007/JHEP12(2022)137",
    journal = "JHEP",
    volume = "12",
    pages = "137",
    year = "2022"
}

@article{Sohn:2023fte,
    author = "Sohn, Wuhyun and Fergusson, James R. and Shellard, E. P. S.",
    title = "{High-resolution CMB bispectrum estimator with flexible modal bases}",
    eprint = "2305.14646",
    archivePrefix = "arXiv",
    primaryClass = "astro-ph.CO",
    doi = "10.1103/PhysRevD.108.063504",
    journal = "Phys. Rev. D",
    volume = "108",
    number = "6",
    pages = "063504",
    year = "2023"
}

@article{Werth:2024aui,
    author = "Werth, Denis and Pinol, Lucas and Renaux-Petel, S{\'e}bastien",
    title = "{CosmoFlow: Python Package for Cosmological Correlators}",
    eprint = "2402.03693",
    archivePrefix = "arXiv",
    primaryClass = "astro-ph.CO",
    doi = "10.1088/1361-6382/ad6740",
    journal = "Class. Quant. Grav.",
    volume = "41",
    number = "17",
    pages = "175015",
    year = "2024"
}

@article{Pinol:2023oux,
    author = "Pinol, Lucas and Renaux-Petel, S{\'e}bastien and Werth, Denis",
    title = "{The cosmological flow: a systematic approach to primordial correlators}",
    eprint = "2312.06559",
    archivePrefix = "arXiv",
    primaryClass = "astro-ph.CO",
    doi = "10.1088/1475-7516/2025/02/019",
    journal = "JCAP",
    volume = "02",
    pages = "019",
    year = "2025"
}

@article{Werth:2023pfl,
    author = "Werth, Denis and Pinol, Lucas and Renaux-Petel, S{\'e}bastien",
    title = "{Cosmological Flow of Primordial Correlators}",
    eprint = "2302.00655",
    archivePrefix = "arXiv",
    primaryClass = "hep-th",
    doi = "10.1103/PhysRevLett.133.141002",
    journal = "Phys. Rev. Lett.",
    volume = "133",
    number = "14",
    pages = "141002",
    year = "2024"
}

@article{Senatore:2009gt,
    author = "Senatore, Leonardo and Smith, Kendrick M. and Zaldarriaga, Matias",
    title = "{Non-Gaussianities in Single Field Inflation and their Optimal Limits from the WMAP 5-year Data}",
    eprint = "0905.3746",
    archivePrefix = "arXiv",
    primaryClass = "astro-ph.CO",
    doi = "10.1088/1475-7516/2010/01/028",
    journal = "JCAP",
    volume = "01",
    pages = "028",
    year = "2010"
}

@article{Babich:2004gb,
    author = "Babich, Daniel and Creminelli, Paolo and Zaldarriaga, Matias",
    title = "{The Shape of non-Gaussianities}",
    eprint = "astro-ph/0405356",
    archivePrefix = "arXiv",
    reportNumber = "HUTP-04-A022",
    doi = "10.1088/1475-7516/2004/08/009",
    journal = "JCAP",
    volume = "08",
    pages = "009",
    year = "2004"
}

@article{Fergusson:2008ra,
    author = "Fergusson, J. R. and Shellard, E. P. S.",
    title = "{The shape of primordial non-Gaussianity and the CMB bispectrum}",
    eprint = "0812.3413",
    archivePrefix = "arXiv",
    primaryClass = "astro-ph",
    doi = "10.1103/PhysRevD.80.043510",
    journal = "Phys. Rev. D",
    volume = "80",
    pages = "043510",
    year = "2009"
}

@article{Alvarez:2014vva,
    author = "Alvarez, Marcelo and others",
    title = "{Testing Inflation with Large Scale Structure: Connecting Hopes with Reality}",
    eprint = "1412.4671",
    archivePrefix = "arXiv",
    primaryClass = "astro-ph.CO",
    month = "12",
    year = "2014"
}

@article{Meerburg:2019qqi,
    author = "Meerburg, P. Daniel and others",
    title = "{Primordial Non-Gaussianity}",
    eprint = "1903.04409",
    archivePrefix = "arXiv",
    primaryClass = "astro-ph.CO",
    reportNumber = "FERMILAB-PUB-19-140-A",
    journal = "Bull. Am. Astron. Soc.",
    volume = "51",
    number = "3",
    pages = "107",
    year = "2019"
}

@article{Schlegel:2019eqc,
    author = "Schlegel, David J. and others",
    title = "{Astro2020 APC White Paper: The MegaMapper: a z {\ensuremath{>}} 2 Spectroscopic Instrument for the Study of Inflation and Dark Energy}",
    eprint = "1907.11171",
    archivePrefix = "arXiv",
    primaryClass = "astro-ph.IM",
    reportNumber = "FERMILAB-FN-1081-AE-SCD",
    journal = "Bull. Am. Astron. Soc.",
    volume = "51",
    number = "7",
    pages = "229",
    year = "2019"
}

@article{Munoz:2015eqa,
    author = {Mu{\~n}oz, Julian B. and Ali-Ha{\"\i}moud, Yacine and Kamionkowski, Marc},
    title = "{Primordial non-gaussianity from the bispectrum of 21-cm fluctuations in the dark ages}",
    eprint = "1506.04152",
    archivePrefix = "arXiv",
    primaryClass = "astro-ph.CO",
    doi = "10.1103/PhysRevD.92.083508",
    journal = "Phys. Rev. D",
    volume = "92",
    number = "8",
    pages = "083508",
    year = "2015"
}

@article{Arkani-Hamed:2023bsv,
    author = "Arkani-Hamed, Nima and Baumann, Daniel and Hillman, Aaron and Joyce, Austin and Lee, Hayden and Pimentel, Guilherme L.",
    title = "{Kinematic Flow and the Emergence of Time}",
    eprint = "2312.05300",
    archivePrefix = "arXiv",
    primaryClass = "hep-th",
    doi = "10.1103/dsjm-tckw",
    journal = "Phys. Rev. Lett.",
    volume = "135",
    number = "3",
    pages = "031602",
    year = "2025"
}

@article{Hillman:2023ezp,
    author = "Hillman, Aaron",
    title = "{A Subtraction Scheme for Feynman Integrals}",
    eprint = "2311.03439",
    archivePrefix = "arXiv",
    primaryClass = "hep-th",
    month = "11",
    year = "2023"
}

@article{Guth:1980zm,
    author = "Guth, Alan H.",
    editor = "Fang, Li-Zhi and Ruffini, R.",
    title = "{The Inflationary Universe: A Possible Solution to the Horizon and Flatness Problems}",
    reportNumber = "SLAC-PUB-2576",
    doi = "10.1103/PhysRevD.23.347",
    journal = "Phys. Rev. D",
    volume = "23",
    pages = "347--356",
    year = "1981"
}

@article{Linde:1981mu,
    author = "Linde, Andrei D.",
    editor = "Fang, Li-Zhi and Ruffini, R.",
    title = "{A New Inflationary Universe Scenario: A Possible Solution of the Horizon, Flatness, Homogeneity, Isotropy and Primordial Monopole Problems}",
    reportNumber = "LEBEDEV-81-229",
    doi = "10.1016/0370-2693(82)91219-9",
    journal = "Phys. Lett. B",
    volume = "108",
    pages = "389--393",
    year = "1982"
}

@article{Albrecht:1982wi,
    author = "Albrecht, Andreas and Steinhardt, Paul J.",
    editor = "Fang, Li-Zhi and Ruffini, R.",
    title = "{Cosmology for Grand Unified Theories with Radiatively Induced Symmetry Breaking}",
    reportNumber = "UPR-0185T",
    doi = "10.1103/PhysRevLett.48.1220",
    journal = "Phys. Rev. Lett.",
    volume = "48",
    pages = "1220--1223",
    year = "1982"
}

@article{Starobinsky:1982ee,
    author = "Starobinsky, Alexei A.",
    title = "{Dynamics of Phase Transition in the New Inflationary Universe Scenario and Generation of Perturbations}",
    doi = "10.1016/0370-2693(82)90541-X",
    journal = "Phys. Lett. B",
    volume = "117",
    pages = "175--178",
    year = "1982"
}

@article{Maldacena:2002vr,
    author = "Maldacena, Juan Martin",
    title = "{Non-Gaussian features of primordial fluctuations in single field inflationary models}",
    eprint = "astro-ph/0210603",
    archivePrefix = "arXiv",
    doi = "10.1088/1126-6708/2003/05/013",
    journal = "JHEP",
    volume = "05",
    pages = "013",
    year = "2003"
}

@article{Hepp:1966eg,
    author = "Hepp, Klaus",
    title = "{Proof of the Bogolyubov-Parasiuk theorem on renormalization}",
    doi = "10.1007/BF01773358",
    journal = "Commun. Math. Phys.",
    volume = "2",
    pages = "301--326",
    year = "1966"
}

@book{Weinzierl:2022eaz,
    author = "Weinzierl, Stefan",
    title = "{Feynman Integrals. A Comprehensive Treatment for Students and Researchers}",
    eprint = "2201.03593",
    archivePrefix = "arXiv",
    primaryClass = "hep-th",
    reportNumber = "MITP/22-001",
    doi = "10.1007/978-3-030-99558-4",
    isbn = "978-3-030-99557-7, 978-3-030-99560-7, 978-3-030-99558-4",
    publisher = "Springer",
    series = "UNITEXT for Physics",
    year = "2022"
}

@article{Arkani-Hamed:2018kmz,
    author = "Arkani-Hamed, Nima and Baumann, Daniel and Lee, Hayden and Pimentel, Guilherme L.",
    title = "{The Cosmological Bootstrap: Inflationary Correlators from Symmetries and Singularities}",
    eprint = "1811.00024",
    archivePrefix = "arXiv",
    primaryClass = "hep-th",
    doi = "10.1007/JHEP04(2020)105",
    journal = "JHEP",
    volume = "04",
    pages = "105",
    year = "2020"
}

@article{Arkani-Hamed:2015bza,
    author = "Arkani-Hamed, Nima and Maldacena, Juan",
    title = "{Cosmological Collider Physics}",
    eprint = "1503.08043",
    archivePrefix = "arXiv",
    primaryClass = "hep-th",
    month = "3",
    year = "2015"
}

@article{Gibbons:1977mu,
    author = "Gibbons, G. W. and Hawking, S. W.",
    title = "{Cosmological Event Horizons, Thermodynamics, and Particle Creation}",
    doi = "10.1103/PhysRevD.15.2738",
    journal = "Phys. Rev. D",
    volume = "15",
    pages = "2738--2751",
    year = "1977"
}

@article{Bros:1994dn,
    author = "Bros, J. and Moschella, U. and Gazeau, J. P.",
    title = "{Quantum field theory in the de Sitter universe}",
    doi = "10.1103/PhysRevLett.73.1746",
    journal = "Phys. Rev. Lett.",
    volume = "73",
    pages = "1746--1749",
    year = "1994"
}

@article{Bros:1995js,
    author = "Bros, Jacques and Moschella, Ugo",
    title = "{Two point functions and quantum fields in de Sitter universe}",
    eprint = "gr-qc/9511019",
    archivePrefix = "arXiv",
    reportNumber = "SACLAY-SPH-T-94-160",
    doi = "10.1142/S0129055X96000123",
    journal = "Rev. Math. Phys.",
    volume = "8",
    pages = "327--392",
    year = "1996"
}

@article{Chen:2017ryl,
    author = "Chen, Xingang and Wang, Yi and Xianyu, Zhong-Zhi",
    title = "{Schwinger-Keldysh Diagrammatics for Primordial Perturbations}",
    eprint = "1703.10166",
    archivePrefix = "arXiv",
    primaryClass = "hep-th",
    doi = "10.1088/1475-7516/2017/12/006",
    journal = "JCAP",
    volume = "12",
    pages = "006",
    year = "2017"
}

@article{Wang:2021qez,
    author = "Wang, Lian-Tao and Xianyu, Zhong-Zhi and Zhong, Yi-Ming",
    title = "{Precision calculation of inflation correlators at one loop}",
    eprint = "2109.14635",
    archivePrefix = "arXiv",
    primaryClass = "hep-ph",
    doi = "10.1007/JHEP02(2022)085",
    journal = "JHEP",
    volume = "02",
    pages = "085",
    year = "2022"
}

@article{Qin:2023ejc,
    author = "Qin, Zhehan and Xianyu, Zhong-Zhi",
    title = "{Closed-form formulae for inflation correlators}",
    eprint = "2301.07047",
    archivePrefix = "arXiv",
    primaryClass = "hep-th",
    doi = "10.1007/JHEP07(2023)001",
    journal = "JHEP",
    volume = "07",
    pages = "001",
    year = "2023"
}

@article{Herderschee:2025znl,
    author = "Herderschee, Aidan",
    title = "{Feynman-like parameterizations of (anti-)de Sitter Witten diagrams for all masses at any loop order}",
    eprint = "2509.02699",
    archivePrefix = "arXiv",
    primaryClass = "hep-th",
    month = "9",
    year = "2025"
}

@article{Borinsky:2023jdv,
    author = "Borinsky, Michael and Munch, Henrik J. and Tellander, Felix",
    title = "{Tropical Feynman integration in the Minkowski regime}",
    eprint = "2302.08955",
    archivePrefix = "arXiv",
    primaryClass = "hep-ph",
    reportNumber = "DESY-23-026",
    doi = "10.1016/j.cpc.2023.108874",
    journal = "Comput. Phys. Commun.",
    volume = "292",
    pages = "108874",
    year = "2023"
}

@article{Soper:1999xk,
    author = "Soper, Davison E.",
    title = "{Techniques for QCD calculations by numerical integration}",
    eprint = "hep-ph/9910292",
    archivePrefix = "arXiv",
    doi = "10.1103/PhysRevD.62.014009",
    journal = "Phys. Rev. D",
    volume = "62",
    pages = "014009",
    year = "2000"
}

@article{Binoth:2005ff,
    author = "Binoth, T. and Guillet, J. Ph. and Heinrich, G. and Pilon, E. and Schubert, C.",
    title = "{An Algebraic/numerical formalism for one-loop multi-leg amplitudes}",
    eprint = "hep-ph/0504267",
    archivePrefix = "arXiv",
    reportNumber = "LAPTH-1097-05, WUE-ITP-2005-003, ZU-TH-08-05",
    doi = "10.1088/1126-6708/2005/10/015",
    journal = "JHEP",
    volume = "10",
    pages = "015",
    year = "2005"
}

@article{Pittau:2021jbs,
    author = "Pittau, Roberto and Webber, Bryan",
    title = "{Direct numerical evaluation of multi-loop integrals without contour deformation}",
    eprint = "2110.12885",
    archivePrefix = "arXiv",
    primaryClass = "hep-ph",
    doi = "10.1140/epjc/s10052-022-10008-6",
    journal = "Eur. Phys. J. C",
    volume = "82",
    number = "1",
    pages = "55",
    year = "2022"
}

@article{Mizera:2021icv,
    author = "Mizera, Sebastian and Telen, Simon",
    title = "{Landau discriminants}",
    eprint = "2109.08036",
    archivePrefix = "arXiv",
    primaryClass = "math-ph",
    doi = "10.1007/JHEP08(2022)200",
    journal = "JHEP",
    volume = "08",
    pages = "200",
    year = "2022"
}

@article{Borinsky:2020rqs,
    author = "Borinsky, Michael",
    title = "{Tropical Monte Carlo quadrature for Feynman integrals}",
    eprint = "2008.12310",
    archivePrefix = "arXiv",
    primaryClass = "math-ph",
    reportNumber = "Nikhef 2020-027",
    doi = "10.4171/aihpd/158",
    journal = "Ann. Inst. H. Poincare D Comb. Phys. Interact.",
    volume = "10",
    number = "4",
    pages = "635--685",
    year = "2023"
}

@article{Baumann:2011nk,
    author = "Baumann, Daniel and Green, Daniel",
    title = "{Signatures of Supersymmetry from the Early Universe}",
    eprint = "1109.0292",
    archivePrefix = "arXiv",
    primaryClass = "hep-th",
    doi = "10.1103/PhysRevD.85.103520",
    journal = "Phys. Rev. D",
    volume = "85",
    pages = "103520",
    year = "2012"
}

@article{Chen:2009zp,
    author = "Chen, Xingang and Wang, Yi",
    title = "{Quasi-Single Field Inflation and Non-Gaussianities}",
    eprint = "0911.3380",
    archivePrefix = "arXiv",
    primaryClass = "hep-th",
    doi = "10.1088/1475-7516/2010/04/027",
    journal = "JCAP",
    volume = "04",
    pages = "027",
    year = "2010"
}

@article{Noumi:2012vr,
    author = "Noumi, Toshifumi and Yamaguchi, Masahide and Yokoyama, Daisuke",
    title = "{Effective field theory approach to quasi-single field inflation and effects of heavy fields}",
    eprint = "1211.1624",
    archivePrefix = "arXiv",
    primaryClass = "hep-th",
    reportNumber = "UT-KOMABA-12-9, TIT-HEP-625",
    doi = "10.1007/JHEP06(2013)051",
    journal = "JHEP",
    volume = "06",
    pages = "051",
    year = "2013"
}

@article{Lee:2016vti,
    author = "Lee, Hayden and Baumann, Daniel and Pimentel, Guilherme L.",
    title = "{Non-Gaussianity as a Particle Detector}",
    eprint = "1607.03735",
    archivePrefix = "arXiv",
    primaryClass = "hep-th",
    doi = "10.1007/JHEP12(2016)040",
    journal = "JHEP",
    volume = "12",
    pages = "040",
    year = "2016"
}

@article{Chen:2016uwp,
    author = "Chen, Xingang and Wang, Yi and Xianyu, Zhong-Zhi",
    title = "{Standard Model Background of the Cosmological Collider}",
    eprint = "1610.06597",
    archivePrefix = "arXiv",
    primaryClass = "hep-th",
    doi = "10.1103/PhysRevLett.118.261302",
    journal = "Phys. Rev. Lett.",
    volume = "118",
    number = "26",
    pages = "261302",
    year = "2017"
}

@article{Planck:2019kim,
    author = "Akrami, Y. and others",
    collaboration = "Planck",
    title = "{Planck 2018 results. IX. Constraints on primordial non-Gaussianity}",
    eprint = "1905.05697",
    archivePrefix = "arXiv",
    primaryClass = "astro-ph.CO",
    doi = "10.1051/0004-6361/201935891",
    journal = "Astron. Astrophys.",
    volume = "641",
    pages = "A9",
    year = "2020"
}

@article{Chen:2016nrs,
    author = "Chen, Xingang and Wang, Yi and Xianyu, Zhong-Zhi",
    title = "{Loop Corrections to Standard Model Fields in Inflation}",
    eprint = "1604.07841",
    archivePrefix = "arXiv",
    primaryClass = "hep-th",
    doi = "10.1007/JHEP08(2016)051",
    journal = "JHEP",
    volume = "08",
    pages = "051",
    year = "2016"
}

@inbook{Gawrilow:2000qhs,
    author = "Gawrilow, Ewgenij and Joswig, Michael",
    title = "{polymake: a Framework for Analyzing Convex Polytopes}",
    doi = "10.1007/978-3-0348-8438-9_2",
    year = "2000",
    booktitle = {Polytopes --- Combinatorics and Computation},
    publisher = {Birkh\"{a}user},
    pages = {43--73}
}

@article{Baumann:2019oyu,
    author = "Baumann, Daniel and Duaso Pueyo, Carlos and Joyce, Austin and Lee, Hayden and Pimentel, Guilherme L.",
    title = "{The cosmological bootstrap: weight-shifting operators and scalar seeds}",
    eprint = "1910.14051",
    archivePrefix = "arXiv",
    primaryClass = "hep-th",
    doi = "10.1007/JHEP12(2020)204",
    journal = "JHEP",
    volume = "12",
    pages = "204",
    year = "2020"
}

@article{Baumann:2020dch,
    author = "Baumann, Daniel and Duaso Pueyo, Carlos and Joyce, Austin and Lee, Hayden and Pimentel, Guilherme L.",
    title = "{The Cosmological Bootstrap: Spinning Correlators from Symmetries and Factorization}",
    eprint = "2005.04234",
    archivePrefix = "arXiv",
    primaryClass = "hep-th",
    doi = "10.21468/SciPostPhys.11.3.071",
    journal = "SciPost Phys.",
    volume = "11",
    pages = "071",
    year = "2021"
}

@article{Qin:2024gtr,
    author = "Qin, Zhehan",
    title = "{Cosmological correlators at the loop level}",
    eprint = "2411.13636",
    archivePrefix = "arXiv",
    primaryClass = "hep-th",
    doi = "10.1007/JHEP03(2025)051",
    journal = "JHEP",
    volume = "03",
    pages = "051",
    year = "2025"
}

@article{Kumar:2017ecc,
    author = "Kumar, Soubhik and Sundrum, Raman",
    title = "{Heavy-Lifting of Gauge Theories By Cosmic Inflation}",
    eprint = "1711.03988",
    archivePrefix = "arXiv",
    primaryClass = "hep-ph",
    reportNumber = "UMD-PP-017-31",
    doi = "10.1007/JHEP05(2018)011",
    journal = "JHEP",
    volume = "05",
    pages = "011",
    year = "2018"
}

@article{Kumar:2019ebj,
    author = "Kumar, Soubhik and Sundrum, Raman",
    title = "{Cosmological Collider Physics and the Curvaton}",
    eprint = "1908.11378",
    archivePrefix = "arXiv",
    primaryClass = "hep-ph",
    reportNumber = "UMD-PP-019-04",
    doi = "10.1007/JHEP04(2020)077",
    journal = "JHEP",
    volume = "04",
    pages = "077",
    year = "2020"
}

@article{Cheung:2007st,
    author = "Cheung, Clifford and Creminelli, Paolo and Fitzpatrick, A. Liam and Kaplan, Jared and Senatore, Leonardo",
    title = "{The Effective Field Theory of Inflation}",
    eprint = "0709.0293",
    archivePrefix = "arXiv",
    primaryClass = "hep-th",
    reportNumber = "IC-2007-032",
    doi = "10.1088/1126-6708/2008/03/014",
    journal = "JHEP",
    volume = "03",
    pages = "014",
    year = "2008"
}

@article{Creminelli:2004yq,
    author = "Creminelli, Paolo and Zaldarriaga, Matias",
    title = "{Single field consistency relation for the 3-point function}",
    eprint = "astro-ph/0407059",
    archivePrefix = "arXiv",
    reportNumber = "HUTP-04-A032",
    doi = "10.1088/1475-7516/2004/10/006",
    journal = "JCAP",
    volume = "10",
    pages = "006",
    year = "2004"
}

@article{Chen:2006nt,
    author = "Chen, Xingang and Huang, Min-xin and Kachru, Shamit and Shiu, Gary",
    title = "{Observational signatures and non-Gaussianities of general single field inflation}",
    eprint = "hep-th/0605045",
    archivePrefix = "arXiv",
    reportNumber = "SLAC-PUB-11840, MAD-TH-06-3, UFIFT-HEP-06-9, SU-ITP-06-12, CU-TP-1147",
    doi = "10.1088/1475-7516/2007/01/002",
    journal = "JCAP",
    volume = "01",
    pages = "002",
    year = "2007"
}

@article{Chen:2009we,
    author = "Chen, Xingang and Wang, Yi",
    title = "{Large non-Gaussianities with Intermediate Shapes from Quasi-Single Field Inflation}",
    eprint = "0909.0496",
    archivePrefix = "arXiv",
    primaryClass = "astro-ph.CO",
    reportNumber = "MIT-CTP-4071",
    doi = "10.1103/PhysRevD.81.063511",
    journal = "Phys. Rev. D",
    volume = "81",
    pages = "063511",
    year = "2010"
}

@article{Baumann:2017jvh,
    author = "Baumann, Daniel and Goon, Garrett and Lee, Hayden and Pimentel, Guilherme L.",
    title = "{Partially Massless Fields During Inflation}",
    eprint = "1712.06624",
    archivePrefix = "arXiv",
    primaryClass = "hep-th",
    doi = "10.1007/JHEP04(2018)140",
    journal = "JHEP",
    volume = "04",
    pages = "140",
    year = "2018"
}

@article{Bekaert:2017khg,
    author = "Bekaert, Xavier and Skvortsov, Evgeny D.",
    title = "{Elementary particles with continuous spin}",
    eprint = "1708.01030",
    archivePrefix = "arXiv",
    primaryClass = "hep-th",
    doi = "10.1142/S0217751X17300198",
    journal = "Int. J. Mod. Phys. A",
    volume = "32",
    number = "23n24",
    pages = "1730019",
    year = "2017"
}

@article{Pimentel:2025rds,
    author = "Pimentel, Guilherme L. and Yang, Chen",
    title = "{Strongly coupled sectors in inflation: gapless theories and unparticles}",
    eprint = "2503.17840",
    archivePrefix = "arXiv",
    primaryClass = "hep-th",
    doi = "10.1007/JHEP04(2026)146",
    journal = "JHEP",
    volume = "04",
    pages = "146",
    year = "2026"
}

@article{Bodas:2020yho,
    author = "Bodas, Arushi and Kumar, Soubhik and Sundrum, Raman",
    title = "{The Scalar Chemical Potential in Cosmological Collider Physics}",
    eprint = "2010.04727",
    archivePrefix = "arXiv",
    primaryClass = "hep-ph",
    reportNumber = "UMD-PP-020-09",
    doi = "10.1007/JHEP02(2021)079",
    journal = "JHEP",
    volume = "02",
    pages = "079",
    year = "2021"
}

@article{Wang:2025qww,
    author = "Wang, Dong-Gang and Zhang, Bowei",
    title = "{Bootstrapping the cosmological collider with resonant features}",
    eprint = "2505.19066",
    archivePrefix = "arXiv",
    primaryClass = "hep-th",
    doi = "10.1007/JHEP09(2025)122",
    journal = "JHEP",
    volume = "09",
    pages = "122",
    year = "2025"
}

@article{Xianyu:2022jwk,
    author = "Xianyu, Zhong-Zhi and Zhang, Hongyu",
    title = "{Bootstrapping one-loop inflation correlators with the spectral decomposition}",
    eprint = "2211.03810",
    archivePrefix = "arXiv",
    primaryClass = "hep-th",
    doi = "10.1007/JHEP04(2023)103",
    journal = "JHEP",
    volume = "04",
    pages = "103",
    year = "2023"
}

@article{Lyth:1996im,
    author = "Lyth, David H.",
    title = "{What would we learn by detecting a gravitational wave signal in the cosmic microwave background anisotropy?}",
    eprint = "hep-ph/9606387",
    archivePrefix = "arXiv",
    reportNumber = "LANCASTER-TH-9612",
    doi = "10.1103/PhysRevLett.78.1861",
    journal = "Phys. Rev. Lett.",
    volume = "78",
    pages = "1861--1863",
    year = "1997"
}

@article {MR3010271,
    AUTHOR = {Nilsson, Lisa and Passare, Mikael},
     TITLE = {Mellin transforms of multivariate rational functions},
   JOURNAL = {J. Geom. Anal.},
  FJOURNAL = {Journal of Geometric Analysis},
    VOLUME = {23},
      YEAR = {2013},
    NUMBER = {1},
     PAGES = {24--46},
      ISSN = {1050-6926,1559-002X},
   MRCLASS = {44A15 (33C60)},
  MRNUMBER = {3010271},
MRREVIEWER = {Osman\ Y\"urekli},
       DOI = {10.1007/s12220-011-9235-7},
       URL = {https://doi.org/10.1007/s12220-011-9235-7},
}

@article{Binoth:2000ps,
    author = "Binoth, T. and Heinrich, G.",
    title = "{An automatized algorithm to compute infrared divergent multi-loop integrals}",
    eprint = "hep-ph/0004013",
    archivePrefix = "arXiv",
    reportNumber = "LAPTH-789-00, LPT-ORSAY-00-37",
    doi = "10.1016/S0550-3213(00)00429-6",
    journal = "Nucl. Phys. B",
    volume = "585",
    pages = "741--759",
    year = "2000"
}

@article{Arkani-Hamed:2019mrd,
    author = "Arkani-Hamed, Nima and He, Song and Lam, Thomas",
    title = "{Stringy canonical forms}",
    eprint = "1912.08707",
    archivePrefix = "arXiv",
    primaryClass = "hep-th",
    doi = "10.1007/JHEP02(2021)069",
    journal = "JHEP",
    volume = "02",
    pages = "069",
    year = "2021"
}

@article {MR2487491,
    AUTHOR = {Postnikov, Alexander},
     TITLE = {Permutohedra, associahedra, and beyond},
   JOURNAL = {Int. Math. Res. Not. IMRN},
  FJOURNAL = {International Mathematics Research Notices. IMRN},
      YEAR = {2009},
    NUMBER = {6},
     PAGES = {1026--1106},
      ISSN = {1073-7928,1687-0247},
   MRCLASS = {05E30},
  MRNUMBER = {2487491},
       DOI = {10.1093/imrn/rnn153},
       URL = {https://doi.org/10.1093/imrn/rnn153},
}

@article{Kaneko:2009qx,
    author = "Kaneko, Toshiaki and Ueda, Takahiro",
    title = "{A Geometric method of sector decomposition}",
    eprint = "0908.2897",
    archivePrefix = "arXiv",
    primaryClass = "hep-ph",
    doi = "10.1016/j.cpc.2010.04.001",
    journal = "Comput. Phys. Commun.",
    volume = "181",
    pages = "1352--1361",
    year = "2010"
}

@article {MR3189470,
    AUTHOR = {Berkesch, Christine and Forsg{\aa}rd, Jens and Passare, Mikael},
     TITLE = {Euler-{M}ellin integrals and {$A$}-hypergeometric functions},
   JOURNAL = {Michigan Math. J.},
  FJOURNAL = {Michigan Mathematical Journal},
    VOLUME = {63},
      YEAR = {2014},
    NUMBER = {1},
     PAGES = {101--123},
      ISSN = {0026-2285,1945-2365},
   MRCLASS = {44A15 (33C05)},
  MRNUMBER = {3189470},
MRREVIEWER = {L.\ Debnath},
       DOI = {10.1307/mmj/1395234361},
       URL = {https://doi.org/10.1307/mmj/1395234361},
}

@article{Giroux:2026tgd,
    author = "Giroux, Mathieu and Mizera, Sebastian and Salvatori, Giulio",
    title = "{SubTropica}",
    eprint = "2604.20954",
    archivePrefix = "arXiv",
    primaryClass = "hep-th",
    month = "4",
    year = "2026"
}

@article{Salvatori:2024nva,
    author = "Salvatori, Giulio",
    title = "{The Tropical Geometry of Subtraction Schemes}",
    eprint = "2406.14606",
    archivePrefix = "arXiv",
    primaryClass = "hep-th",
    month = "6",
    year = "2024"
}

@article{Chen:2018xck,
    author = "Chen, Xingang and Wang, Yi and Xianyu, Zhong-Zhi",
    title = "{Neutrino Signatures in Primordial Non-Gaussianities}",
    eprint = "1805.02656",
    archivePrefix = "arXiv",
    primaryClass = "hep-ph",
    doi = "10.1007/JHEP09(2018)022",
    journal = "JHEP",
    volume = "09",
    pages = "022",
    year = "2018"
}

@article{Gross:2010qma,
    author = "Gross, Eilam and Vitells, Ofer",
    title = "{Trial factors for the look elsewhere effect in high energy physics}",
    eprint = "1005.1891",
    archivePrefix = "arXiv",
    primaryClass = "physics.data-an",
    doi = "10.1140/epjc/s10052-010-1470-8",
    journal = "Eur. Phys. J. C",
    volume = "70",
    pages = "525--530",
    year = "2010"
}

@article{Planck:2018vyg,
    author = "Aghanim, N. and others",
    collaboration = "Planck",
    title = "{Planck 2018 results. VI. Cosmological parameters}",
    eprint = "1807.06209",
    archivePrefix = "arXiv",
    primaryClass = "astro-ph.CO",
    doi = "10.1051/0004-6361/201833910",
    journal = "Astron. Astrophys.",
    volume = "641",
    pages = "A6",
    year = "2020",
    note = "[Erratum: Astron.Astrophys. 652, C4 (2021)]"
}

@article{Weinberg:2005vy,
    author = "Weinberg, Steven",
    title = "{Quantum contributions to cosmological correlations}",
    eprint = "hep-th/0506236",
    archivePrefix = "arXiv",
    reportNumber = "UTTG-01-05",
    doi = "10.1103/PhysRevD.72.043514",
    journal = "Phys. Rev. D",
    volume = "72",
    pages = "043514",
    year = "2005"
}

@article{Senatore:2009cf,
    author = "Senatore, Leonardo and Zaldarriaga, Matias",
    title = "{On Loops in Inflation}",
    eprint = "0912.2734",
    archivePrefix = "arXiv",
    primaryClass = "hep-th",
    doi = "10.1007/JHEP12(2010)008",
    journal = "JHEP",
    volume = "12",
    pages = "008",
    year = "2010"
}

@article{Bunch:1978yq,
    author = "Bunch, T. S. and Davies, P. C. W.",
    title = "{Quantum Field Theory in de Sitter Space: Renormalization by Point Splitting}",
    doi = "10.1098/rspa.1978.0060",
    journal = "Proc. Roy. Soc. Lond. A",
    volume = "360",
    pages = "117--134",
    year = "1978"
}

@article{Sohn:2024xzd,
    author = "Sohn, Wuhyun and Wang, Dong-Gang and Fergusson, James R. and Shellard, E. P. S.",
    title = "{Searching for cosmological collider in the Planck CMB data}",
    eprint = "2404.07203",
    archivePrefix = "arXiv",
    primaryClass = "astro-ph.CO",
    doi = "10.1088/1475-7516/2024/09/016",
    journal = "JCAP",
    volume = "09",
    pages = "016",
    year = "2024"
}

@article{Kumar:2026ogn,
    author = "Kumar, Soubhik and Lu, Qianshu and Xianyu, Zhong-Zhi and Zhang, Yisong",
    title = "{Cosmological Collider Searches beyond the Hubble Scale with Planck Data}",
    eprint = "2603.15728",
    archivePrefix = "arXiv",
    primaryClass = "hep-ph",
    month = "3",
    year = "2026"
}

@article{Kumar:2026dih,
    author = "Kumar, Soubhik and Lu, Qianshu and Xianyu, Zhong-Zhi and Zhang, Yisong",
    title = "{Scalars at the Cosmological Collider: Full Shapes of Tree Diagrams and Bispectrum Searches using Planck Data}",
    eprint = "2604.07434",
    archivePrefix = "arXiv",
    primaryClass = "hep-ph",
    month = "4",
    year = "2026"
}

@article{Suman:2025tpv,
    author = "Suman, Petar and Wang, Dong-Gang and Sohn, Wuhyun and Fergusson, James R. and Shellard, E. P. S.",
    title = "{Searching for Cosmological Collider in the Planck CMB Data II: collider templates and Modal analysis}",
    eprint = "2512.22085",
    archivePrefix = "arXiv",
    primaryClass = "astro-ph.CO",
    month = "12",
    year = "2025"
}

@article{Zhang:2025nzd,
    author = "Zhang, Hongyu",
    title = "{Dimensional regularization of bubble diagrams in de Sitter spacetime}",
    eprint = "2507.19318",
    archivePrefix = "arXiv",
    primaryClass = "hep-th",
    doi = "10.1007/JHEP02(2026)119",
    journal = "JHEP",
    volume = "02",
    pages = "119",
    year = "2026"
}

@article{Aoki:2026vbc,
    author = "Aoki, Shuntaro and Qin, Zhehan and Yamaguchi, Masahide and Zhu, Yuhang",
    title = "{Fermionic Bubble Loop in Cosmological Collider Revisited: Exact signals from spectral and Mellin-Barnes methods}",
    eprint = "2605.28054",
    archivePrefix = "arXiv",
    primaryClass = "hep-th",
    reportNumber = "RIKEN-iTHEMS-Report-26",
    month = "5",
    year = "2026"
}

@article{Bodas:2025vpb,
    author = "Bodas, Arushi and Broadberry, Edward and Sundrum, Raman and Xu, Zhaohui",
    title = "{Charged loops at the cosmological collider with chemical potential}",
    eprint = "2507.22978",
    archivePrefix = "arXiv",
    primaryClass = "hep-ph",
    reportNumber = "FERMILAB-PUB-25-0519-V",
    doi = "10.1007/JHEP01(2026)083",
    journal = "JHEP",
    volume = "01",
    pages = "083",
    year = "2026"
}

@article{Sleight:2019mgd,
    author = "Sleight, Charlotte",
    title = "{A Mellin Space Approach to Cosmological Correlators}",
    eprint = "1906.12302",
    archivePrefix = "arXiv",
    primaryClass = "hep-th",
    doi = "10.1007/JHEP01(2020)090",
    journal = "JHEP",
    volume = "01",
    pages = "090",
    year = "2020"
}

@article{Sleight:2019hfp,
    author = "Sleight, Charlotte and Taronna, Massimo",
    title = "{Bootstrapping Inflationary Correlators in Mellin Space}",
    eprint = "1907.01143",
    archivePrefix = "arXiv",
    primaryClass = "hep-th",
    reportNumber = "PUPT-2590",
    doi = "10.1007/JHEP02(2020)098",
    journal = "JHEP",
    volume = "02",
    pages = "098",
    year = "2020"
}

@article{Meerburg:2016zdz,
    author = {Meerburg, P. Daniel and M{\"u}nchmeyer, Moritz and Mu{\~n}oz, Julian B. and Chen, Xingang},
    title = "{Prospects for Cosmological Collider Physics}",
    eprint = "1610.06559",
    archivePrefix = "arXiv",
    primaryClass = "astro-ph.CO",
    doi = "10.1088/1475-7516/2017/03/050",
    journal = "JCAP",
    volume = "03",
    pages = "050",
    year = "2017"
}

@article{Arkani-Hamed:2022cqe,
    author = "Arkani-Hamed, Nima and Hillman, Aaron and Mizera, Sebastian",
    title = "{Feynman polytopes and the tropical geometry of UV and IR divergences}",
    eprint = "2202.12296",
    archivePrefix = "arXiv",
    primaryClass = "hep-th",
    doi = "10.1103/PhysRevD.105.125013",
    journal = "Phys. Rev. D",
    volume = "105",
    number = "12",
    pages = "125013",
    year = "2022"
}

@article{Qin:2022fbv,
    author = "Qin, Zhehan and Xianyu, Zhong-Zhi",
    title = "{Helical inflation correlators: partial Mellin-Barnes and bootstrap equations}",
    eprint = "2208.13790",
    archivePrefix = "arXiv",
    primaryClass = "hep-th",
    doi = "10.1007/JHEP04(2023)059",
    journal = "JHEP",
    volume = "04",
    pages = "059",
    year = "2023"
}

@article{Qin:2023nhv,
    author = "Qin, Zhehan and Xianyu, Zhong-Zhi",
    title = "{Nonanalyticity and on-shell factorization of inflation correlators at all loop orders}",
    eprint = "2308.14802",
    archivePrefix = "arXiv",
    primaryClass = "hep-th",
    doi = "10.1007/JHEP01(2024)168",
    journal = "JHEP",
    volume = "01",
    pages = "168",
    year = "2024"
}

@article{Baumann:2024mvm,
    author = "Baumann, Daniel and Goodhew, Harry and Lee, Hayden",
    title = "{Kinematic flow for cosmological loop integrands}",
    eprint = "2410.17994",
    archivePrefix = "arXiv",
    primaryClass = "hep-th",
    doi = "10.1007/JHEP07(2025)131",
    journal = "JHEP",
    volume = "07",
    pages = "131",
    year = "2025"
}

@article{Arkani-Hamed:2023kig,
    author = "Arkani-Hamed, Nima and Baumann, Daniel and Hillman, Aaron and Joyce, Austin and Lee, Hayden and Pimentel, Guilherme L.",
    title = "{Differential equations for cosmological correlators}",
    eprint = "2312.05303",
    archivePrefix = "arXiv",
    primaryClass = "hep-th",
    doi = "10.1007/JHEP09(2025)009",
    journal = "JHEP",
    volume = "09",
    pages = "009",
    year = "2025"
}

@article{DiPietro:2021sjt,
    author = "Di Pietro, Lorenzo and Gorbenko, Victor and Komatsu, Shota",
    title = "{Analyticity and unitarity for cosmological correlators}",
    eprint = "2108.01695",
    archivePrefix = "arXiv",
    primaryClass = "hep-th",
    reportNumber = "CERN-TH-2021-118",
    doi = "10.1007/JHEP03(2022)023",
    journal = "JHEP",
    volume = "03",
    pages = "023",
    year = "2022"
}

@article{Hogervorst:2021uvp,
    author = "Hogervorst, Matthijs and Penedones, Jo{\~a}o and Vaziri, Kamran Salehi",
    title = "{Towards the non-perturbative cosmological bootstrap}",
    eprint = "2107.13871",
    archivePrefix = "arXiv",
    primaryClass = "hep-th",
    doi = "10.1007/JHEP02(2023)162",
    journal = "JHEP",
    volume = "02",
    pages = "162",
    year = "2023"
}

@article{Loparco:2023rug,
    author = "Loparco, Manuel and Penedones, Joao and Salehi Vaziri, Kamran and Sun, Zimo",
    title = {{The K{\"a}ll{\'e}n-Lehmann representation in de Sitter spacetime}},
    eprint = "2306.00090",
    archivePrefix = "arXiv",
    primaryClass = "hep-th",
    doi = "10.1007/JHEP12(2023)159",
    journal = "JHEP",
    volume = "12",
    pages = "159",
    year = "2023"
}

@article{Bhowmick:2025mxh,
    author = "Bhowmick, Supritha and Lee, Mang Hei Gordon and Ghosh, Diptimoy and Ullah, Farman",
    title = "{Singularities in cosmological loop correlators}",
    eprint = "2503.21880",
    archivePrefix = "arXiv",
    primaryClass = "hep-th",
    doi = "10.1007/JHEP02(2026)116",
    journal = "JHEP",
    volume = "02",
    pages = "116",
    year = "2026"
}

@article{Jain:2025maa,
    author = "Jain, Diksha and Pajer, Enrico and Tong, Xi",
    title = "{Unitary and Analytic Renormalisation of Cosmological Correlators}",
    eprint = "2509.02696",
    archivePrefix = "arXiv",
    primaryClass = "hep-th",
    doi = "10.1007/JHEP03(2026)225",
    journal = "JHEP",
    volume = "03",
    pages = "225",
    year = "2026"
}

@article{Liu:2026jzn,
    author = "Liu, Haoyuan and Qin, Zhehan and Wu, Jiayi and Xianyu, Zhong-Zhi and Zhang, Hongyu",
    title = "{On-Shell Bootstrap of Loop Inflation Correlators with Spectral Dispersion}",
    eprint = "2606.02686",
    archivePrefix = "arXiv",
    primaryClass = "hep-th",
    month = "6",
    year = "2026"
}

@article{Belrhali:2026ygh,
    author = "Belrhali, Nathan and Poisson, Arthur and Renaux-Petel, S{\'e}bastien",
    title = "{Laplace Space for Cosmological Correlators}",
    eprint = "2606.27309",
    archivePrefix = "arXiv",
    primaryClass = "hep-th",
    month = "6",
    year = "2026"
}

@article{Belrhali:2026jqe,
    author = "Belrhali, Nathan and Poisson, Arthur and Renaux-Petel, S{\'e}bastien",
    title = "{Massive Cosmological Correlators from Flat Space: a Laplace-Space Approach}",
    eprint = "2606.27311",
    archivePrefix = "arXiv",
    primaryClass = "hep-th",
    month = "6",
    year = "2026"
}

@article{Cardona:2017tsw,
    author = "Cardona, Carlos",
    title = "{Mellin-(Schwinger) representation of One-loop Witten diagrams in AdS}",
    eprint = "1708.06339",
    archivePrefix = "arXiv",
    primaryClass = "hep-th",
    reportNumber = "NCTS-TH-1713",
    month = "8",
    year = "2017"
}

@article{Heckelbacher:2022hbq,
    author = "Heckelbacher, Till and Sachs, Ivo and Skvortsov, Evgeny and Vanhove, Pierre",
    title = "{Analytical evaluation of cosmological correlation functions}",
    eprint = "2204.07217",
    archivePrefix = "arXiv",
    primaryClass = "hep-th",
    reportNumber = "IPhT-t22/02, LMU-ASC 13/22",
    doi = "10.1007/JHEP08(2022)139",
    journal = "JHEP",
    volume = "08",
    pages = "139",
    year = "2022"
}

@article{Lu:2019tjj,
    author = "Lu, Shiyun and Wang, Yi and Xianyu, Zhong-Zhi",
    title = "{A Cosmological Higgs Collider}",
    eprint = "1907.07390",
    archivePrefix = "arXiv",
    primaryClass = "hep-th",
    doi = "10.1007/JHEP02(2020)011",
    journal = "JHEP",
    volume = "02",
    pages = "011",
    year = "2020"
}

@article{Chen:2016hrz,
    author = "Chen, Xingang and Wang, Yi and Xianyu, Zhong-Zhi",
    title = "{Standard Model Mass Spectrum in Inflationary Universe}",
    eprint = "1612.08122",
    archivePrefix = "arXiv",
    primaryClass = "hep-th",
    doi = "10.1007/JHEP04(2017)058",
    journal = "JHEP",
    volume = "04",
    pages = "058",
    year = "2017"
}

@article{Assassi:2013gxa,
    author = "Assassi, Valentin and Baumann, Daniel and Green, Daniel and McAllister, Liam",
    title = "{Planck-Suppressed Operators}",
    eprint = "1304.5226",
    archivePrefix = "arXiv",
    primaryClass = "hep-th",
    doi = "10.1088/1475-7516/2014/01/033",
    journal = "JCAP",
    volume = "01",
    pages = "033",
    year = "2014"
}

@article{Kumar:2018jxz,
    author = "Kumar, Soubhik and Sundrum, Raman",
    title = "{Seeing Higher-Dimensional Grand Unification In Primordial Non-Gaussianities}",
    eprint = "1811.11200",
    archivePrefix = "arXiv",
    primaryClass = "hep-ph",
    reportNumber = "UMD-PP-018-09",
    doi = "10.1007/JHEP04(2019)120",
    journal = "JHEP",
    volume = "04",
    pages = "120",
    year = "2019"
}

@article{Wang:2019gbi,
    author = "Wang, Lian-Tao and Xianyu, Zhong-Zhi",
    title = "{In Search of Large Signals at the Cosmological Collider}",
    eprint = "1910.12876",
    archivePrefix = "arXiv",
    primaryClass = "hep-ph",
    doi = "10.1007/JHEP02(2020)044",
    journal = "JHEP",
    volume = "02",
    pages = "044",
    year = "2020"
}

@article{Gelfand:1990bua,
    author = "Gelfand, I. M. and Kapranov, M. M. and Zelevinsky, A. V.",
    title = "{Generalized Euler integrals and A-hypergeometric functions}",
    doi = "10.1016/0001-8708(90)90048-R",
    journal = "Adv. Math.",
    volume = "84",
    number = "2",
    pages = "255--271",
    year = "1990"
}

@article{Matsubara-Heo:2023ylc,
    author = "Matsubara-Heo, Saiei-Jaeyeong and Mizera, Sebastian and Telen, Simon",
    title = "{Four lectures on Euler integrals}",
    eprint = "2306.13578",
    archivePrefix = "arXiv",
    primaryClass = "math-ph",
    doi = "10.21468/SciPostPhysLectNotes.75",
    journal = "SciPost Phys. Lect. Notes",
    volume = "75",
    pages = "1",
    year = "2023"
}

@article{Grimm:2024tbg,
    author = "Grimm, Thomas W. and Hoefnagels, Arno",
    title = "{Reductions of GKZ systems and applications to cosmological correlators}",
    eprint = "2409.13815",
    archivePrefix = "arXiv",
    primaryClass = "hep-th",
    doi = "10.1007/JHEP04(2025)196",
    journal = "JHEP",
    volume = "04",
    pages = "196",
    year = "2025"
}

@article{Chernikov:1968zm,
    author = "Chernikov, N. A. and Tagirov, E. A.",
    title = "{Quantum theory of scalar field in de Sitter space-time}",
    journal = "Ann. Inst. H. Poincare Phys. Theor. A",
    volume = "9",
    number = "2",
    pages = "109--141",
    year = "1968"
}

@article{Borinsky:2025asc,
    author = "Borinsky, Michael and Fraaije, Mathijs",
    title = "{Tropical sampling from Feynman measures}",
    eprint = "2504.09613",
    archivePrefix = "arXiv",
    primaryClass = "hep-th",
    doi = "10.1016/j.cpc.2025.109846",
    journal = "Comput. Phys. Commun.",
    volume = "317",
    pages = "109846",
    year = "2025"
}

@article{Pimentel:2026kqc,
    author = "Pimentel, Guilherme L. and Westerdijk, Tom",
    title = "{On Cosmological Correlators at One Loop}",
    eprint = "2601.00952",
    archivePrefix = "arXiv",
    primaryClass = "hep-th",
    month = "1",
    year = "2026"
}

@article{Henn:2026lfz,
    author = "Henn, Johannes and Mei, Jiajie and Yang, Qinglin",
    title = "{A compact analytic formula for the one-loop triangle cosmological correlator}",
    eprint = "2608.17877",
    archivePrefix = "arXiv",
    primaryClass = "hep-th",
    reportNumber = "MPP-2026-138",
    month = "8",
    year = "2026"
}

@article{Kalaja:2020mkq,
    author = "Kalaja, Alba and Meerburg, P. Daniel and Pimentel, Guilherme L. and Coulton, William R.",
    title = "{Fundamental limits on constraining primordial non-Gaussianity}",
    eprint = "2011.09461",
    archivePrefix = "arXiv",
    primaryClass = "astro-ph.CO",
    doi = "10.1088/1475-7516/2021/04/050",
    journal = "JCAP",
    volume = "04",
    pages = "050",
    year = "2021"
}

@article{vonManteuffel:2014qoa,
    author = "von Manteuffel, Andreas and Panzer, Erik and Schabinger, Robert M.",
    title = "{A quasi-finite basis for multi-loop Feynman integrals}",
    eprint = "1411.7392",
    archivePrefix = "arXiv",
    primaryClass = "hep-ph",
    reportNumber = "MITP-14-076",
    doi = "10.1007/JHEP02(2015)120",
    journal = "JHEP",
    volume = "02",
    pages = "120",
    year = "2015"
}

\end{document}